\documentclass[review]{elsarticle}

\usepackage{lineno,hyperref}
\modulolinenumbers[5]

\journal{Icarus}

\usepackage{tikz}

\biboptions{authoryear}

\usepackage[super]{nth}
\usepackage{fancyhdr}

\begin{document}

\begin{frontmatter}

\title{Star catalog position and proper motion corrections in
asteroid astrometry II: The Gaia era}


\author[JPL]{Siegfried Eggl\corref{mycorrespondingauthor}\fnref{uw}}
\cortext[mycorrespondingauthor]{Corresponding author}
\ead{eggl@uw.edu}

\author[JPL]{Davide Farnocchia}
\author[JPL]{Alan B. Chamberlin}
\author[JPL]{Steven R. Chesley}

\address[JPL]{Jet Propulsion Laboratory, California Institute of Technology, 4800 Oak Grove Drive, 91101 Pasadena, CA, USA}
\fntext[uw]{Present address: LSST/DIRAC Institute, Department of Astronomy, University of Washington, 15th Ave NE, Seattle, WA 98195, USA}




\begin{abstract}
Astrometric positions of moving objects in the Solar System have been measured using a variety of star catalogs in the past.
Previous work has shown that systematic errors in star catalogs can affect the accuracy of astrometric observations. That, in turn, can influence the resulting orbit fits for minor planets. In order to quantify these systematic errors, we compare the positions and proper motion of stellar sources in the most utilized star catalogs to the second release of the Gaia star catalog. The accuracy of Gaia astrometry allows us to unambiguously identify local biases and derive a scheme that can be used to correct past astrometric observations of solar system objects.
Here we provide a substantially improved debiasing scheme for 26 astrometric catalogs that were extensively used in minor planet astrometry.
Revised corrections near the galactic center eliminate artifacts that could be traced back to reference catalogs used in previous debiasing schemes.
Median differences in stellar positions between catalogs now tend to be on the order of several tens of milliarcseconds (mas) but can be as large as 175 mas. Median stellar proper motion corrections scatter around 0.3 mas/yr and range from 1--4 mas/yr for star catalogs with and without proper motion, respectively.
The tables in this work are meant to be applied to existing optical observations. They are not intended to correct new astrometric measurments as those should make use of the Gaia astrometric catalog.
Since previous debiasing schemes already reduced systematics in past observations to a large extent, corrections beyond the current work may not be needed in the foreseeable future.

.   %Zonal differences are .
% Correcting past observations based on Gaia DR 2 will only slightly improve our ability to predict the future whereabouts of Solar System objects.
\end{abstract}


\end{frontmatter}
%%%
%\linenumbers
%%%
\section{Introduction}
Over more than two centuries astronomers have collected approximately 200 million astrometric observations of minor planets in the Solar System\footnote{Minor Planet Center (MPC), http://www.minorplanetcenter.net/}.
This dataset %trove 
of astrometric measurements is essential to estimate the trajectories of Solar System Objects (SSOs) such as near-Earth asteroids \citep{milani10}. The vast majority of astrometry used for orbit determination of SSOs is optical, i.e. pairs of Right Ascension (RA) and Declination (DEC) measurements at given epochs. In contrast, only a tiny fraction of the SSO population has been observed by radar\footnote{About one in a thousand asteroids, see NASA Jet Propulsion Laboratory (JPL), https://echo.jpl.nasa.gov/asteroids/index.html/}. Improving the quality of optical astrometry is a worthwhile endeavor, especially if that means extending the arc of observations for objects of interest.  
Over the years a variety of star catalogs has been used to measure astrometric positions of asteroids to a common frame of reference such as the ICRF \citep{icrf2} (Table \ref{tab:catalogs}).
With improving quality of astrometric measurements it became evident that some observations of asteroids showed systematic errors that deviated from the expected scatter around the nominal trajectory \citep{carpino03}.  
Some of those systematic errors could be traced back to the use of specific astrometric catalogs in the observation reduction process. To assess these systematic errors one can compare stellar positions and proper motion between various star catalogs. 
\cite{chesley10} used the 2MASS \citep{2mass} catalog as a reference to identify local systematics in positions of stars in the USNO catalogs \citep{usnoa2,usnoa1,usnob1}. Correcting for regional biases in star positions led to better ephemeris predictions and improved the error statistics of astrometric observations of asteroids.

Selecting the most accurate stars in the PPM XL catalog based on 2MASS astrometry, \cite{farnocchia15} improved the scheme by \cite{chesley10} and debiased practically all of the prevalent star catalogs used for asteroid astrometry. \cite{farnocchia15} included proper motion in the debiasing process which led to widespread improvements in the post-fit residuals of asteroids, most notably (99942) Apophis, (101955) Bennu and (6489) Golevka. 
The present work follows a similar approach with the advantage of having Gaia data, described in section \ref{sec:gaiadr2}, as a reference. 
Following a brief outline of the debiasing process (section \ref{sec:debias}) we calculate position and proper motion corrections for stellar catalogs that have been used extensively in minor planet astrometry (sections \ref{sec:cats} - \ref{sec:interp}). 
%The resulting catalog corrections are available at %\url{ftp://ssd.jpl.nasa.gov/pub/ssd/debias/debias_2018.tgz} and %\url{ftp://ssd.jpl.nasa.gov/pub/ssd/debias/debias_hires_2018.tgz}. 
In section \ref{sec:eph} we then assess the quality of ephemeris predictions based on the new debiasing scheme.
A discussion of our results concludes this article.
\begin{table}\tiny
\begin{center}
\begin{tabular}{llccccc}
\hline
Catalog & \multicolumn{2}{c}{MPC designation}  & \multicolumn{2}{c}{Asteroid observations} & VizieR & Reference\\
             & ADES  & flag & count &  \%   &  CDS ID &\\
\hline
USNO A2.0   & USNOA2  & c  & 40,964,725 & 21.929 & I/252   &\citet{usnoa2}\\
2MASS       & 2MASS  & L  & 36,656,505 & 19.623 &  II/246   & \citet{2mass}\\
UCAC 2      &  UCAC2 & r  & 30,808,742 & 16.493 & I/289    &\citet{ucac2}\\
UCAC 4      & UCAC4  & q  & 25,280,408 & 13.533 & I/322A       &\citet{ucac4}\\
USNO B1.0   & USNOB1  & o  & 16,827,916 & 9.008 & I/284       &\citet{usnob1}\\
SST RC4     &  SSTRC4 & R  & 14,143,319 & 7.571 & N/A       & \citet{sstrc5} \\
Gaia DR 1    & Gaia1  & U  & 7,878,286 & 4.217 & I/337         &\citet{gaiadr1}\\
UCAC 3      &  UCAC3 & u  & 3,460,491 & 1.852 & I/315         &\citet{ucac3}\\
USNO A1.0   & USNOA1 & a  & 2,196,215 & 1.175 & N/A         &\citet{usnoa1}\\
Unspecified     & N/A & N/A & 1,995,344 & 1.068 & N/A       &N/A\\
USNO SA2.0  & USNOSA2 & d  & 1,757,923 & 0.941 & N/A        &\citet{usnoa2}\\
SDSS DR 7    & SDSS7  & N  & 985,016 & 0.527 & II/294          &\citet{sdss7}\\
GSC 1.1     &  GSC1.1 & i  & 645,795 & 0.345 & I/220          &\citet{gsc1.1}\\
UCAC 1      & UCAC1  & e  & 516,449 & 0.276 &   I/268        &\citet{ucac1}\\
GSC ACT     & GSCACT  & m  & 474,122 & 0.253 & I/255         &\citet{gsc_act}\\
CMC 14      & CMC14  & w  & 426,781 & 0.228 & I/304          &\citet{cmc14}\\
Tycho 2     & Tyc2  & g  & 414,400 & 0.221 &  I/259          &\citet{tycho2}\\
USNO SA1.0  & USNOSA1  & b  & 349,099 & 0.186 & N/A        &\citet{usnoa1}\\
PPM XL       & PPM XL  & t  & 302,432 & 0.162 & I/317         &\citet{ppmxl}\\
GSC (unspec.)& GSC & z & 292,814 & 0.156    & N/A          &N/A\\
ACT         & ACT  & l & 112,431 & 0.060 & I/246             &\citet{act}\\
Other catalogs & N/A &N/A& 97,256 & $0.052$ & N/A & N/A\\
NOMAD       & NOMAD  & v  & 82,997 & 0.044 & N/A           &\citet{nomad}\\
PPM         & PPM  & p & 47,227 & 0.025 & I/146,I/193              &\citet{ppm}\\
%Yale        &  Yale & K & 37,672 & 0.020 & I/238A             &\citet{yale}\\
URAT 1      & URAT1  & S & 37,105 & 0.019 & I/329            &\citet{urat1}\\
%SAO         & SAO  & C & 28,964 & 0.015 & I/131A              &\citet{sao}\\
GSC 1.2     & GSC1.2& j  & 17,269 & 0.009 & I/254            &\citet{gsc1.2}\\
%MPOSC3      & MPOSC3  & P & 14,757 & 0.008 &  N/A          &\citet{mposc3}\\
CMC 15      & CMC15& Q  & 11,869 & 0.006 &  I/327            &\citet{cmc15}\\
%AC          & AC  & A  & 7,071 & 0.004 & I/96                &\citet{ac}\\
SDSS DR 8    & SDSS8  & n &  6,157 & 0.003 & II/306            &\citet{sdss8}\\
%AGK 3        & AGK3  & D &  4,659 & 0.002 &  I/61B            &\citet{agk3}\\
UCAC 5   & UCAC5 & W & 1,618 & $<$0.001 & I/340 & \citet{ucac5} \\
%Hipparcos 2& Hip2 & x & 118 & $<$0.001 & I/311 & \citet{hip2}  \\
%GSC-2.2 & k & & 745 & 0.00 & \citet{gsc2.2}\\
%GSC-2.3 & M & & 86 & 0.00 & \citet{gsc2.3}\\

\hline
\end{tabular}
\end{center}
\caption{Star catalogs considered in this work. This table accounts for all the astrometry
  available up to March 23, 2018.}
\label{tab:catalogs}
\end{table}
%%%
\section{The Gaia catalogs} 
\label{sec:gaiadr2}
Since 2013 the Gaia mission has been collecting astrometric information on roughly 1.7 billion sources \citep{gaiadr2}. A first data release based on a partially completed survey featured around a billion sources largely without proper motion \citep{gaiadr1}.
With the second Gaia data release (Gaia DR 2) published on April 25, 2018 this information has become publicly accessible through the Gaia Archive\footnote{Gaia Archive, https://gea.esac.esa.int/archive/}. 
%The high uniformity and excellent quality of Gaia DR 2 parallaxes and proper motions could be achieved by relying exclusively on data from the Gaia satellite. 
About 1.3 billion sources between 3$<$Gmag$<$21 come with a full five-parameter astrometric solution, i.e. right ascension and declination positions on the sky $(\alpha,\delta)$, parallaxes, and proper motion. Uncertainties in the parallaxes range between 0.04 milliarcseconds (mas) for sources with Gmag$<$15 and 0.7 mas at Gmag$=$20. The corresponding uncertainties in the proper motion components start at 0.06 mas/yr for Gmag$<$15 and reach 1.2 mas/yr for Gmag$=$20. The Gaia astrometric catalog is, thus, a reference of unprecedented accuracy well suited to identify systematic errors in other star catalogs.
%%%
\section{Star position and proper motion corrections} 
\label{sec:debias}
Similar to \citet{chesley10} and \citet{farnocchia15} we compare the various star catalogs that were used for minor planet astrometry in the past to a reference catalog, in this case Gaia DR 2. 
% Potential catalog systematics are known to vary across the sky \citep{farnocchia15}. (XXX I would remove this sentence, also this was known to Chesley et al. 2010 too XXX)
By dividing the celestial sphere into 49,152
equal-area tiles ($\sim$0.8 square degrees each) using a HEALPix tesselation
\citep{healpix} and comparing stellar positions and proper motion within a given tile we can quantify local systematic differences.
To this end we extract stellar positions and proper motion at epoch J2000 from the reference and the test catalog.
Given the variability in the number of stars per tile in each catalog, identification of common stars requires some attention. We first search for spatial correlation between astrometric sources within  1'' of the reference position. 
If candidates are found we apply the identification and outlier rejection criterion described in \citet{farnocchia15} to
ensure spurious matches are excluded from the sample.
The median of the differences in the astrometric quantities of each identified star per tile yields the local corrections for position $(\Delta \alpha_{J2000}, \Delta \delta_{J2000})$, and proper motion $(\Delta \mu_{\alpha}, \Delta \mu_{\delta}$) at epoch J2000.
Astrometric observations of minor planets can then be corrected for catalog systematics by subtracting
 \begin{eqnarray*}
\Delta\alpha & = & \Delta \alpha_{J2000} + \Delta \mu_{\alpha} \cdot \Delta t,\\
\Delta\delta & = & \Delta \delta_{J2000} + \Delta \mu_{\delta} \cdot \Delta t,\\
\end{eqnarray*}
with $\Delta t=t - 2451545.0$, the difference between the epoch of observation and J2000 in JD.
All differential quantities $\Delta{\alpha}$, $\Delta{\alpha}_{J2000}$, and $\Delta \mu_{\alpha}$ include the spherical metric factor $\cos\delta$.
%
%In the following sections we briefly introduce the set of astrometric catalogs that has not been discussed in \citet{farnocchia15}.
%%%
\section{Added astrometric catalogs} 
\label{sec:cats}
In addition to the 20 catalogs discussed in \citet{farnocchia15} and Gaia DR 1 we added the following five catalogs to be compared to Gaia DR 2:
% CMC 15, %Hipparcos 2, 
% SDSS DR 8, SST RC 5, UCAC 5, URAT 1  and PPM XL itself to the comparison with Gaia DR 2.
\begin{itemize}
\item CMC 15 is the \nth{15} installment of the Carlsberg Meridian Catalogue generated through the Carlsberg Meridian Telescope, formerly the Carlsberg Automatic Meridian Circle \citep{cmc15}. CMC 15 has been compiled between March 1999 and March 2011, encompasses around 134,000,000 entries, and claims a positional accuracy between 35 to 100 mas for a declination range from -40 and +50 degrees.
% Hipparcos 2 was a rereduction of the Hipparcos catalog published in 1997 providing a substantial improvement - about a factor of two - on the already excellent astrometry (target accuracy 2 mas) delivered by ESA's Hipparcos mission \citep{hip2}. One of the limitations of the Hippachos 2 catalog, and the main reason why it has not been more popular in minor planet astrometry, is its relatively small size. Hipparcos 2 contains merely 117,955 sources.    

\item The \nth{8} data release of the Sloan Digital Sky Survey (SDSS) added roughly 4 times the number of sources published in data release 7, namely about 325 million \citep{sdss8}. The fact that astrometric data together with proper motion is available has made SDSS DR 7 \citep{sdss7} attractive for minor planet astrometry. Almost 1 million observations of minor planets were reduced with SDSS DR 7 (Table \ref{tab:catalogs}). SDSS DR 8, although offering a significant increase in source density, has not been as popular, which is partly due to known issues with the astrometry \citep{sdss8_astrometry}.         

\item The Space Surveillance Telescope (SST) has been relying on a proprietary catalog based on selected sources from a variety of other astrometric and photometric catalogs for the reduction of its 14+ million observations of asteroids in the Solar System. Although not in the public domain, the \nth{5} version of the SST catalog (SST-RC5) has been kindly made available for this study \citep{sstrc5}. Most of the astrometric observations considered in this work have been submitted under SST-RC4. Astrometric standard stars in SST-RC5 are largely the same as in the \nth{4} installment of the SST catalog, however \citep{sstrc5}. Hence, we use the same corrections to debias astrometric obsevations submitted under SST-RC4 and SST-RC5. 

\item UCAC 5, is the \nth{5} installment of the US Naval Observatory CCD Astrograph Catalog (UCAC) \citep{ucac5}. By combining UCAC 5 with Gaia DR1 data new proper motions on the Gaia coordinate system for over 107 million stars were obtained with typical accuracies of 1--2 mas/yr for R=11--15 mag, and about 5 mas/yr through R=16 mag.

\item URAT, a follow-up project to the UCAC series featured and increased limiting magnitude that allowed for a roughly 4-fold increase in the average number of stars per square degree as compared to UCAC. Additionally, stars as bright as about \nth{3} magnitude were added to the catalog. The URAT 1 catalog \citep{urat1} has its mean epoch between 2012.3 and 2014.6 supporting a magnitude range from 3 to 18.5 in R-band with a positional precision of 5 to 40 mas. It covers most of the northern hemisphere and some areas down to -24.8 deg in declination.
\end{itemize}
%%%
\section{Catalog Comparison Results}
\begin{figure}
\begin{tabular}{ll}
\includegraphics[width=0.5\linewidth]{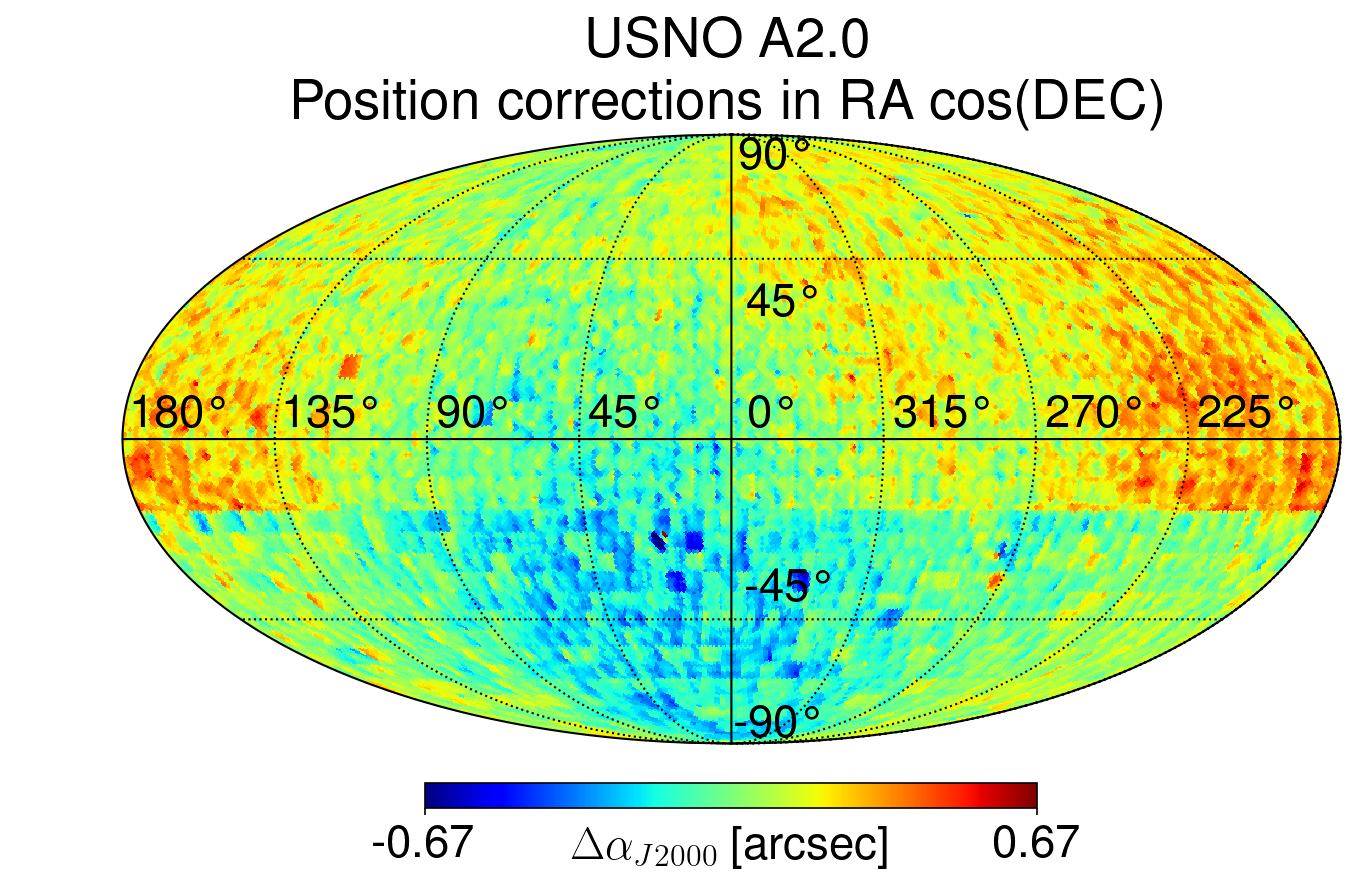} &
\includegraphics[width=0.5\linewidth]{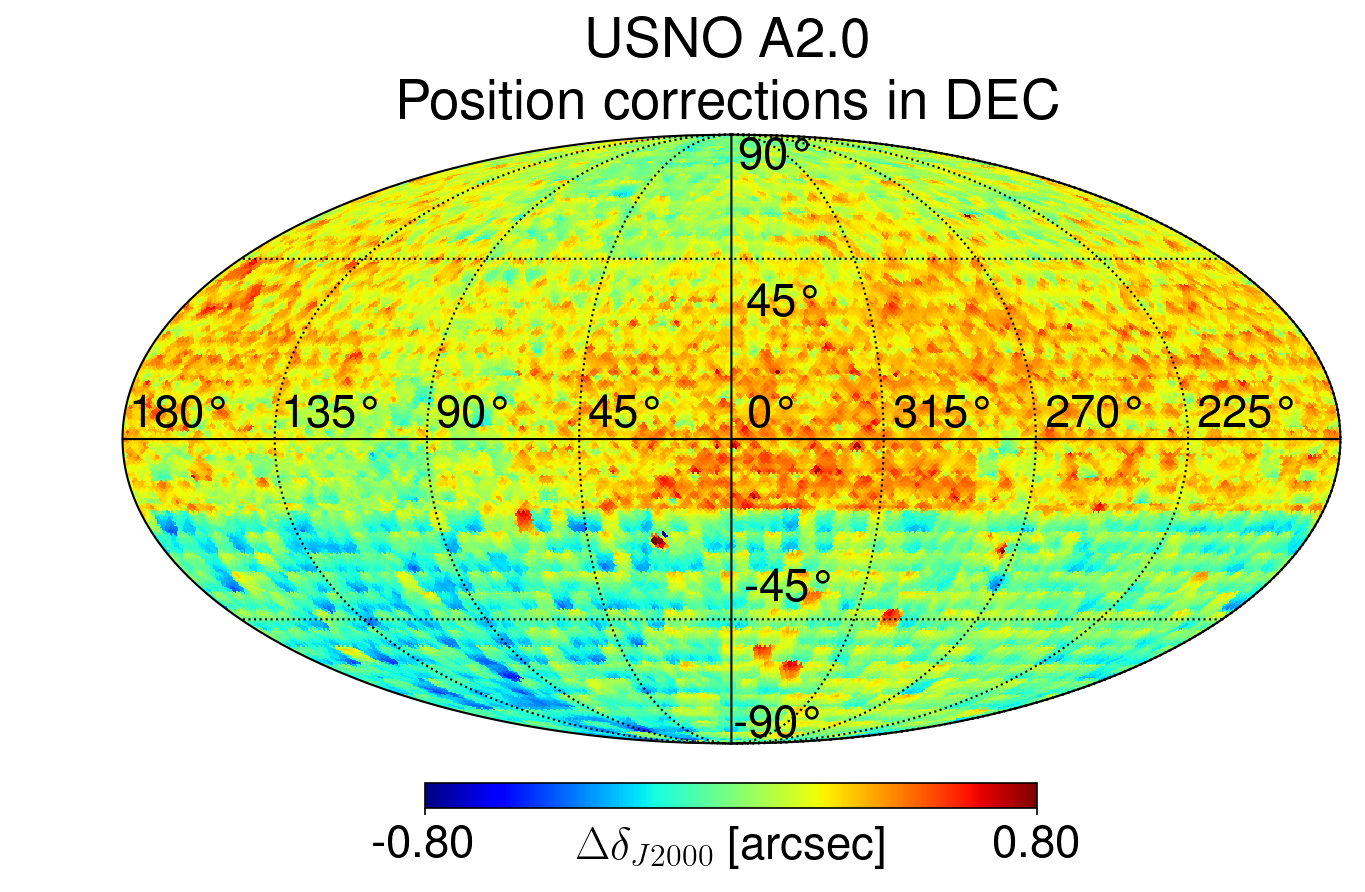} \\
\includegraphics[width=0.5\linewidth]{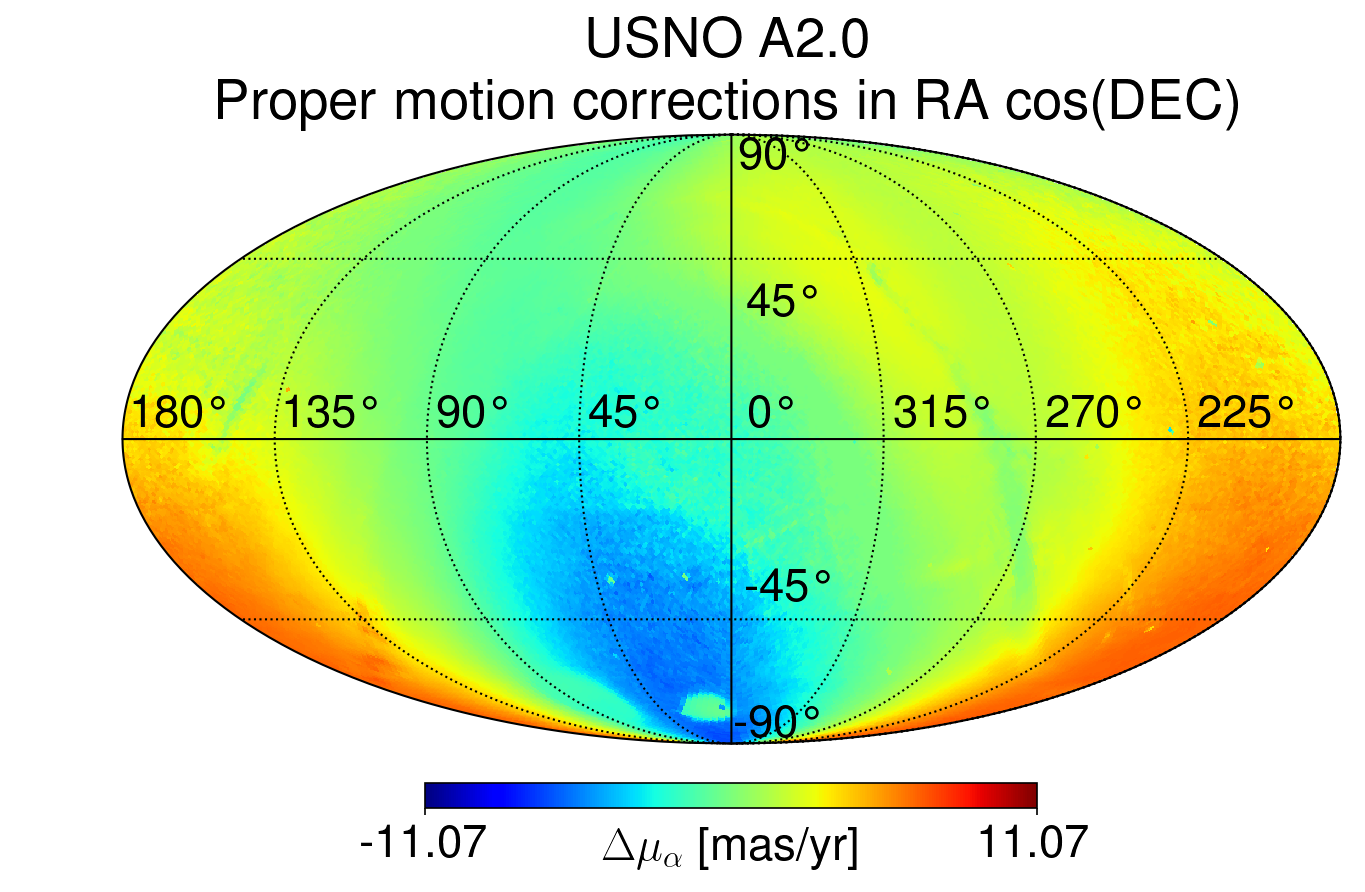} &
\includegraphics[width=0.5\linewidth]{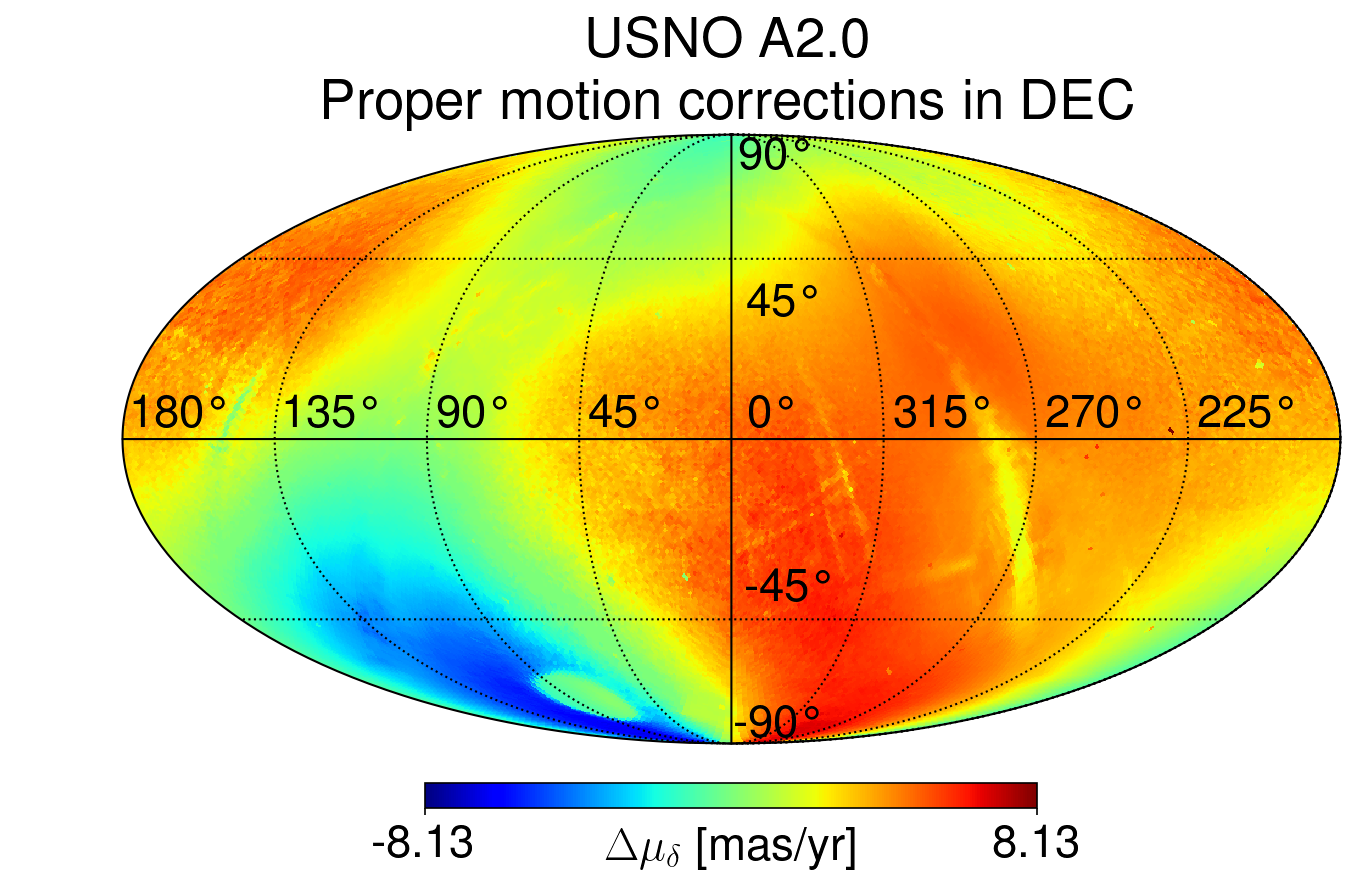}
\end{tabular}
\caption{USNO A2.0 catalog systematics with respect to Gaia DR 2. The top panels display local differences in right ascension and declination, the bottom panels show the systematics caused by the missing proper motion.   \label{fig:usnoa2}}
\end{figure}

\begin{figure}
\includegraphics[width=\linewidth]{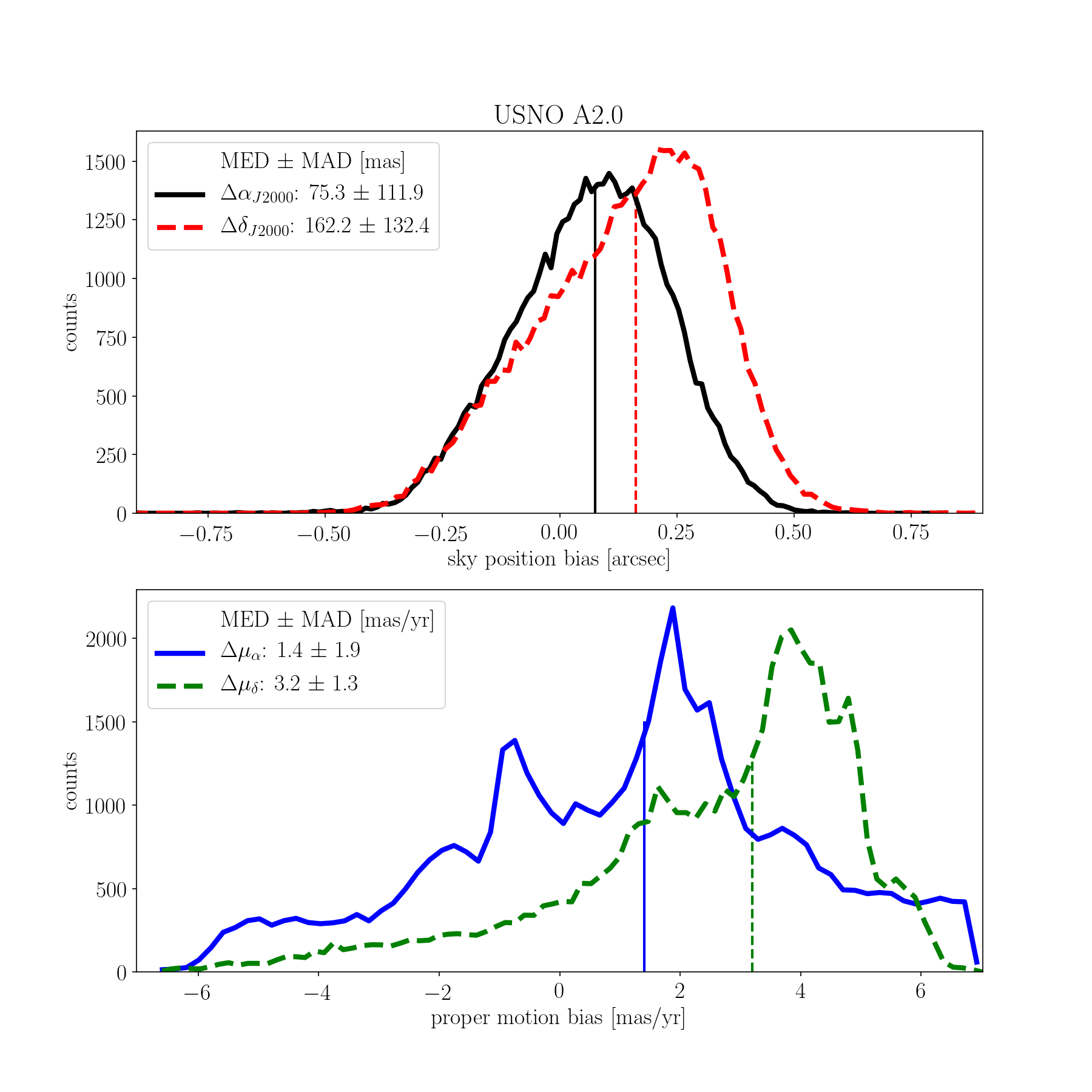} 
\caption{Summary statistics of USNO A2.0 catalog systematics with respect to Gaia DR 2. The top panel displays the distribution of local differences in right ascension and declination. The signature of the missing proper motion is shown in the bottom panel. The legend contains the median (MED) and median absolute deviation (MAD) for the respective quantities.\label{fig:usnoa2_hist}}
\end{figure}

\begin{figure}
\begin{tabular}{ll}
\includegraphics[width=0.5\linewidth]{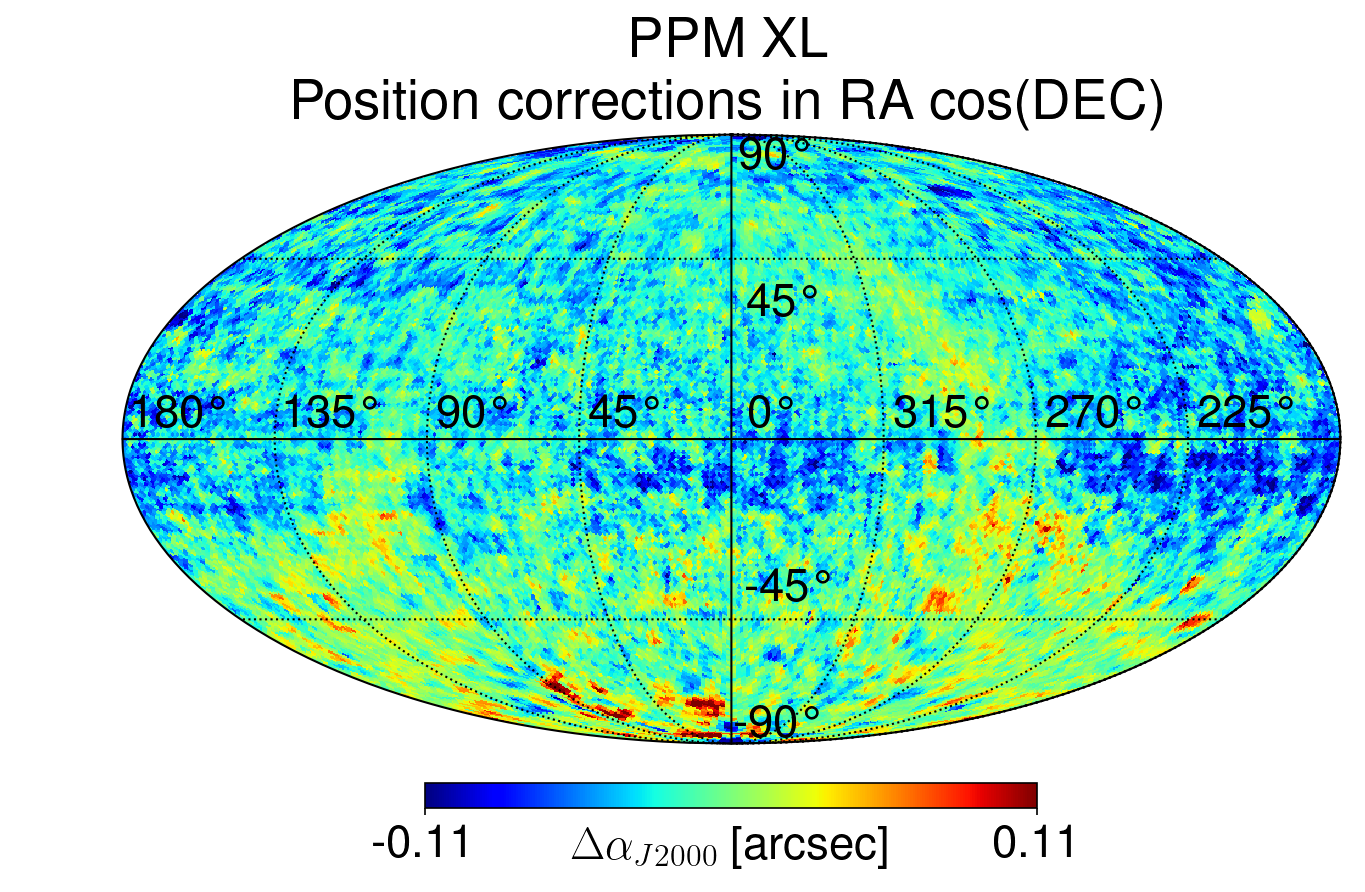} &
\includegraphics[width=0.5\linewidth]{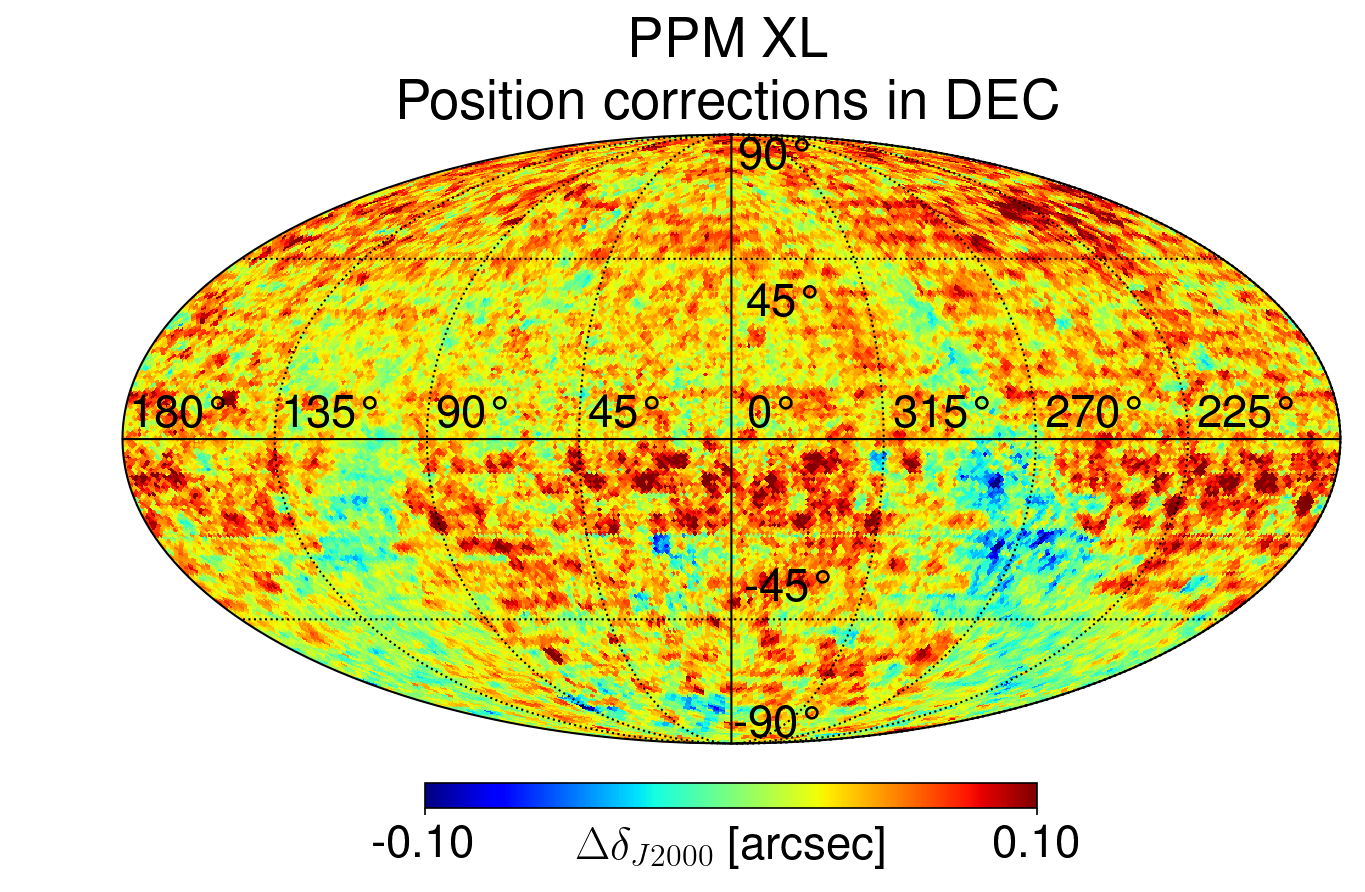} \\
\includegraphics[width=0.5\linewidth]{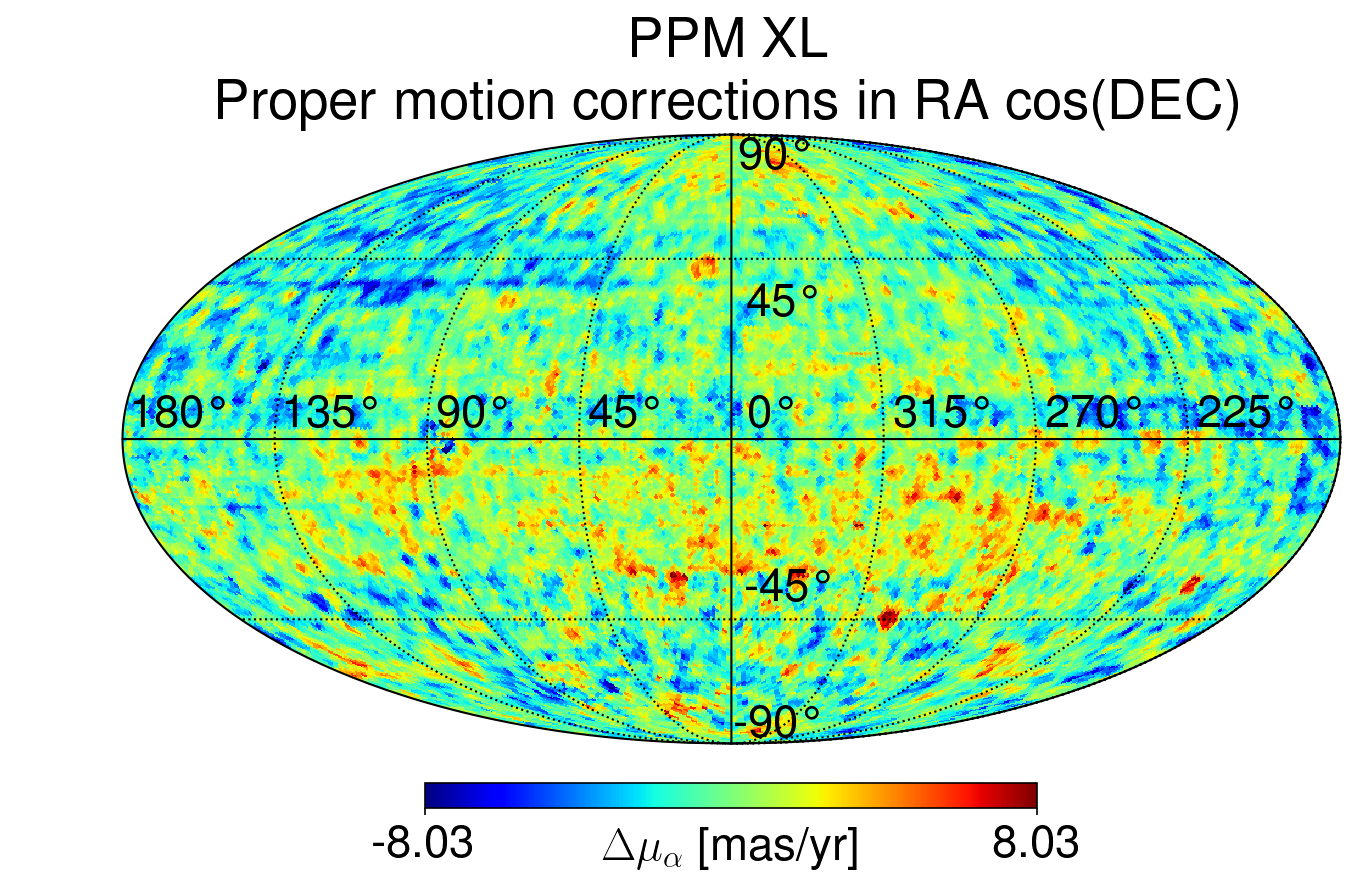} &
\includegraphics[width=0.5\linewidth]{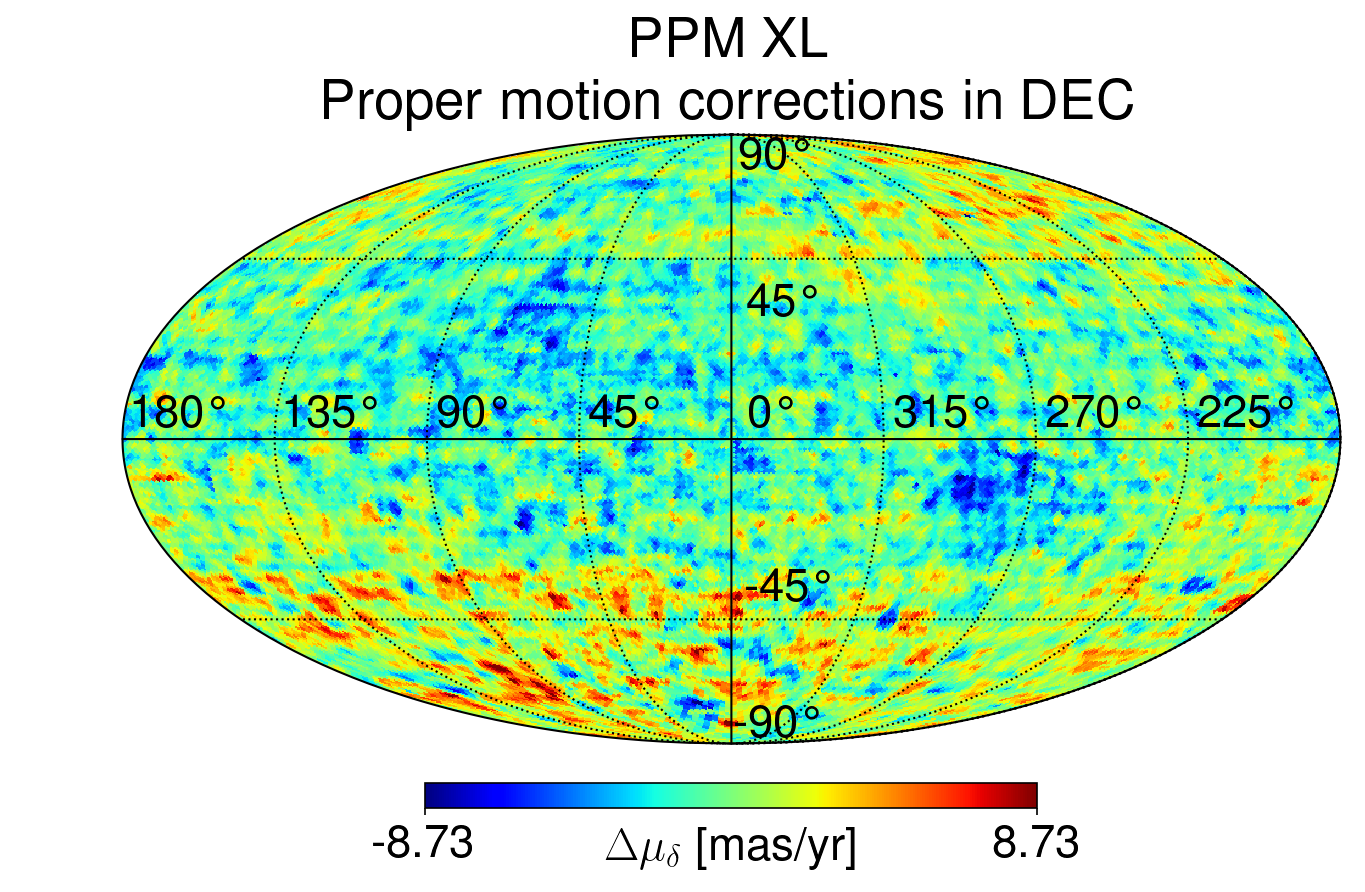}\\
\end{tabular}
\caption{Same as Figure \ref{fig:usnoa2} but for the
PPM XL catalog. 
% In addition, the number of stars per HEALPix tile in PPM XL (bottom left) and fraction of matching stars between PPM XL and Gaia DR 2 (bottom right) are shown.
%Note the drop in the fraction of matching stars near the galactic center.
Note the larger corrections near the galactic center ($\alpha\approx$ 270 deg, $\delta\approx$-30 deg).
\label{fig:PPMXL}}
\end{figure}

With more than 40 million entries in the Minor Planet Center database the USNO A2.0 catalog has been one of the most popular choices for reducing astrometric observations of minor planets.
Figures \ref{fig:usnoa2} and \ref{fig:usnoa2_hist} show the proposed corrections derived from a comparison with Gaia DR 2.
The regional biases in stellar positions first found by \citet{chesley10} and later confirmed by \citet{farnocchia15} are clearly visible in the top panels of Figure \ref{fig:usnoa2}.
Figure \ref{fig:usnoa2_hist} shows the distribution of the corrections collected over the entire sky. The visual impression from Figure \ref{fig:usnoa2}, namely that there is an overall bias towards higher declination values in stellar positions, is confirmed in the distribution of $\Delta \delta_{J2000}$. The lack of proper motion as well as a median bias of 162 mas in declination with a mode closer to a quarter arcsecond all suggest that astrometric observations reduced with USNO A2.0 could benefit from a correction. 
Table~\ref{tab:cat_stats} reports the size of the corrections in terms of the all sky median (MED) and median absolute deviation (MAD) for each studied catalog. 
Of particular interest is the comparison between PPM XL and Gaia DR 2, since a subset of the former was used as a reference for stellar positions and proper motion in \citet{farnocchia15}.
The corresponding graphs are presented in Figure \ref{fig:PPMXL}.
The range of position corrections is only about one sixth of that in USNO A2.0 and the subset of 2MASS stars used for astrometry also has a relatively small positional bias as can be gathered from Table \ref{tab:cat_stats}.
A comparison between Gaia DR 2 and UCAC 4 (Figure \ref{fig:ucac4}) reveals that
the relatively large corrections seen by \citet[][Figures 9 to 12]{farnocchia15} near the galactic center ($\alpha\approx$ 270 deg, $\delta\approx$-30 deg) appear to actually be associated to the PPM XL reference, rather than the UCAC catalogs. This is indicated by the signature seen near the galactic center in all four panels of Figure \ref{fig:PPMXL} that is now substantially absent from all four panels of Figure \ref{fig:ucac4}.

% There is, however, a significant non-uniformity in the corrections close to the Galactic center ($\alpha\approx$ 270 deg, $\delta\approx$-30 deg) both in position and proper motion, which leads to a drop in the fraction of stars that could be matched between Gaia DR 2 and PPM XL. Despite the careful selection of the subset of stars that was used as a reference in \citet{farnocchia15}---2MASS has very little bias---the non-uniformity in PPM XL affected the previous debiasing scheme. This signature is most evident in the corrections for the UCAC catalog series as presented in \citet[][Figures 9 to 12]{farnocchia15}.
% %
% A comparison between Gaia DR 2 and UCAC 4 (Figure \ref{fig:ucac4}) does not show any regional bias near the Galactic center in UCAC 4, although proper motion bias is generally higher in the vicinity of the Milky Way.
%

% Comparing the distributions of PPM XL and UCAC 4 catalog corrections in Figures \ref{fig:PPM XL_hist} and \ref{fig:ucac4_hist} one can see that the latter shows less bias. Similar conclusions hold for UCAC 2 as can be seen in Table \ref{tab:cat_stats}. Both UCAC 2 and UCAC 4 were used extensively in minor planet astrometry totalling around 56 million observations.

With the exception of the artificial structures close to the Galactic center introduced through PPM XL, our results for UCAC, USNO and CMC catalogs are very similar to those found in \citet{farnocchia15}. 
The corresponding Figures are provided in the Appendix.
Among the newly added catalogs UCAC 5 continues the tradition of the UCAC family with an increase in the overall quality of the catalog. Local corrections are mostly due to differences between stellar positions in Gaia DR 2 and Gaia DR 1. The latter were ingested into UCAC 5. URAT 1 has a higher source density compared to UCAC 4, but it also exhibits larger deviations from Gaia DR 2 in both stellar positions and proper motion. 
We confirm the known issues with astrometric accuracy around the north celestial pole in SDSS DR 8 \citep{sdss8_astrometry}, which were addressed in the following data releases.  
Apart from minor systematics stemming from survey incompleteness, Gaia DR 1 had excellent stellar positions at the epoch of its publication \citep{gaiadr1}. Since Gaia DR 1 largely lacks proper motion, the differences to Gaia DR 2 become significant at the epoch of comparison (J2000), however. This is reflected in Table \ref{tab:cat_stats}. The quality of stellar positions, even if measured through the Gaia satellite, degrades relatively quickly with time if no proper motion corrections are applied. Minor planet astrometrists are, thus, encouraged to switch to Gaia DR 2 as soon as possible. 
We debias all observations where corrections are available with the exception of those reduced with ACT, Gaia DR 1, Tycho-2 and UCAC-5, since the latter show no large scale structures in local position and proper motion systematics.
\begin{figure}
\includegraphics[width=\linewidth]{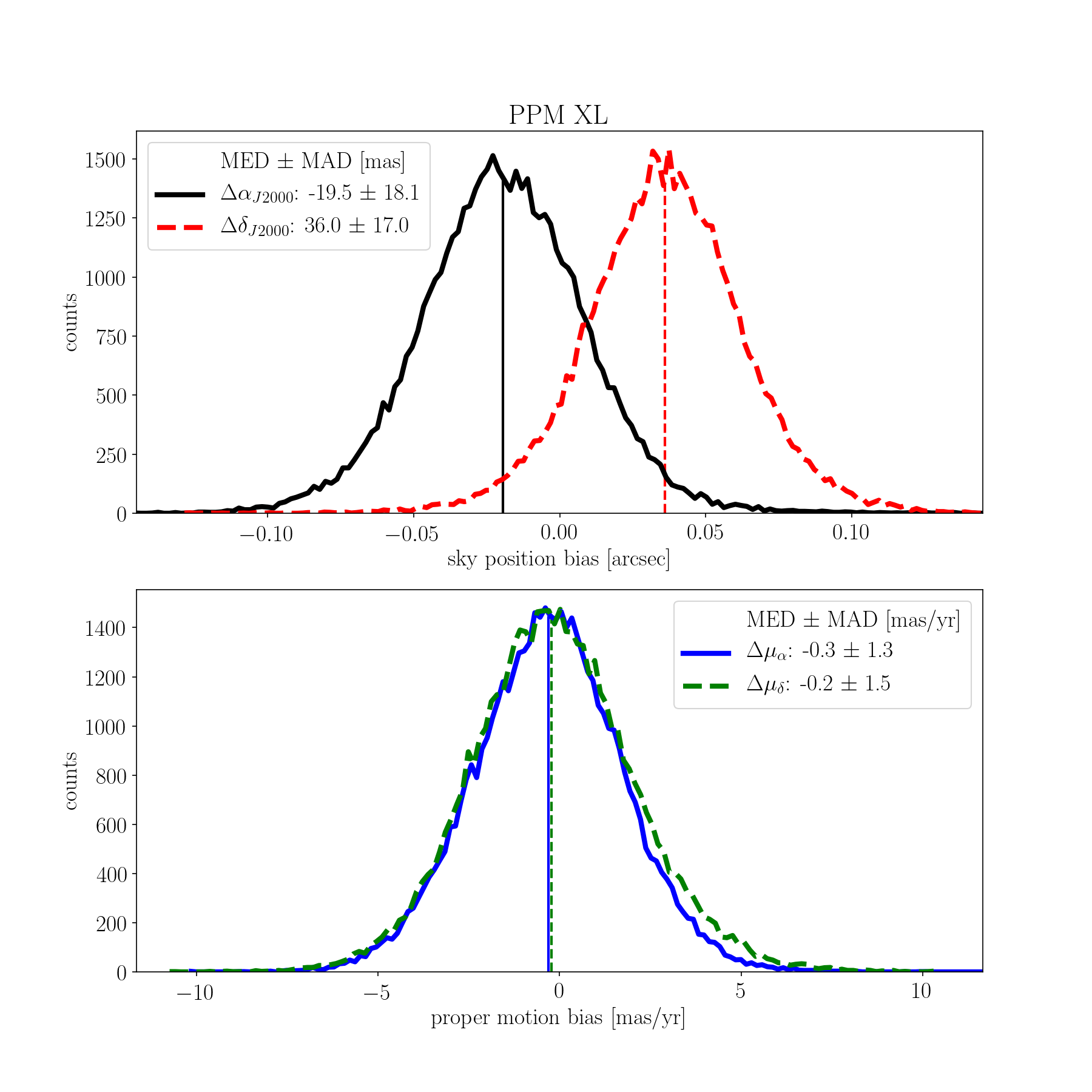} 
\caption{Same as Figure \ref{fig:usnoa2_hist} for the PPM XL catalog.  \label{fig:ppmxl_hist}}
\end{figure}
\begin{figure}
\begin{tabular}{ll}
\includegraphics[width=0.5\linewidth]{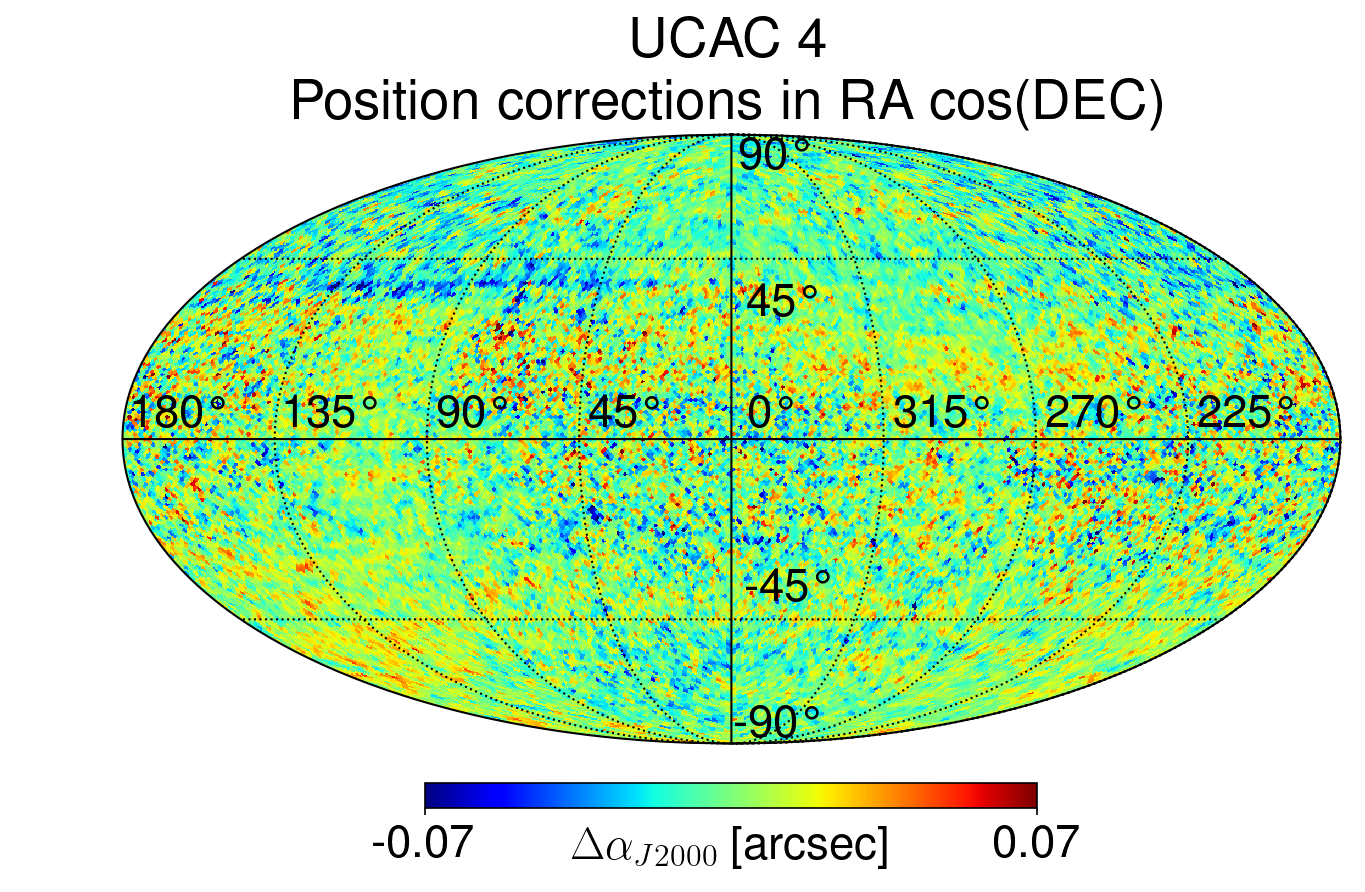} &
\includegraphics[width=0.5\linewidth]{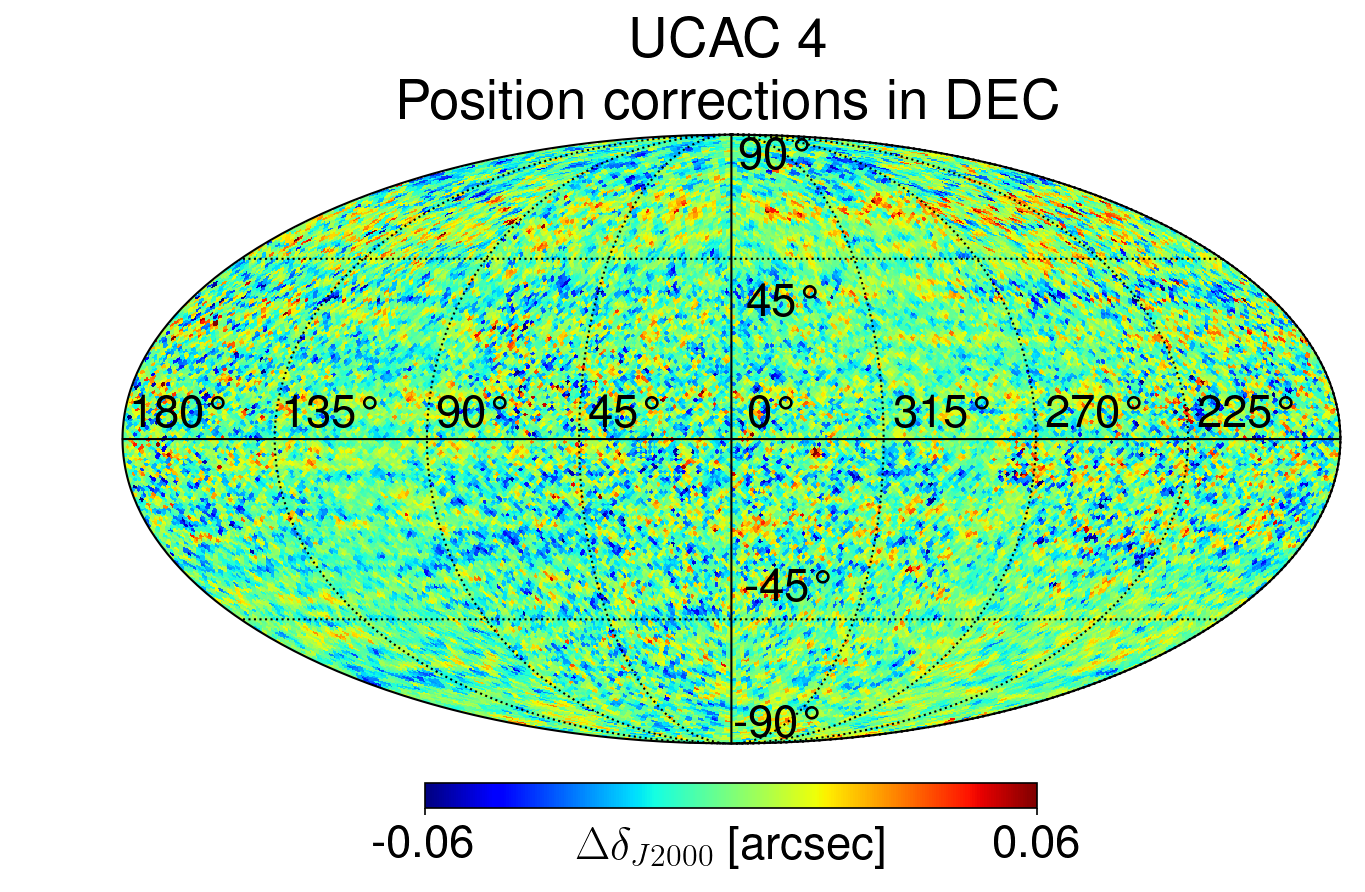} \\
\includegraphics[width=0.5\linewidth]{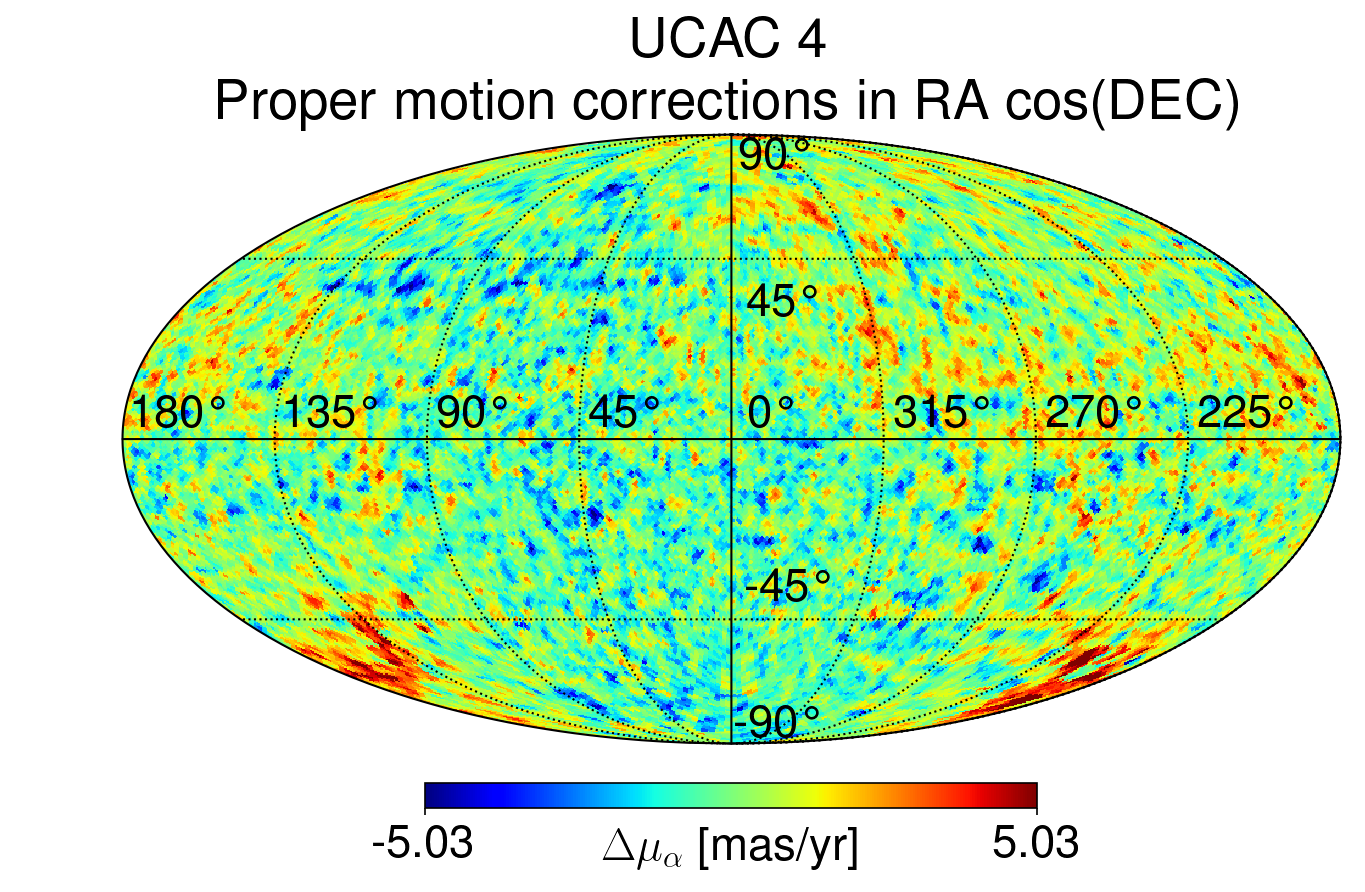} &
\includegraphics[width=0.5\linewidth]{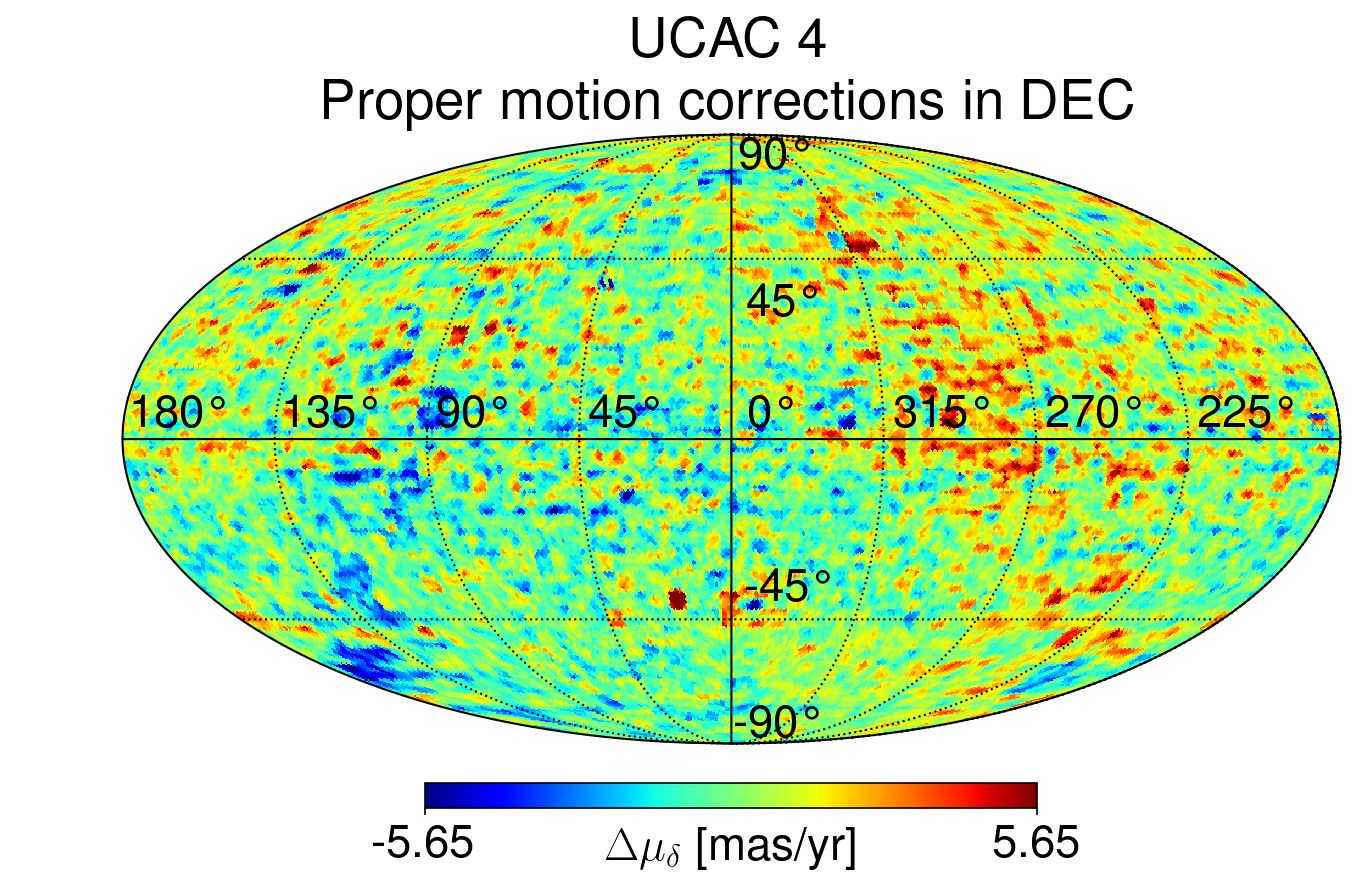}\\
\end{tabular}
\caption{Same as Figure \ref{fig:PPMXL} but for the UCAC 4 catalog. Systematics are small in positions, but traces of stellar proper motion errors in the galactic plane are visible. \label{fig:ucac4}}
\end{figure}
\begin{figure}
\includegraphics[width=\linewidth]{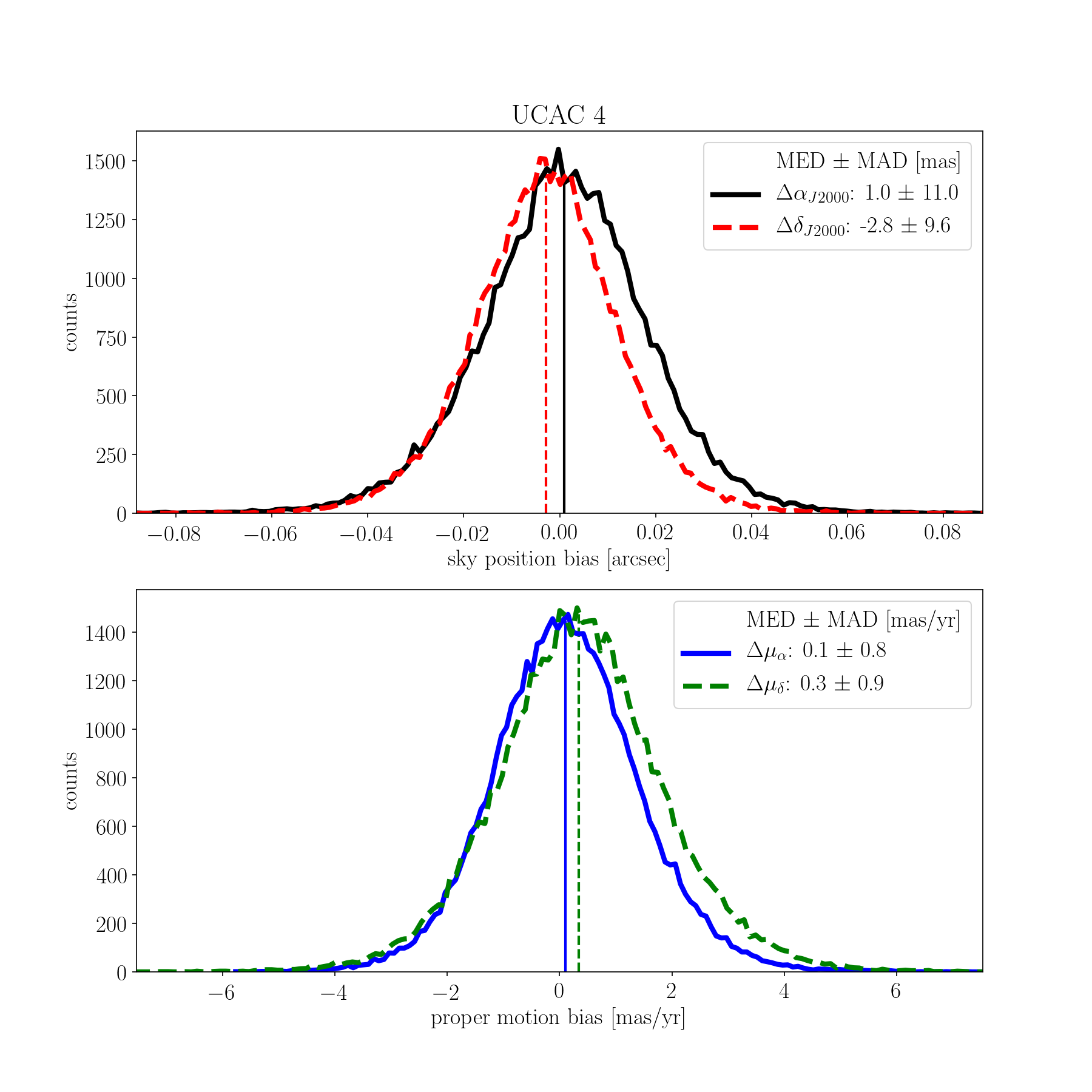} 
\caption{Same as Figure \ref{fig:usnoa2_hist} for the UCAC 4 catalog. \label{fig:ucac4_hist}}
\end{figure}
\begin{table}\tiny
\begin{center}
\begin{tabular}{|l|cc|cc|cc|cc|c|c|}
  \hline
Catalog & \multicolumn{2}{c}{$\Delta{\alpha}_{J2000}$ [mas]} & \multicolumn{2}{c}{$\Delta{\delta}_{J2000}$ [mas]} & \multicolumn{2}{c}{$\Delta\mu_{\alpha}$ [mas/yr]} &
\multicolumn{2}{c}{$\Delta\mu_{\delta}$ [mas/yr]} & Sky [\%] & proper\\
& MED & MAD & MED & MAD & MED & MAD &MED & MAD &coverage & motion\\ 
%& [arcsec] & [arcsec]  & [mas/yr] & [mas/yr] & \\
\hline
2MASS&  -2.4& 15.5& 11.6& 13.6& 1.51& 2.50& 3.76& 1.50& 100 & no \\ 
%AC &219.2& 304.0& 292.5& 246.1& 0.28& 1.93& 3.54& 1.55& 28  & no\\ 
ACT&  -0.8& 11.8& 0.2& 10.9& -0.05& 1.03& 0.01& 0.98& 100 & yes\\ 
CMC 14& 2.0& 16.7& -18.3& 13.8& 1.53& 2.45& 4.22& 0.98& 61& no \\ 
CMC 15&  1.5& 19.0& -19.8& 15.7& 1.52& 2.53& 4.09& 1.17&70& no \\ 
GSC 1.1& -45.5& 266.8& 17.1& 235.1& 1.07& 2.81& 3.87& 1.48&100& no \\ 
GSC 1.2& 23.6& 127.7& -16.2& 119.3& 1.06& 2.81& 3.87& 1.48& 100& no \\ 
GSC ACT& 29.4& 82.2& 44.8& 69.9& 1.05& 2.76& 3.84& 1.45& 100& no\\ 
Gaia DR 1& -22.5& 30.5& -49.2& 20.2& 1.48& 2.00& 3.27& 1.35& 100 & no \\ 
%Hipparcos 2& 0.1& 7.1& 0.1& 6.1& 0.12& 0.76& 0.01& 0.66& 100& yes\\ 
NOMAD& -19.8& 62.3& 154.8& 80.8& 1.19& 1.78& 2.66& 1.29& 100& yes\\ 
PPM& 12.6& 131.2& -68.4& 130.0& -0.21& 2.77& -0.48& 2.84& 100& yes\\ 
PPM XL& -19.5& 18.1& 36.0& 17.0& -0.30& 1.34& -0.23& 1.46& 100& yes\\ 
SDSS DR 7& -5.4& 16.4& -7.6& 16.2& 0.29& 0.35& 0.67& 0.43& 13& yes\\ 
SDSS DR 8& -4.7& 22.0& -5.2& 24.0& 0.30& 0.47& 0.75& 0.51& 16& yes\\ 
SST RC 4\&5&-15.2& 15.0& 25.0& 14.4& 0.27& 1.08& 0.10& 1.13& 100& yes\\ 
Tycho 2& -1.0& 10.1& -1.6& 9.8& -0.08& 0.82& -0.31& 0.82& 100& yes\\ 
UCAC 1& -4.0& 14.0& 12.5& 13.1& 2.70& 3.06& 2.94& 4.39& 39& yes\\ 
UCAC 2& -0.8& 10.3& 3.1& 9.2& -0.30& 1.31& -0.60& 1.27& 88& yes\\ 
UCAC 3& -2.3& 13.3& -1.6& 10.9& -0.75& 1.82& 0.84& 1.87& 100& yes\\ 
UCAC 4& 1.0& 11.0& -2.8& 9.6& 0.10& 0.84& 0.34& 0.94& 100& yes\\ 
UCAC 5& 1.6& 4.8& 1.0& 3.7& -0.11& 0.32& -0.06& 0.25& 100& yes\\ 
URAT 1& -23.0& 26.9& -43.7& 20.4& 0.42& 1.17& 0.70& 1.14& 61& yes\\ 
USNO A1.0&-17.3& 261.4& -22.9& 231.1& 1.39& 1.92& 3.21& 1.34& 100& no\\ 
USNO A2.0& 75.3& 111.9& 162.2& 132.4& 1.40& 1.91& 3.20& 1.34& 100& no\\ 
USNO B1.0& -25.4& 79.3& 175.2& 76.8& 1.26& 2.00& 3.02& 1.36& 100& yes\\ 
USNO SA1.0& -20.3& 260.8& -17.1& 230.8& 1.27& 1.96& 3.27& 1.34&100& no\\
USNO SA2.0& 72.7& 111.8& 164.1& 131.5& 1.35& 1.96& 3.25& 1.34& 100& no\\ 
\hline
\end{tabular}
\end{center}
\caption{The columns show the median (MED) and median absolute deviation (MAD) of corrections in
  position (right ascension and declination) and proper motion (right ascension and declination) over the entire sky as well as the fraction of sky coverage for each catalog. The last column states whether or not a catalog contains data on stellar proper motion.}
\label{tab:cat_stats}
\end{table}
%%%
\section{Interpolation of Debiasing Tables}
\label{sec:interp}
\begin{figure}
\centering\includegraphics[width=0.7\linewidth]{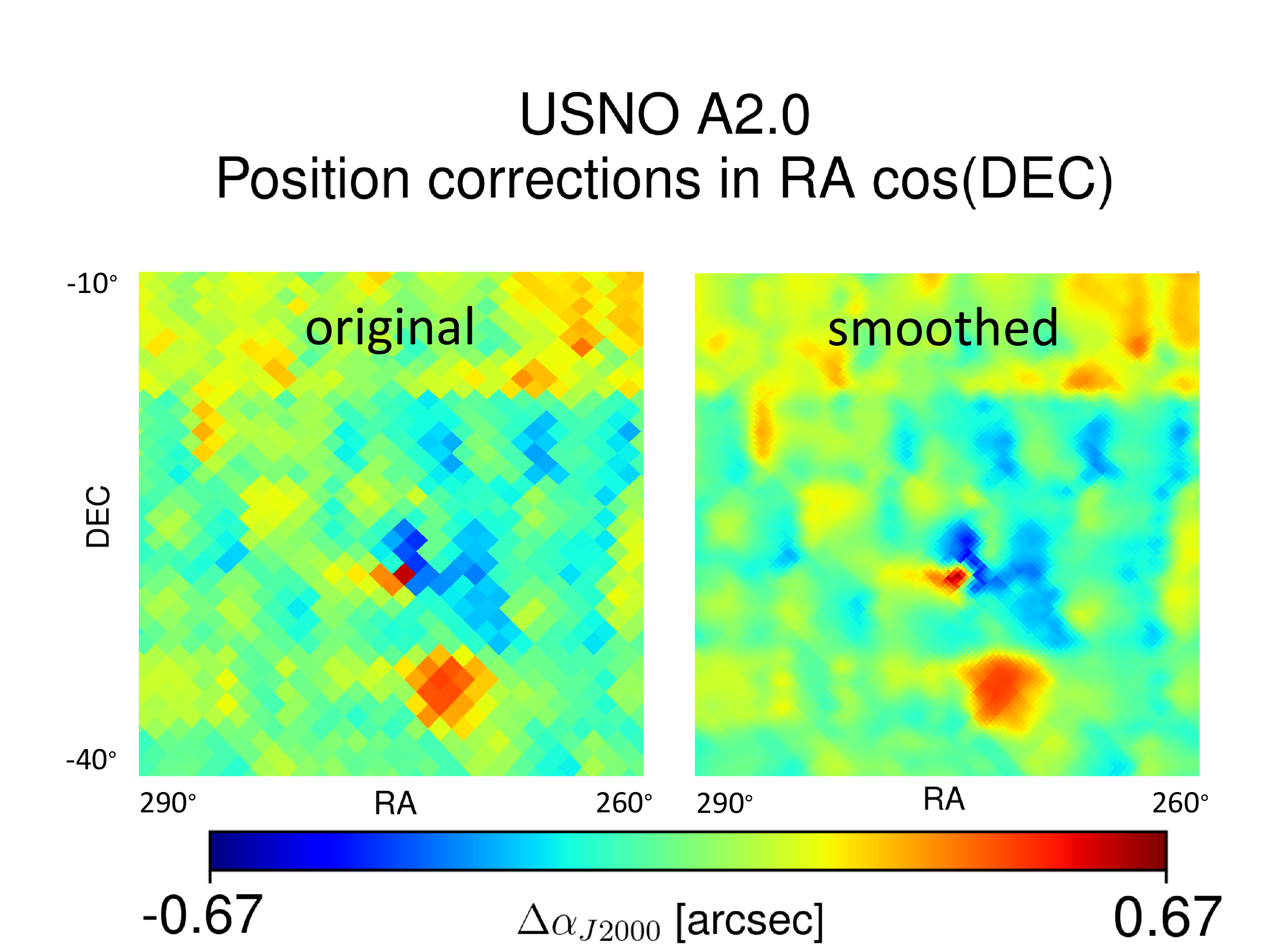} 
\caption{Zoom into a region with a large jump between positive and negative catalog corrections in $\Delta \alpha_{J2000}$ for the USNO A2.0 catalog. Smoothing can avoid quasi-discontinuous transitions between HEALPix tiles with opposite bias values.  \label{fig:smooth}}
\end{figure}

Due to the discrete nature of calculating corrections to astrometric measurments on HEALPix tiles, transitions between neighboring corrections are not guaranteed to be smooth. 
As an example, Figure \ref{fig:smooth} shows that adjacent USNO A2.0 tiles can have significantly different bias values, with variations upwards of 0.6 arcsec.
% Figure \ref{fig:smooth} shows, for instance, that in USNO A1.0 tiles with positive and negative corrections on the order of an arcsecond can have common borders. 
Corrections for minor planet observations that cross such a border would see an artificial jump in the corresponding coordinate. A finer HEALPix tessellation could be used to mitigate such a problem. However, a finer tessellation would mean that fewer stars per tile are available to estimate the magnitude of the correction. 
In order to circumvent reducing the sample size of stars per tile, we use radial basis functions \citep[RBFs,][]{schaback1995} to interpolate data from neighboring tiles onto a finer HEALPix tessellation. This reduces artificial discontinuities along tile borders while at the same time retaining a sufficient number of stars in each of the stencil tiles.   
%%%
\section{Ephemeris Prediction Test}
\label{sec:eph}
\begin{figure}
\begin{tabular}{cc}
\includegraphics[width=0.5\linewidth]{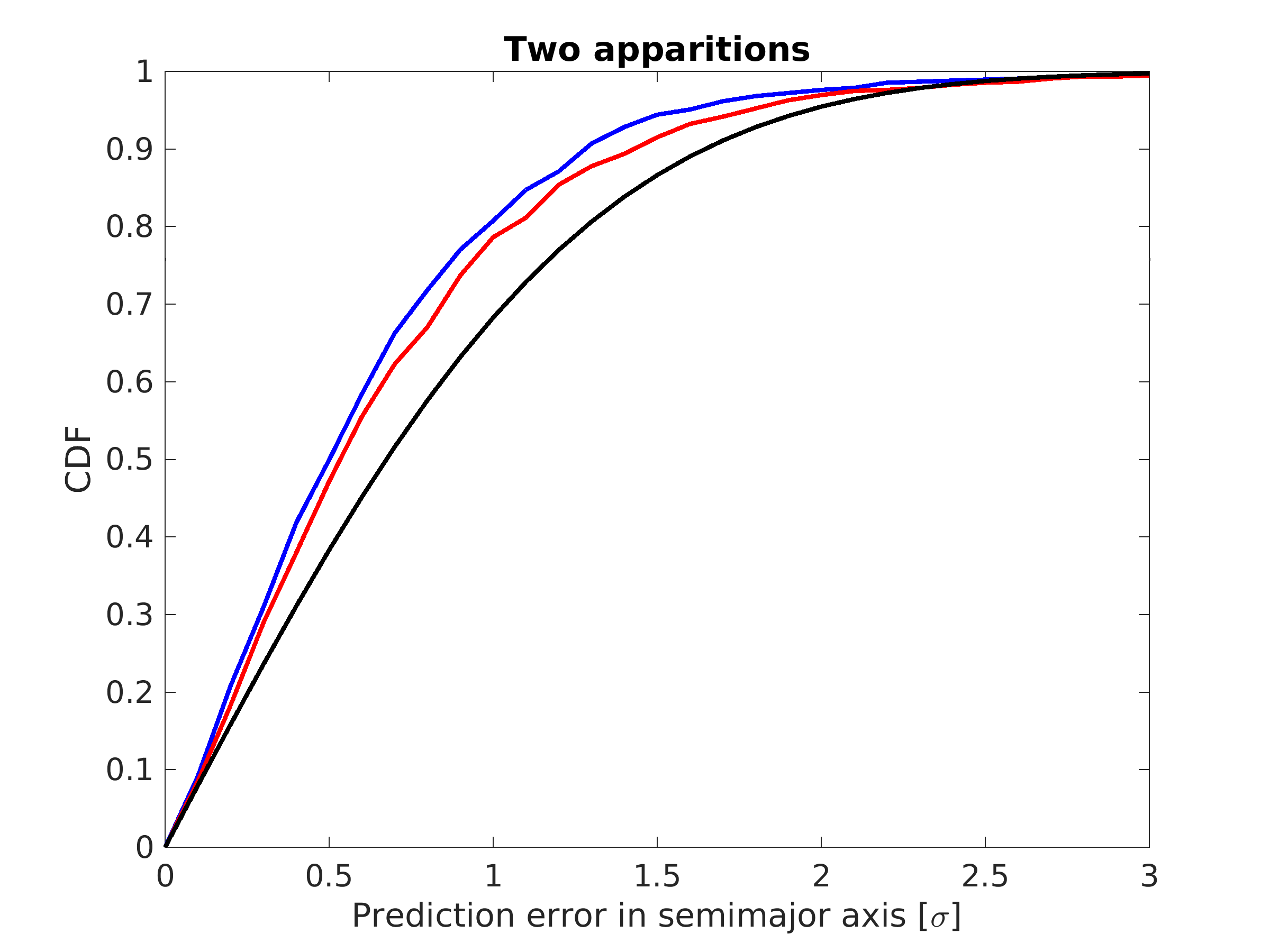} &
\includegraphics[width=0.5\linewidth]{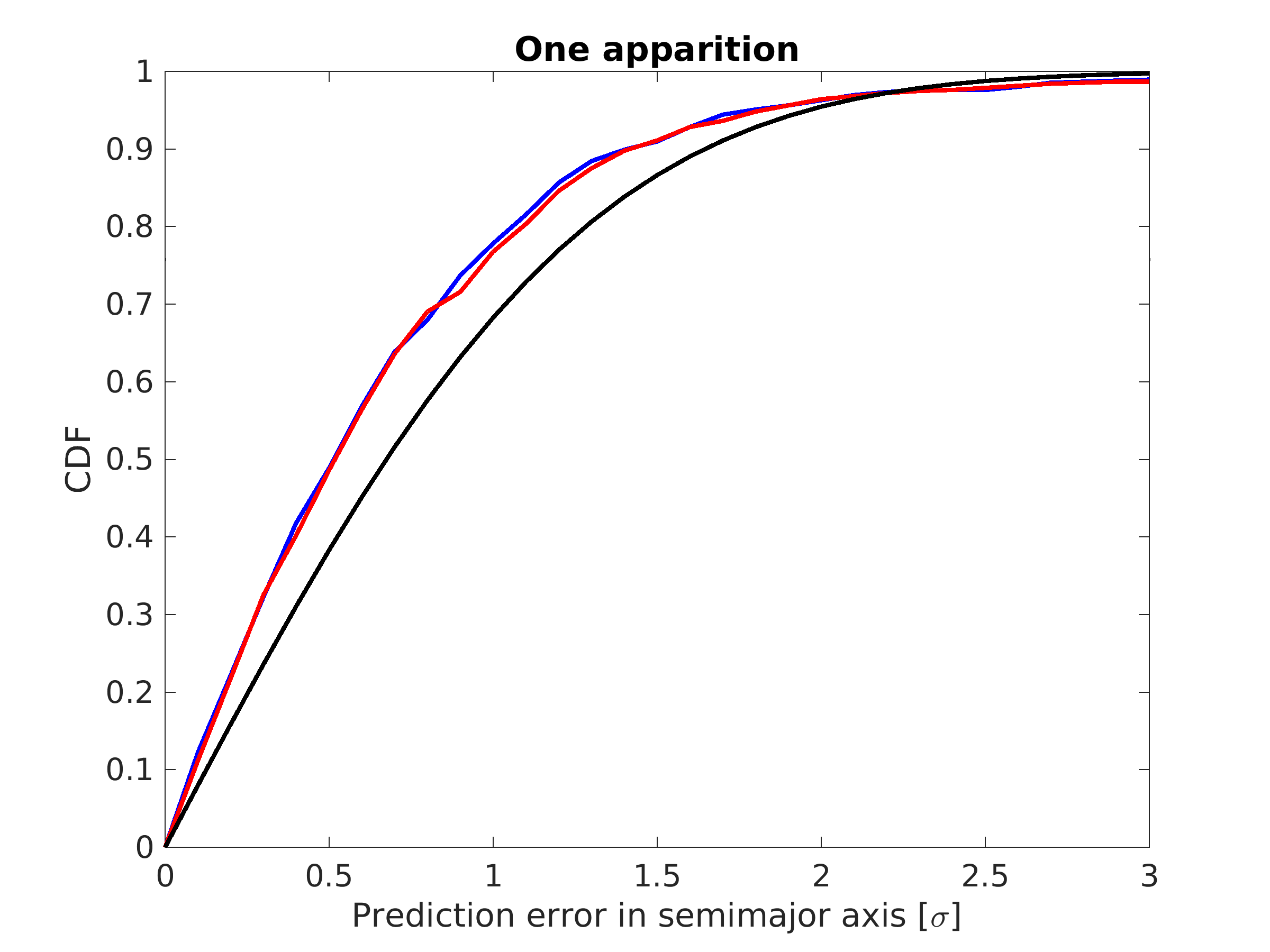}
\end{tabular}
\caption{Cumulative distributions of orbit prediction errors for the debiasing schemes presented in this work (blue line), \citet{farnocchia15} (red line) and the theoretical Gaussian (black line) are shown. The first and second apparition have been used in the left graph and the penultimate apparition in the right graph. \label{fig:eph_pred}}
\end{figure}
We follow the example of \citet[][sec. 6]{chesley10} and \citet[][sec. 4.3]{farnocchia15} and evaluate the impact of the new debiasing scheme on asteroid orbit determination and prediction. To achieve that we compare the prediction performance for an ensemble of 700 near-Earth objects (NEOs), each of which had at least five apparitions since 1998. Every one of those apparitions has at least ten observations over 15 days.
Using the astrometric observations from either the first two apparitions or the penultimate apparition, respectively, we construct predictions close to the center of the third apparition, near the middle of the true observational arc. 
The predictions are then compared to the reference solution that uses the full arc. The analysis is performed in a consistent way such that a given debiasing scheme is used for both the short arcs derived from a subset of apparitions and the full arc. All astrometric data is weighted according to the scheme presented in \citet{verevs2017}.

Figure \ref{fig:eph_pred} shows the corresponding results.
The prediction errors in the orbital semimajor axes of 700 NEOs derived from astrometry debiased with respect to Gaia DR 2 are only slightly smaller that those using the debiasing scheme presented in \citet{farnocchia15}. These results suggest that most of the statistical bias was de facto eliminated with the previous debiasing approach.
Using interpolated debiasing tables as discussed in the previous section did not significantly affect the above prediction statistics, either. Given the larger size of the interpolated table and the potential loss in computational efficiency we recommend using interpolated values only when quasi-discontinuities in the former debiasing table are an obstacle to high-fidelity orbit determination.  

\section{Conclusions}
\label{sec:conclusions}
The Gaia astrometric catalog published with the second Gaia data release has become the new standard for minor planet astrometry. 
Its uniformity and high quality in both stellar positions and proper motion allow us to improve prior astrometric observations of minor planets by correcting for potential systematics in other astrometric catalogs. 
In this work we have used Gaia DR 2 as a reference to calculate debiasing tables for 26 astrometric catalogs. New tables for six catalogs (PPM XL, SST RC5, Gaia DR 1, CMC 15, URAT, UCAC 5 and SDSS 8) that had not been studied in \citet{farnocchia15} are included in this work.
The quality of catalog corrections presented here is significantly improved compared to previous work, in particular for the UCAC series.  
The switch to the new debiasing scheme based on Gaia DR 2 delivers only incremental improvement of the quality of near-Earth asteroid orbits over \citet{farnocchia15}, however.
This suggests that further debiasing attempts may not be necessary. 
The procedure suggested in this work can help mitigate some of the systematics introduced by the use of star catalogs other than Gaia. We would like to emphasize that 
the tables are meant as an a posteriori correction to existing astrometric data. They should not be used to debias new astrometric measurments.
We strongly recommend contemporary Solar System astrometry be conducted with Gaia DR 2 and its successors. 

\section{Data Availability}
Tables containing standard resolution and interpolated catalog corrections derived in this work are publicly available at \url{ftp://ssd.jpl.nasa.gov/pub/ssd/debias/debias_2018.tgz}
and 
\url{ftp://ssd.jpl.nasa.gov/pub/ssd/debias/debias_hires2018.tgz}.

\section*{Acknowledgments}
 This research has received funding from the Jet Propulsion Laboratory through the California Institute of Technology postdoctoral fellowship program, under a contract with the National Aeronautics and Space Administration.
 VizieR catalogue access tools, CDS, Strasbourg, France, have been essential to this research project \citep{vizier}.
 This work made use of the IAU Minor Planet Center observations database.
%%%\section*{References}

\bibliography{refs}

\begin{thebibliography}{35}
\expandafter\ifx\csname natexlab\endcsname\relax\def\natexlab#1{#1}\fi
\providecommand{\url}[1]{\texttt{#1}}
\providecommand{\href}[2]{#2}
\providecommand{\path}[1]{#1}
\providecommand{\DOIprefix}{doi:}
\providecommand{\ArXivprefix}{arXiv:}
\providecommand{\URLprefix}{URL: }
\providecommand{\Pubmedprefix}{pmid:}
\providecommand{\doi}[1]{\href{http://dx.doi.org/#1}{\path{#1}}}
\providecommand{\Pubmed}[1]{\href{pmid:#1}{\path{#1}}}
\providecommand{\bibinfo}[2]{#2}
\ifx\xfnm\relax \def\xfnm[#1]{\unskip,\space#1}\fi
%Type = Article
\bibitem[{Abazajian et~al.(2009)Abazajian, Adelman-McCarthy, Ag{\"u}eros,
  Allam, Prieto, An, Anderson, Anderson, Annis, Bahcall et~al.}]{sdss7}
\bibinfo{author}{Abazajian, K.N.}, \bibinfo{author}{Adelman-McCarthy, J.K.},
  \bibinfo{author}{Ag{\"u}eros, M.A.}, \bibinfo{author}{Allam, S.S.},
  \bibinfo{author}{Prieto, C.A.}, \bibinfo{author}{An, D.},
  \bibinfo{author}{Anderson, K.S.}, \bibinfo{author}{Anderson, S.F.},
  \bibinfo{author}{Annis, J.}, \bibinfo{author}{Bahcall, N.A.}, et~al.,
  \bibinfo{year}{2009}.
\newblock \bibinfo{title}{{The seventh data release of the Sloan Digital Sky
  Survey}}.
\newblock \bibinfo{journal}{The Astrophysical Journal Supplement Series}
  \bibinfo{volume}{182}, \bibinfo{pages}{543}.
%Type = Article
\bibitem[{{Adelman-McCarthy} and {et al.}(2011)}]{sdss8}
\bibinfo{author}{{Adelman-McCarthy}, J.K.}, \bibinfo{author}{{et al.}},
  \bibinfo{year}{2011}.
\newblock \bibinfo{title}{{VizieR Online Data Catalog: The SDSS Photometric
  Catalog, Release 8 (Adelman-McCarthy+, 2011)}}.
\newblock \bibinfo{journal}{VizieR Online Data Catalog} ,
  \bibinfo{pages}{II/306}.
%Type = Article
\bibitem[{Aihara et~al.(2011)Aihara, Prieto, An, Anderson, Aubourg, Balbinot,
  Beers et~al.}]{sdss8_astrometry}
\bibinfo{author}{Aihara, H.}, \bibinfo{author}{Prieto, C.A.},
  \bibinfo{author}{An, D.}, \bibinfo{author}{Anderson, S.F.},
  \bibinfo{author}{Aubourg, {\'{E}}.}, \bibinfo{author}{Balbinot, E.},
  \bibinfo{author}{Beers, T.C.}, et~al., \bibinfo{year}{2011}.
\newblock \bibinfo{title}{{ERRATUM}: {\textquotedblleft}{THE} {EIGHTH} {DATA}
  {RELEASE} {OF} {THE} {SLOAN} {DIGITAL} {SKY} {SURVEY}: {FIRST} {DATA} {FROM}
  {SDSS}-{III}{\textquotedblright} (2011, {ApJS}, 193, 29)}.
\newblock \bibinfo{journal}{The Astrophysical Journal Supplement Series}
  \bibinfo{volume}{195}, \bibinfo{pages}{26}.
\newblock \DOIprefix\doi{10.1088/0067-0049/195/2/26}.
%Type = Article
\bibitem[{Carpino et~al.(2003)Carpino, Milani and Chesley}]{carpino03}
\bibinfo{author}{Carpino, M.}, \bibinfo{author}{Milani, A.},
  \bibinfo{author}{Chesley, S.R.}, \bibinfo{year}{2003}.
\newblock \bibinfo{title}{Error statistics of asteroid optical astrometric
  observations}.
\newblock \bibinfo{journal}{Icarus} \bibinfo{volume}{166},
  \bibinfo{pages}{248--270}.
%Type = Article
\bibitem[{Chesley et~al.(2010)Chesley, Baer and Monet}]{chesley10}
\bibinfo{author}{Chesley, S.R.}, \bibinfo{author}{Baer, J.},
  \bibinfo{author}{Monet, D.G.}, \bibinfo{year}{2010}.
\newblock \bibinfo{title}{Treatment of star catalog biases in asteroid
  astrometric observations}.
\newblock \bibinfo{journal}{Icarus} \bibinfo{volume}{210},
  \bibinfo{pages}{158--181}.
%Type = Misc
\bibitem[{{Copenhagen University} et~al.(2006){Copenhagen University},
  {Institute}, {Cambridge, UK} and {Real Instituto Y Observatorio de La
  Armada}}]{cmc14}
\bibinfo{author}{{Copenhagen University}, O.}, \bibinfo{author}{{Institute},
  A.O.}, \bibinfo{author}{{Cambridge, UK}}, \bibinfo{author}{{Real Instituto Y
  Observatorio de La Armada}, F.E.S.}, \bibinfo{year}{2006}.
\newblock \bibinfo{title}{Carlsberg meridian catalog 14 (cmc14)}.
%Type = Misc
\bibitem[{{Copenhagen University} et~al.(2011){Copenhagen University},
  {Institute}, {Cambridge, UK} and {Real Instituto Y Observatorio de La
  Armada}}]{cmc15}
\bibinfo{author}{{Copenhagen University}, O.}, \bibinfo{author}{{Institute},
  A.O.}, \bibinfo{author}{{Cambridge, UK}}, \bibinfo{author}{{Real Instituto Y
  Observatorio de La Armada}, F.E.S.}, \bibinfo{year}{2011}.
\newblock \bibinfo{title}{Carlsberg meridian catalog 15 (cmc15)}.
%Type = Article
\bibitem[{Farnocchia et~al.(2015)Farnocchia, Chesley, Chamberlin and
  Tholen}]{farnocchia15}
\bibinfo{author}{Farnocchia, D.}, \bibinfo{author}{Chesley, S.},
  \bibinfo{author}{Chamberlin, A.}, \bibinfo{author}{Tholen, D.},
  \bibinfo{year}{2015}.
\newblock \bibinfo{title}{Star catalog position and proper motion corrections
  in asteroid astrometry}.
\newblock \bibinfo{journal}{Icarus} \bibinfo{volume}{245},
  \bibinfo{pages}{94--111}.
%Type = Article
\bibitem[{{Gaia Collaboration} et~al.(2018){Gaia Collaboration}, {Brown},
  {Vallenari}, {Prusti}, {de Bruijne}, {Babusiaux}, {Bailer-Jones}, {Biermann}
  and {et al.}}]{gaiadr2}
\bibinfo{author}{{Gaia Collaboration}}, \bibinfo{author}{{Brown}, A.G.A.},
  \bibinfo{author}{{Vallenari}, A.}, \bibinfo{author}{{Prusti}, T.},
  \bibinfo{author}{{de Bruijne}, J.H.J.}, \bibinfo{author}{{Babusiaux}, C.},
  \bibinfo{author}{{Bailer-Jones}, C.A.L.}, \bibinfo{author}{{Biermann}, M.},
  \bibinfo{author}{{et al.}}, \bibinfo{year}{2018}.
\newblock \bibinfo{title}{{Gaia Data Release 2. Summary of the contents and
  survey properties}}.
\newblock \bibinfo{journal}{\aap} \bibinfo{volume}{616}, \bibinfo{pages}{A1}.
\newblock \DOIprefix\doi{10.1051/0004-6361/201833051},
  \href{http://arxiv.org/abs/1804.09365}{\tt arXiv:1804.09365}.
%Type = Article
\bibitem[{Gorski et~al.(2005)Gorski, Hivon, Banday, Wandelt, Hansen, Reinecke
  and Bartelmann}]{healpix}
\bibinfo{author}{Gorski, K.M.}, \bibinfo{author}{Hivon, E.},
  \bibinfo{author}{Banday, A.J.}, \bibinfo{author}{Wandelt, B.D.},
  \bibinfo{author}{Hansen, F.K.}, \bibinfo{author}{Reinecke, M.},
  \bibinfo{author}{Bartelmann, M.}, \bibinfo{year}{2005}.
\newblock \bibinfo{title}{Healpix: a framework for high-resolution
  discretization and fast analysis of data distributed on the sphere}.
\newblock \bibinfo{journal}{The Astrophysical Journal} \bibinfo{volume}{622},
  \bibinfo{pages}{759}.
%Type = Article
\bibitem[{H{\o}g et~al.(2009)H{\o}g, Fabricius, Makarov, Urban, Corbin, Wycoff,
  Bastian, Schwenkendick, Wicenec and Turon}]{tycho2}
\bibinfo{author}{H{\o}g, E.}, \bibinfo{author}{Fabricius, C.},
  \bibinfo{author}{Makarov, V.}, \bibinfo{author}{Urban, S.},
  \bibinfo{author}{Corbin, T.}, \bibinfo{author}{Wycoff, G.},
  \bibinfo{author}{Bastian, U.}, \bibinfo{author}{Schwenkendick, P.},
  \bibinfo{author}{Wicenec, A.}, \bibinfo{author}{Turon, C.},
  \bibinfo{year}{2009}.
\newblock \bibinfo{title}{{The Tycho-2 Catalogue of the 2.5 million brightest
  stars. Commentary}}.
\newblock \bibinfo{journal}{Astronomy and Astrophysics} \bibinfo{volume}{500},
  \bibinfo{pages}{583--588}.
%Type = Article
\bibitem[{Lasker et~al.(1999)Lasker, Russel and Jenkner}]{gsc_act}
\bibinfo{author}{Lasker, B.}, \bibinfo{author}{Russel, J.},
  \bibinfo{author}{Jenkner, H.}, \bibinfo{year}{1999}.
\newblock \bibinfo{title}{{The HST Guide Star Catalog, Version GSC-ACT (Lasker+
  1996-99)}}.
\newblock \bibinfo{journal}{{VizieR Online Data Catalog}}
  \bibinfo{volume}{I/255}.
%Type = Article
\bibitem[{Lasker et~al.(1996a)Lasker, Russel, Jenkner, Sturch, McLean and
  Shara}]{usnoa1}
\bibinfo{author}{Lasker, B.}, \bibinfo{author}{Russel, J.},
  \bibinfo{author}{Jenkner, H.}, \bibinfo{author}{Sturch, C.},
  \bibinfo{author}{McLean, B.}, \bibinfo{author}{Shara, M.},
  \bibinfo{year}{1996}a.
\newblock \bibinfo{title}{{The 491,848,883 Sources in USNO-A1.0}}.
\newblock \bibinfo{journal}{{Bulletin of the American Astronomical Society}}
  \bibinfo{volume}{28}, \bibinfo{pages}{905}.
%Type = Article
\bibitem[{Lasker et~al.(1996b)Lasker, Russel, Jenkner, Sturch, McLean and
  Shara}]{gsc1.1}
\bibinfo{author}{Lasker, B.}, \bibinfo{author}{Russel, J.},
  \bibinfo{author}{Jenkner, H.}, \bibinfo{author}{Sturch, C.},
  \bibinfo{author}{McLean, B.}, \bibinfo{author}{Shara, M.},
  \bibinfo{year}{1996}b.
\newblock \bibinfo{title}{{The HST Guide Star Catalog Version 1.1}}.
\newblock \bibinfo{journal}{{The Association of Universities for Research in
  Astronomy, Inc}} .
%Type = Article
\bibitem[{Lindegren et~al.(2016)Lindegren, Lammers, Bastian, Hern{\'a}ndez,
  Klioner, Hobbs, Bombrun, Michalik, Ramos-Lerate, Butkevich et~al.}]{gaiadr1}
\bibinfo{author}{Lindegren, L.}, \bibinfo{author}{Lammers, U.},
  \bibinfo{author}{Bastian, U.}, \bibinfo{author}{Hern{\'a}ndez, J.},
  \bibinfo{author}{Klioner, S.}, \bibinfo{author}{Hobbs, D.},
  \bibinfo{author}{Bombrun, A.}, \bibinfo{author}{Michalik, D.},
  \bibinfo{author}{Ramos-Lerate, M.}, \bibinfo{author}{Butkevich, A.}, et~al.,
  \bibinfo{year}{2016}.
\newblock \bibinfo{title}{Gaia data release 1-astrometry: one billion
  positions, two million proper motions and parallaxes}.
\newblock \bibinfo{journal}{Astronomy \& Astrophysics} \bibinfo{volume}{595},
  \bibinfo{pages}{A4}.
%Type = Article
\bibitem[{{Ma} et~al.(2009){Ma}, {Arias}, {Bianco}, {Boboltz}, {Bolotin},
  {Charlot}, {Engelhardt}, {Fey}, {Gaume}, {Gontier}, {Heinkelmann}, {Jacobs},
  {Kurdubov}, {Lambert}, {Malkin}, {Nothnagel}, {Petrov}, {Skurikhina},
  {Sokolova}, {Souchay}, {Sovers}, {Tesmer}, {Titov}, {Wang}, {Zharov},
  {Barache}, {Boeckmann}, {Collioud}, {Gipson}, {Gordon}, {Lytvyn}, {MacMillan}
  and {Ojha}}]{icrf2}
\bibinfo{author}{{Ma}, C.}, \bibinfo{author}{{Arias}, E.F.},
  \bibinfo{author}{{Bianco}, G.}, \bibinfo{author}{{Boboltz}, D.A.},
  \bibinfo{author}{{Bolotin}, S.L.}, \bibinfo{author}{{Charlot}, P.},
  \bibinfo{author}{{Engelhardt}, G.}, \bibinfo{author}{{Fey}, A.L.},
  \bibinfo{author}{{Gaume}, R.A.}, \bibinfo{author}{{Gontier}, A.M.},
  \bibinfo{author}{{Heinkelmann}, R.}, \bibinfo{author}{{Jacobs}, C.S.},
  \bibinfo{author}{{Kurdubov}, S.}, \bibinfo{author}{{Lambert}, S.B.},
  \bibinfo{author}{{Malkin}, Z.M.}, \bibinfo{author}{{Nothnagel}, A.},
  \bibinfo{author}{{Petrov}, L.}, \bibinfo{author}{{Skurikhina}, E.},
  \bibinfo{author}{{Sokolova}, J.R.}, \bibinfo{author}{{Souchay}, J.},
  \bibinfo{author}{{Sovers}, O.J.}, \bibinfo{author}{{Tesmer}, V.},
  \bibinfo{author}{{Titov}, O.A.}, \bibinfo{author}{{Wang}, G.},
  \bibinfo{author}{{Zharov}, V.E.}, \bibinfo{author}{{Barache}, C.},
  \bibinfo{author}{{Boeckmann}, S.}, \bibinfo{author}{{Collioud}, A.},
  \bibinfo{author}{{Gipson}, J.M.}, \bibinfo{author}{{Gordon}, D.},
  \bibinfo{author}{{Lytvyn}, S.O.}, \bibinfo{author}{{MacMillan}, D.S.},
  \bibinfo{author}{{Ojha}, R.}, \bibinfo{year}{2009}.
\newblock \bibinfo{title}{{The Second Realization of the International
  Celestial Reference Frame by Very Long Baseline Interferometry}}.
\newblock \bibinfo{journal}{IERS Technical Note} \bibinfo{volume}{35}.
%Type = Book
\bibitem[{Milani and Gronchi(2010)}]{milani10}
\bibinfo{author}{Milani, A.}, \bibinfo{author}{Gronchi, G.},
  \bibinfo{year}{2010}.
\newblock \bibinfo{title}{Theory of orbit determination}.
\newblock \bibinfo{publisher}{Cambridge University Press}.
%Type = Inproceedings
\bibitem[{Monet(1998)}]{usnoa2}
\bibinfo{author}{Monet, D.G.}, \bibinfo{year}{1998}.
\newblock \bibinfo{title}{{The 526,280,881 objects in the USNO-A2. 0 catalog}},
  in: \bibinfo{booktitle}{Bulletin of the American Astronomical Society}, p.
  \bibinfo{pages}{1427}.
%Type = Misc
\bibitem[{Monet(2018)}]{sstrc5}
\bibinfo{author}{Monet, D.G.}, \bibinfo{year}{2018}.
\newblock \bibinfo{title}{{SST RC 5 star catalog}}.
\newblock \bibinfo{howpublished}{private communication}.
%Type = Article
\bibitem[{Monet et~al.(2003)Monet, Levine, Canzian, Ables, Bird, Dahn, Guetter,
  Harris, Henden, Leggett et~al.}]{usnob1}
\bibinfo{author}{Monet, D.G.}, \bibinfo{author}{Levine, S.E.},
  \bibinfo{author}{Canzian, B.}, \bibinfo{author}{Ables, H.D.},
  \bibinfo{author}{Bird, A.R.}, \bibinfo{author}{Dahn, C.C.},
  \bibinfo{author}{Guetter, H.H.}, \bibinfo{author}{Harris, H.C.},
  \bibinfo{author}{Henden, A.A.}, \bibinfo{author}{Leggett, S.K.}, et~al.,
  \bibinfo{year}{2003}.
\newblock \bibinfo{title}{{The USNO-B Catalog}}.
\newblock \bibinfo{journal}{The Astronomical Journal} \bibinfo{volume}{125},
  \bibinfo{pages}{984}.
%Type = Article
\bibitem[{Morrison et~al.(2001)Morrison, Roeser, McLean, Bucciarelli and
  Lasker}]{gsc1.2}
\bibinfo{author}{Morrison, J.}, \bibinfo{author}{Roeser, S.},
  \bibinfo{author}{McLean, B.}, \bibinfo{author}{Bucciarelli, B.},
  \bibinfo{author}{Lasker, B.}, \bibinfo{year}{2001}.
\newblock \bibinfo{title}{{The guide star catalog, version 1.2: An astrometric
  recalibration and other refinements}}.
\newblock \bibinfo{journal}{The Astronomical Journal} \bibinfo{volume}{121},
  \bibinfo{pages}{1752}.
%Type = Article
\bibitem[{Ochsenbein et~al.(2000)Ochsenbein, Bauer and Marcout}]{vizier}
\bibinfo{author}{Ochsenbein, F.}, \bibinfo{author}{Bauer, P.},
  \bibinfo{author}{Marcout, J.}, \bibinfo{year}{2000}.
\newblock \bibinfo{title}{The vizier database of astronomical catalogues}.
\newblock \bibinfo{journal}{Astronomy and Astrophysics Supplement Series}
  \bibinfo{volume}{143}, \bibinfo{pages}{23--32}.
%Type = Inproceedings
\bibitem[{Roeser and Bastian(1991)}]{ppm}
\bibinfo{author}{Roeser, S.}, \bibinfo{author}{Bastian, U.},
  \bibinfo{year}{1991}.
\newblock \bibinfo{title}{{PPM Star Catalogue North. Positions and proper
  motions of 181.731 stars north of-2.5 degrees declination for equinox and
  epoch. J2000. 0. Vol. 1: Zones+ 80 deg. to+ 30 deg.; Vol. 2: Zones+ 20 deg.
  to-0 deg.}}, in: \bibinfo{booktitle}{Predictability, Stability, and Chaos in
  N-Body Dynamical Systems}.
%Type = Article
\bibitem[{Roeser et~al.(2010)Roeser, Demleitner and Schilbach}]{ppmxl}
\bibinfo{author}{Roeser, S.}, \bibinfo{author}{Demleitner, M.},
  \bibinfo{author}{Schilbach, E.}, \bibinfo{year}{2010}.
\newblock \bibinfo{title}{{The PPMXL catalog of positions and proper motions on
  the ICRS. Combining USNO-B1. 0 and the Two Micron All Sky Survey (2MASS)}}.
\newblock \bibinfo{journal}{The Astronomical Journal} \bibinfo{volume}{139},
  \bibinfo{pages}{2440}.
%Type = Article
\bibitem[{Schaback(1995)}]{schaback1995}
\bibinfo{author}{Schaback, R.}, \bibinfo{year}{1995}.
\newblock \bibinfo{title}{Creating surfaces from scattered data using radial
  basis functions}.
\newblock \bibinfo{journal}{Mathematical methods for curves and surfaces}
  \bibinfo{volume}{477}.
%Type = Article
\bibitem[{Skrutskie et~al.(2006)Skrutskie, Cutri, Stiening, Weinberg,
  Schneider, Carpenter, Beichman, Capps, Chester, Elias et~al.}]{2mass}
\bibinfo{author}{Skrutskie, M.}, \bibinfo{author}{Cutri, R.},
  \bibinfo{author}{Stiening, R.}, \bibinfo{author}{Weinberg, M.},
  \bibinfo{author}{Schneider, S.}, \bibinfo{author}{Carpenter, J.},
  \bibinfo{author}{Beichman, C.}, \bibinfo{author}{Capps, R.},
  \bibinfo{author}{Chester, T.}, \bibinfo{author}{Elias, J.}, et~al.,
  \bibinfo{year}{2006}.
\newblock \bibinfo{title}{{The two micron all sky survey (2MASS)}}.
\newblock \bibinfo{journal}{The Astronomical Journal} \bibinfo{volume}{131},
  \bibinfo{pages}{1163}.
%Type = Article
\bibitem[{Urban et~al.(1998)Urban, Corbin and Wycoff}]{act}
\bibinfo{author}{Urban, S.E.}, \bibinfo{author}{Corbin, T.E.},
  \bibinfo{author}{Wycoff, G.L.}, \bibinfo{year}{1998}.
\newblock \bibinfo{title}{{The {ACT} Reference Catalog}}.
\newblock \bibinfo{journal}{The Astronomical Journal} \bibinfo{volume}{115},
  \bibinfo{pages}{2161--2166}.
\newblock \URLprefix \url{https://doi.org/10.1086%2F300344},
  \DOIprefix\doi{10.1086/300344}.
%Type = Article
\bibitem[{Vere{\v{s}} et~al.(2017)Vere{\v{s}}, Farnocchia, Chesley and
  Chamberlin}]{verevs2017}
\bibinfo{author}{Vere{\v{s}}, P.}, \bibinfo{author}{Farnocchia, D.},
  \bibinfo{author}{Chesley, S.R.}, \bibinfo{author}{Chamberlin, A.B.},
  \bibinfo{year}{2017}.
\newblock \bibinfo{title}{Statistical analysis of astrometric errors for the
  most productive asteroid surveys}.
\newblock \bibinfo{journal}{Icarus} \bibinfo{volume}{296},
  \bibinfo{pages}{139--149}.
%Type = Article
\bibitem[{{Zacharias} et~al.(2017){Zacharias}, {Finch} and {Frouard}}]{ucac5}
\bibinfo{author}{{Zacharias}, N.}, \bibinfo{author}{{Finch}, C.},
  \bibinfo{author}{{Frouard}, J.}, \bibinfo{year}{2017}.
\newblock \bibinfo{title}{{VizieR Online Data Catalog: UCAC5 Catalogue
  (Zacharias+ 2017)}}.
\newblock \bibinfo{journal}{VizieR Online Data Catalog} ,
  \bibinfo{pages}{I/340}.
%Type = Article
\bibitem[{Zacharias et~al.(2010)Zacharias, Finch, Girard, Hambly, Wycoff,
  Zacharias, Castillo, Corbin, DiVittorio, Dutta, Gaume, Gauss, Germain, Hall,
  Hartkopf, Hsu, Holdenried, Makarov, Martines, Mason, Monet, Rafferty, Rhodes,
  Siemers, Smith, Tilleman, Urban, Wieder, Winter and Young}]{ucac3}
\bibinfo{author}{Zacharias, N.}, \bibinfo{author}{Finch, C.},
  \bibinfo{author}{Girard, T.}, \bibinfo{author}{Hambly, N.},
  \bibinfo{author}{Wycoff, G.}, \bibinfo{author}{Zacharias, M.I.},
  \bibinfo{author}{Castillo, D.}, \bibinfo{author}{Corbin, T.},
  \bibinfo{author}{DiVittorio, M.}, \bibinfo{author}{Dutta, S.},
  \bibinfo{author}{Gaume, R.}, \bibinfo{author}{Gauss, S.},
  \bibinfo{author}{Germain, M.}, \bibinfo{author}{Hall, D.},
  \bibinfo{author}{Hartkopf, W.}, \bibinfo{author}{Hsu, D.},
  \bibinfo{author}{Holdenried, E.}, \bibinfo{author}{Makarov, V.},
  \bibinfo{author}{Martines, M.}, \bibinfo{author}{Mason, B.},
  \bibinfo{author}{Monet, D.}, \bibinfo{author}{Rafferty, T.},
  \bibinfo{author}{Rhodes, A.}, \bibinfo{author}{Siemers, T.},
  \bibinfo{author}{Smith, D.}, \bibinfo{author}{Tilleman, T.},
  \bibinfo{author}{Urban, S.}, \bibinfo{author}{Wieder, G.},
  \bibinfo{author}{Winter, L.}, \bibinfo{author}{Young, A.},
  \bibinfo{year}{2010}.
\newblock \bibinfo{title}{{The thrid US Naval Observatory CCD Astrograph
  Catalog UCAC 3}}.
\newblock \bibinfo{journal}{The Astronomical Journal} \bibinfo{volume}{139},
  \bibinfo{pages}{2184--2199}.
\newblock \URLprefix
  \url{{https://doi.org/10.1088%2F0004-6256%2F139%2F6%2F2184}},
  \DOIprefix\doi{10.1088/0004-6256/139/6/2184}.
%Type = Article
\bibitem[{{Zacharias} et~al.(2015){Zacharias}, {Finch}, {Subasavage},
  {Bredthauer}, {Crockett}, {Divittorio}, {Ferguson}, {Harris}, {Harris},
  {Henden}, {Kilian}, {Munn}, {Rafferty}, {Rhodes}, {Schultheiss}, {Tilleman}
  and {Wieder}}]{urat1}
\bibinfo{author}{{Zacharias}, N.}, \bibinfo{author}{{Finch}, C.},
  \bibinfo{author}{{Subasavage}, J.}, \bibinfo{author}{{Bredthauer}, G.},
  \bibinfo{author}{{Crockett}, C.}, \bibinfo{author}{{Divittorio}, M.},
  \bibinfo{author}{{Ferguson}, E.}, \bibinfo{author}{{Harris}, F.},
  \bibinfo{author}{{Harris}, H.}, \bibinfo{author}{{Henden}, A.},
  \bibinfo{author}{{Kilian}, C.}, \bibinfo{author}{{Munn}, J.},
  \bibinfo{author}{{Rafferty}, T.}, \bibinfo{author}{{Rhodes}, A.},
  \bibinfo{author}{{Schultheiss}, M.}, \bibinfo{author}{{Tilleman}, T.},
  \bibinfo{author}{{Wieder}, G.}, \bibinfo{year}{2015}.
\newblock \bibinfo{title}{{The First U.S. Naval Observatory Robotic Astrometric
  Telescope Catalog}}.
\newblock \bibinfo{journal}{\aj} \bibinfo{volume}{150}, \bibinfo{pages}{101}.
\newblock \DOIprefix\doi{10.1088/0004-6256/150/4/101},
  \href{http://arxiv.org/abs/1508.04637}{\tt arXiv:1508.04637}.
%Type = Article
\bibitem[{Zacharias et~al.(2013)Zacharias, Finch, Girard, Henden, Bartlett,
  Monet and Zacharias}]{ucac4}
\bibinfo{author}{Zacharias, N.}, \bibinfo{author}{Finch, C.T.},
  \bibinfo{author}{Girard, T.M.}, \bibinfo{author}{Henden, A.},
  \bibinfo{author}{Bartlett, J.L.}, \bibinfo{author}{Monet, D.G.},
  \bibinfo{author}{Zacharias, M.I.}, \bibinfo{year}{2013}.
\newblock \bibinfo{title}{{The fourth US Naval Observatory CCD Astrograph
  Catalog UCAC 4}}.
\newblock \bibinfo{journal}{The Astronomical Journal} \bibinfo{volume}{145},
  \bibinfo{pages}{44}.
\newblock \URLprefix \url{https://doi.org/10.1088%2F0004-6256%2F145%2F2%2F44},
  \DOIprefix\doi{10.1088/0004-6256/145/2/44}.
%Type = Inproceedings
\bibitem[{Zacharias et~al.(2004a)Zacharias, Monet, Levine, Urban, Gaume and
  Wycoff}]{nomad}
\bibinfo{author}{Zacharias, N.}, \bibinfo{author}{Monet, D.G.},
  \bibinfo{author}{Levine, S.E.}, \bibinfo{author}{Urban, S.E.},
  \bibinfo{author}{Gaume, R.}, \bibinfo{author}{Wycoff, G.L.},
  \bibinfo{year}{2004}a.
\newblock \bibinfo{title}{The naval observatory merged astrometric dataset
  (nomad)}, in: \bibinfo{booktitle}{American Astronomical Society Meeting
  Abstracts}, p. \bibinfo{pages}{1418}.
%Type = Article
\bibitem[{Zacharias et~al.(2000)Zacharias, Urban, Zacharias, Hall, Wycoff,
  Rafferty, Germain, Holdenried, Pohlman, Gauss, Monet and Winter}]{ucac1}
\bibinfo{author}{Zacharias, N.}, \bibinfo{author}{Urban, S.E.},
  \bibinfo{author}{Zacharias, M.I.}, \bibinfo{author}{Hall, D.M.},
  \bibinfo{author}{Wycoff, G.L.}, \bibinfo{author}{Rafferty, T.J.},
  \bibinfo{author}{Germain, M.E.}, \bibinfo{author}{Holdenried, E.R.},
  \bibinfo{author}{Pohlman, J.W.}, \bibinfo{author}{Gauss, F.S.},
  \bibinfo{author}{Monet, D.G.}, \bibinfo{author}{Winter, L.},
  \bibinfo{year}{2000}.
\newblock \bibinfo{title}{{The First {US} Naval Observatory {CCD} Astrograph
  Catalog}}.
\newblock \bibinfo{journal}{The Astronomical Journal} \bibinfo{volume}{120},
  \bibinfo{pages}{2131--2147}.
\newblock \URLprefix \url{https://doi.org/10.1086%2F301563},
  \DOIprefix\doi{10.1086/301563}.
%Type = Techreport
\bibitem[{Zacharias et~al.(2004b)Zacharias, Urban, Zacharias, Wycoff, Hall,
  Monet and Rafferty}]{ucac2}
\bibinfo{author}{Zacharias, N.}, \bibinfo{author}{Urban, S.E.},
  \bibinfo{author}{Zacharias, M.I.}, \bibinfo{author}{Wycoff, G.L.},
  \bibinfo{author}{Hall, D.M.}, \bibinfo{author}{Monet, D.G.},
  \bibinfo{author}{Rafferty, T.J.}, \bibinfo{year}{2004}b.
\newblock \bibinfo{title}{{The second US Naval Observatory CCD Astrograph
  Catalog (UCAC2)}}.
\newblock \bibinfo{type}{Technical Report} \bibinfo{number}{astro-ph/0403060}.
  US Naval Observatory.
\newblock \URLprefix \url{http://cds.cern.ch/record/720865}.

\end{thebibliography}

%%%
 \section*{Appendix}
\begin{figure}
 \begin{tabular}{ll}
 \includegraphics[width=0.5\linewidth]{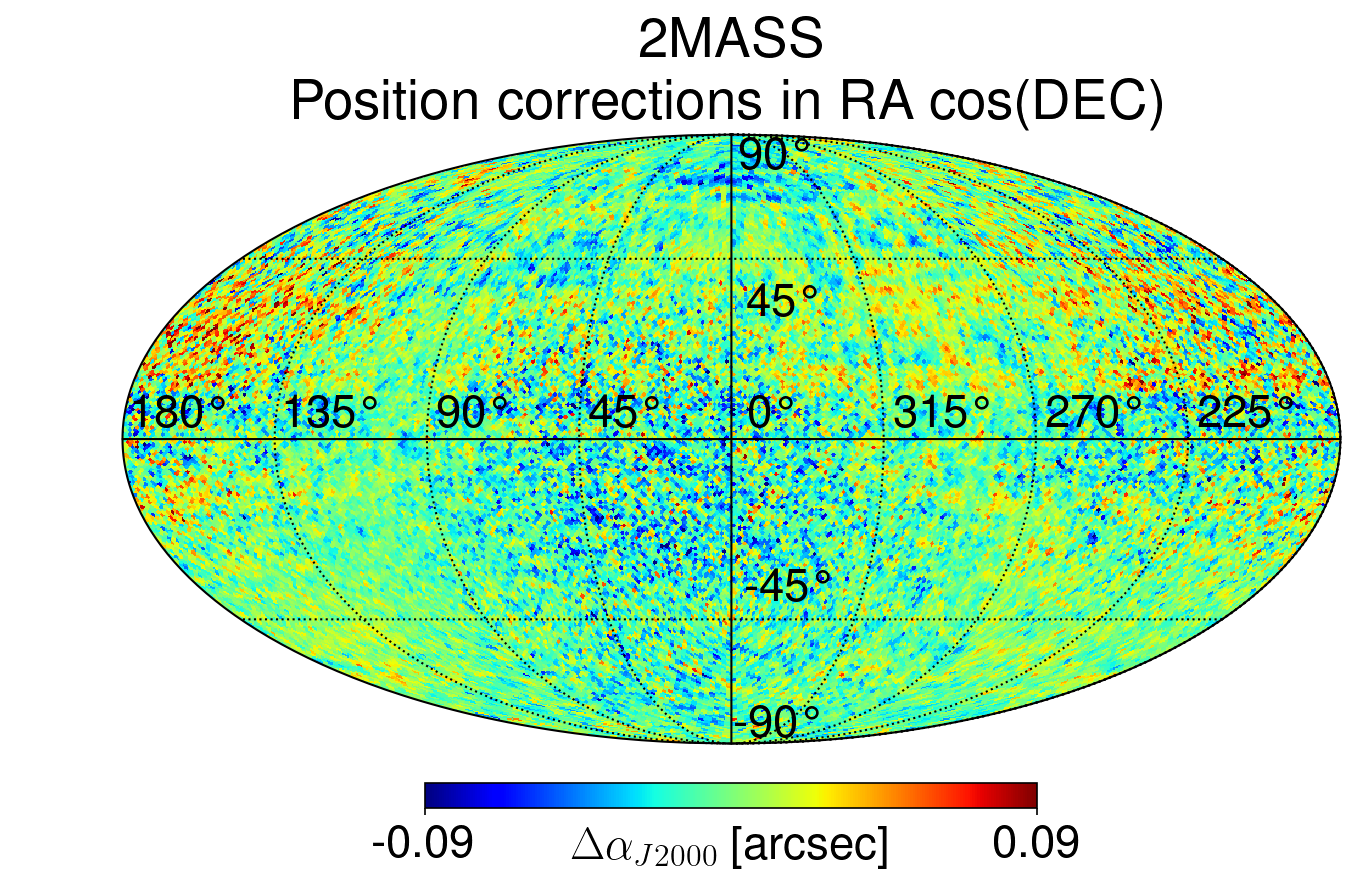} &
 \includegraphics[width=0.5\linewidth]{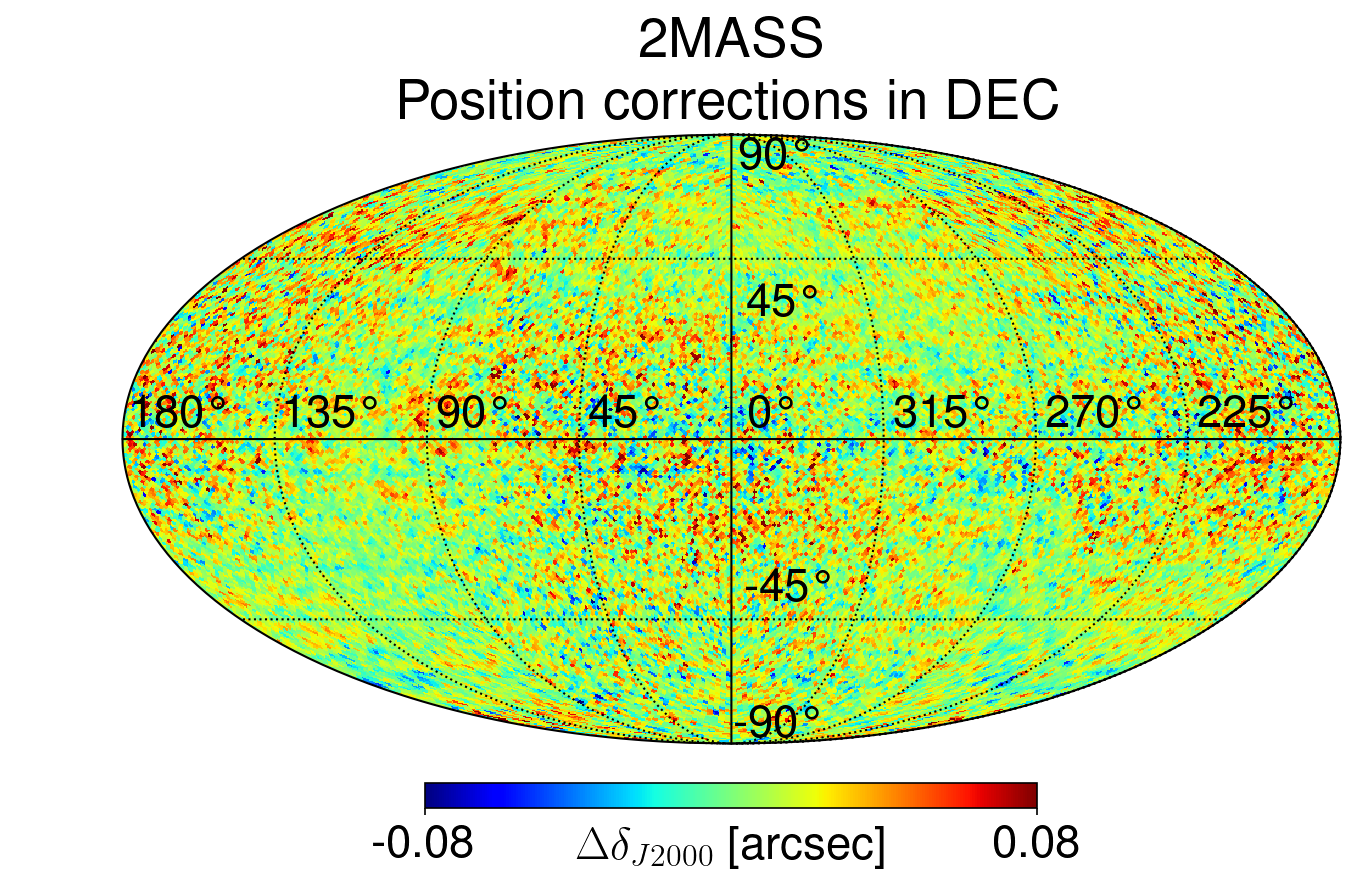} \\
 \includegraphics[width=0.5\linewidth]{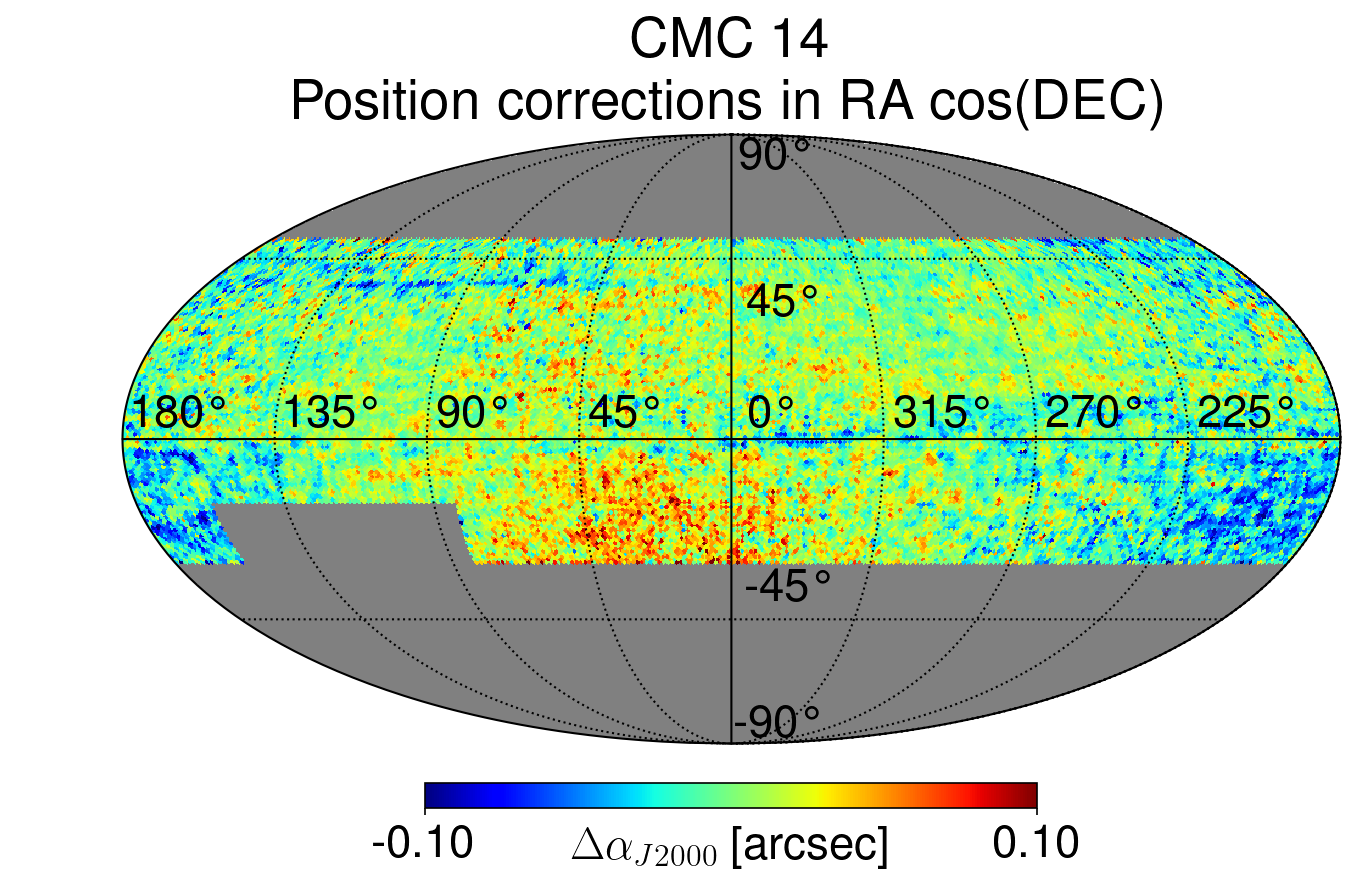} &
 \includegraphics[width=0.5\linewidth]{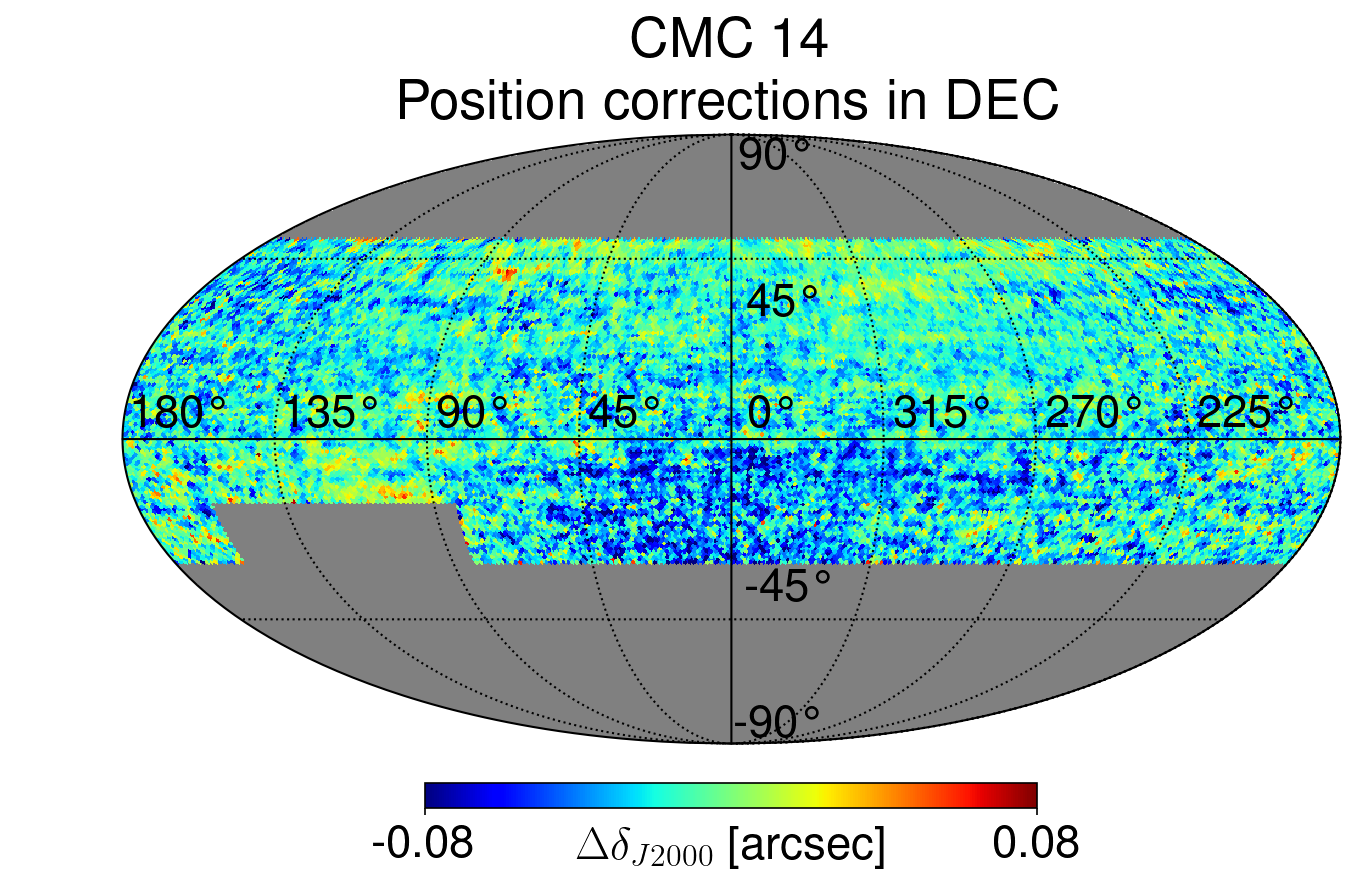} \\
 \includegraphics[width=0.5\linewidth]{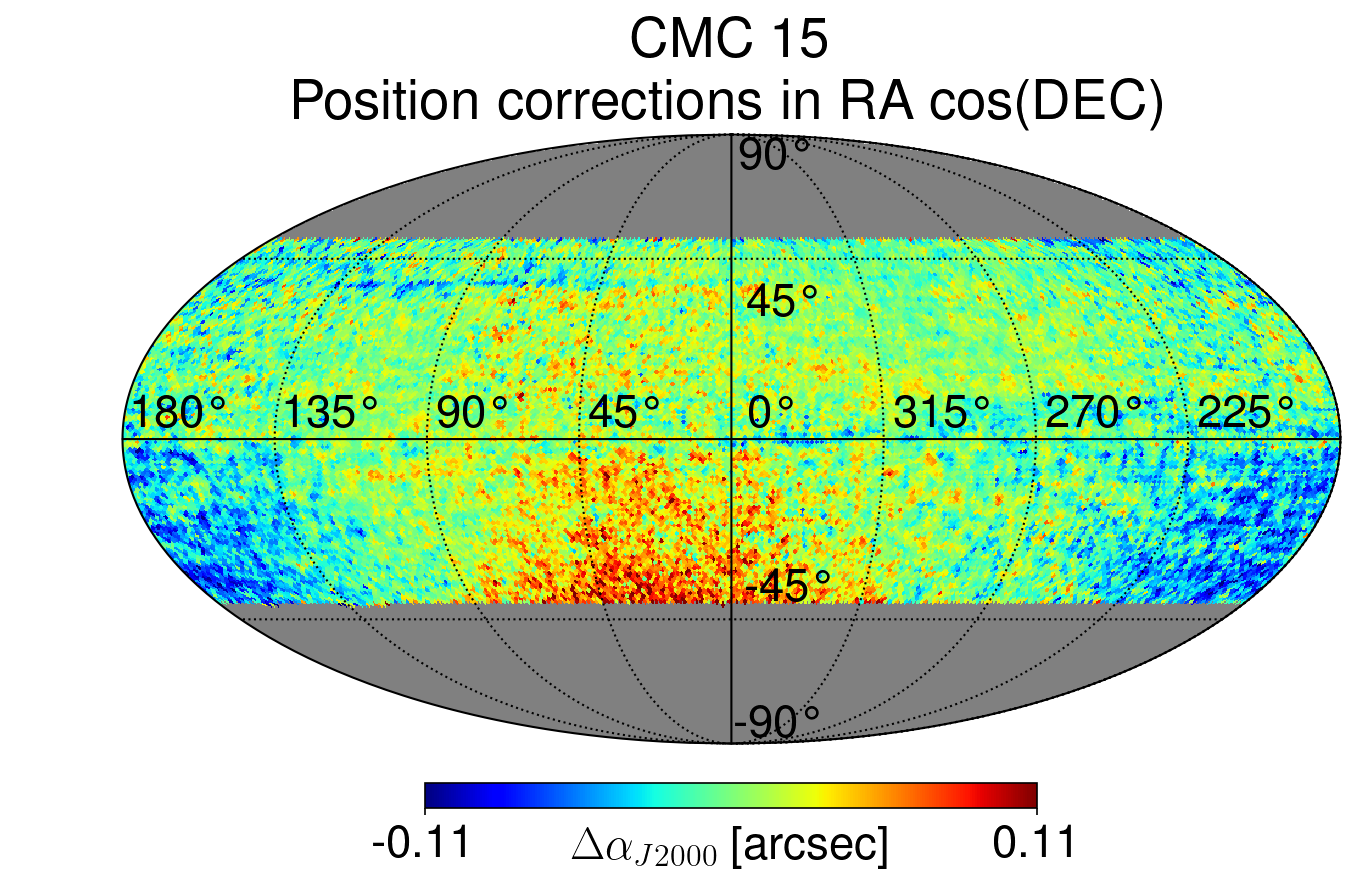} &
 \includegraphics[width=0.5\linewidth]{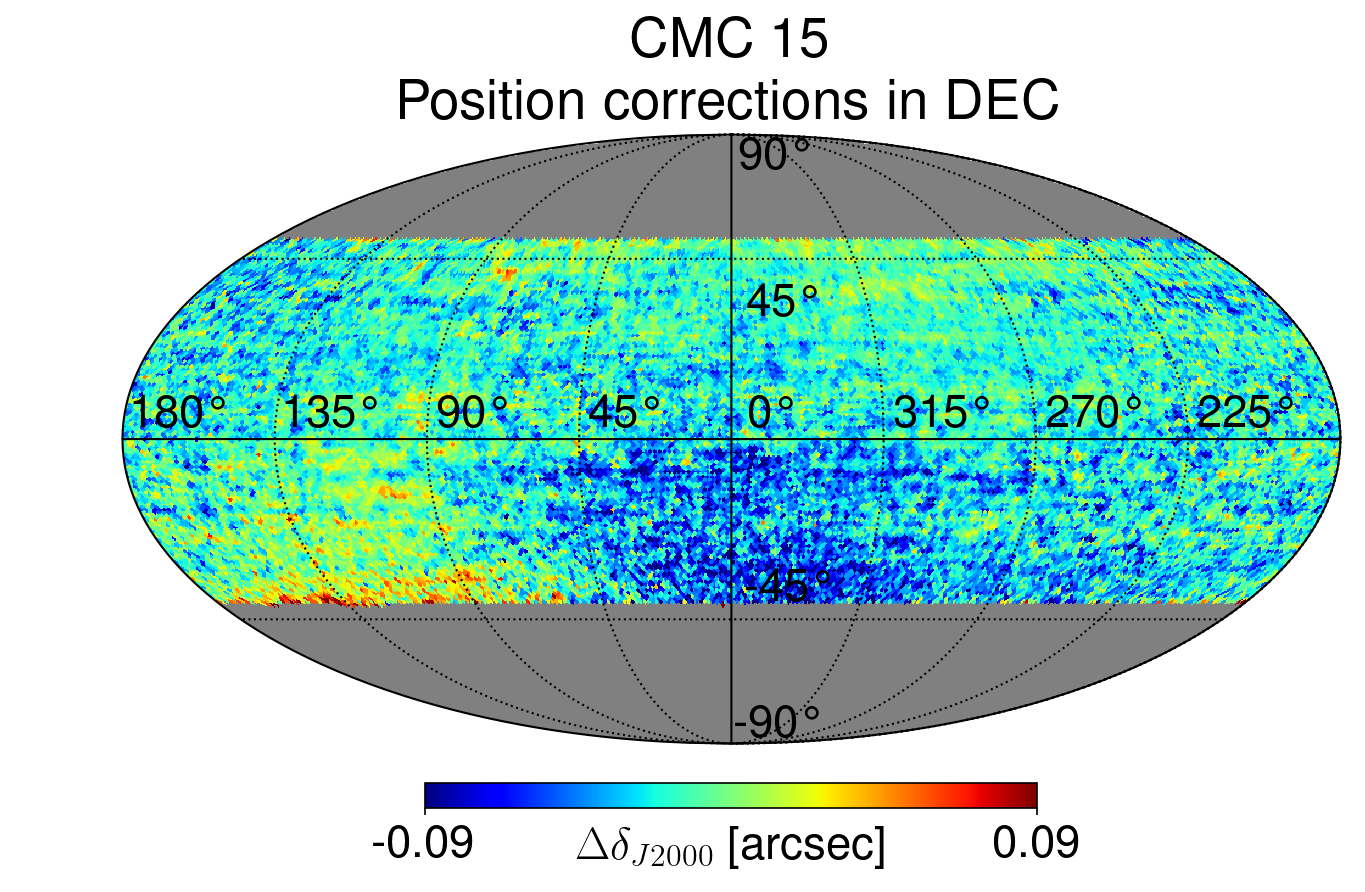} \\
 \end{tabular}
 \caption{Corrections in stellar positions for the 2MASS catalog (top panels), CMC 14
 (center panels) and CMC 15 (bottom panels). None of the above catalogs come with proper motion.\label{fig:2mass}}
 \end{figure}

\begin{figure}
 \begin{tabular}{ll}
 \includegraphics[width=0.5\linewidth]{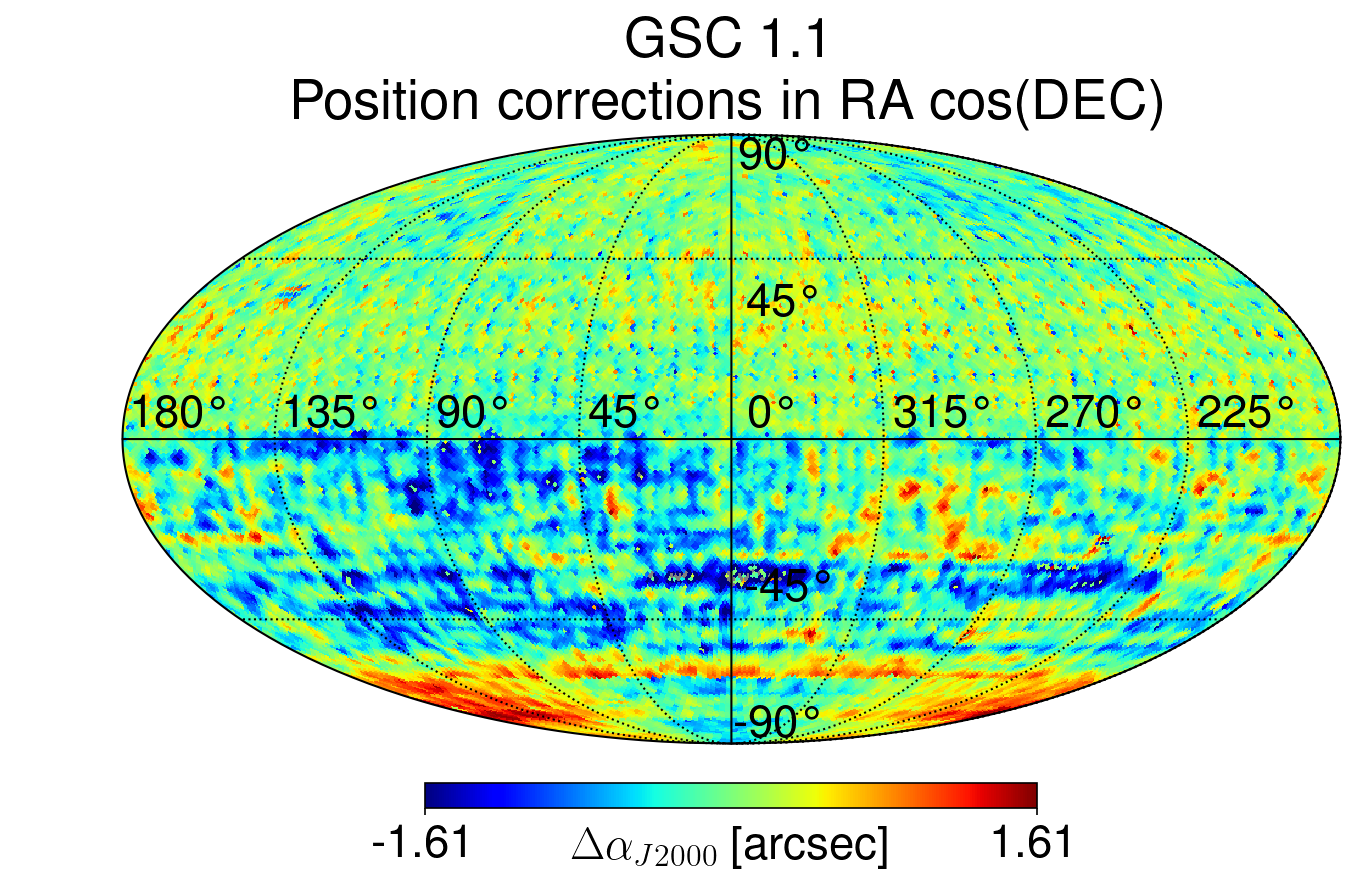} &
 \includegraphics[width=0.5\linewidth]{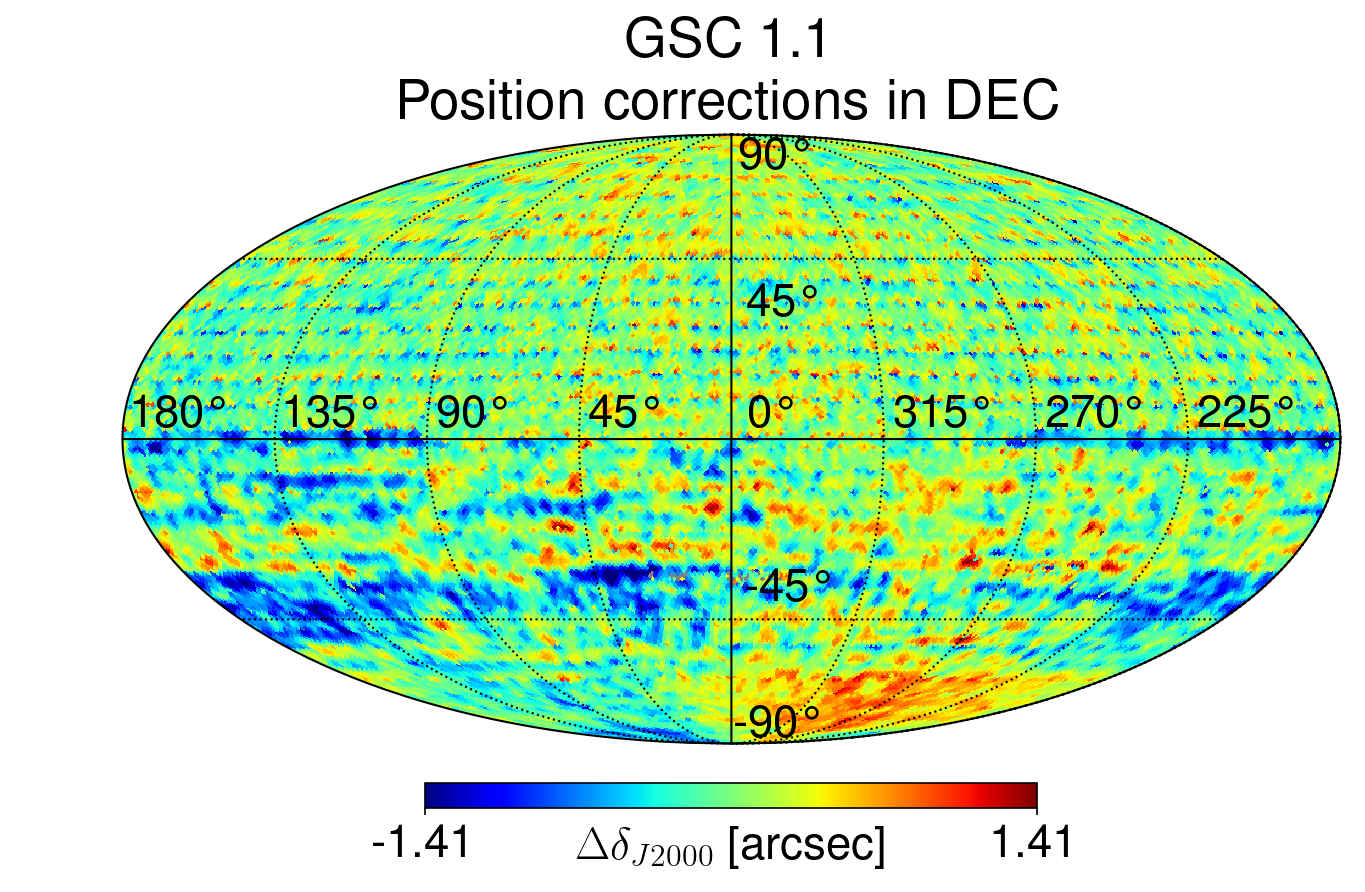} \\
 \includegraphics[width=0.5\linewidth]{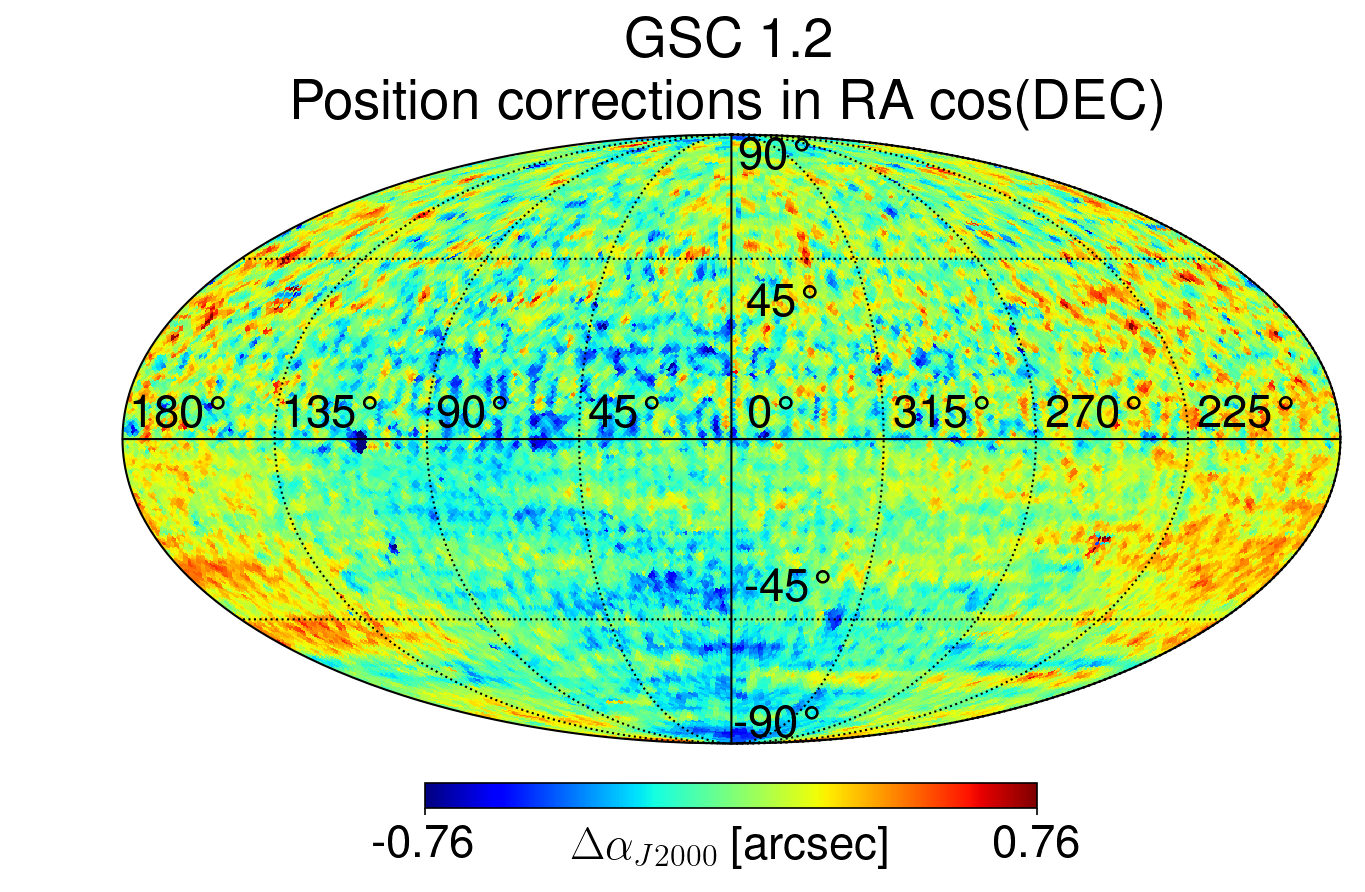} &
 \includegraphics[width=0.5\linewidth]{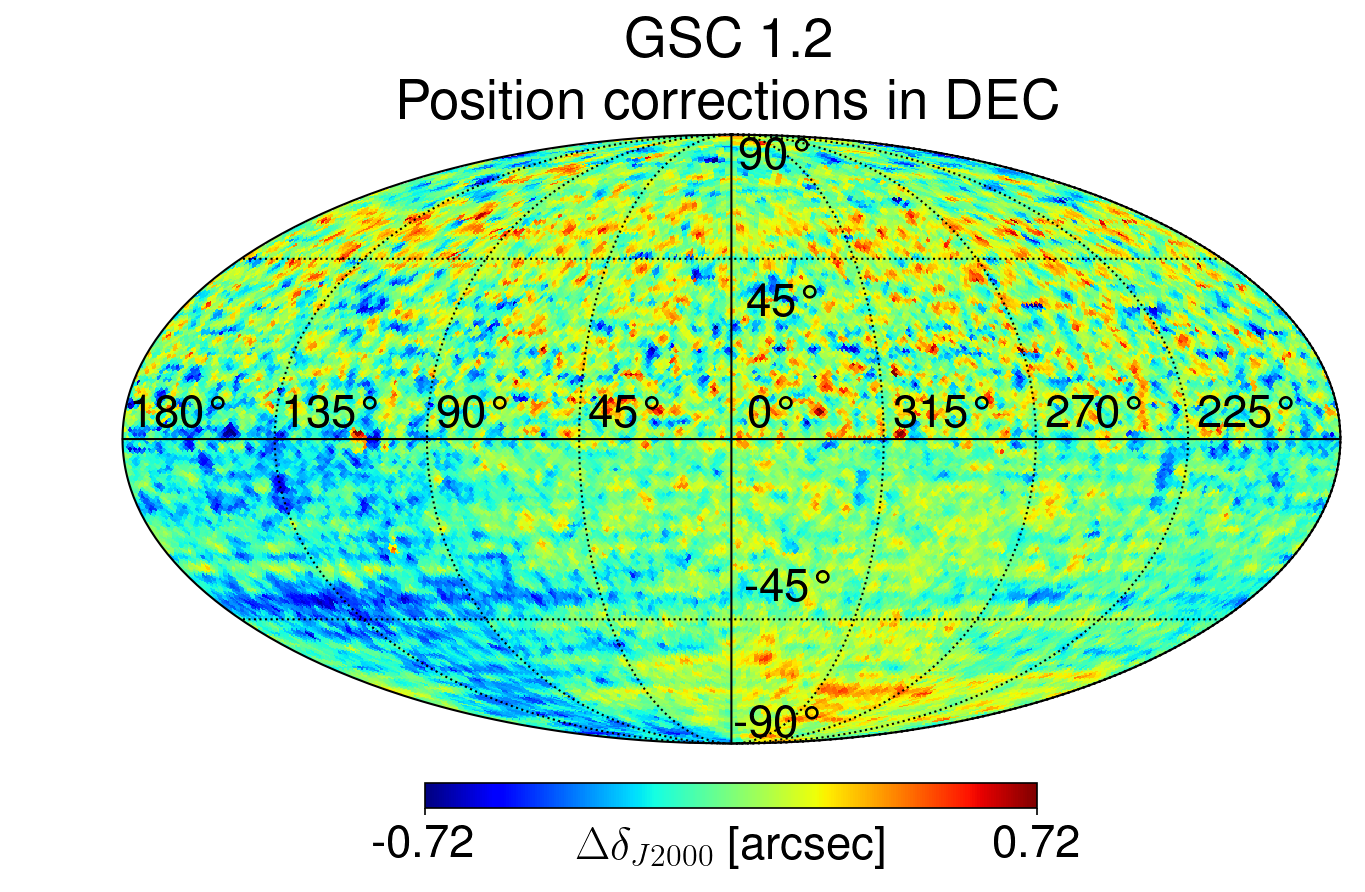} \\
 \includegraphics[width=0.5\linewidth]{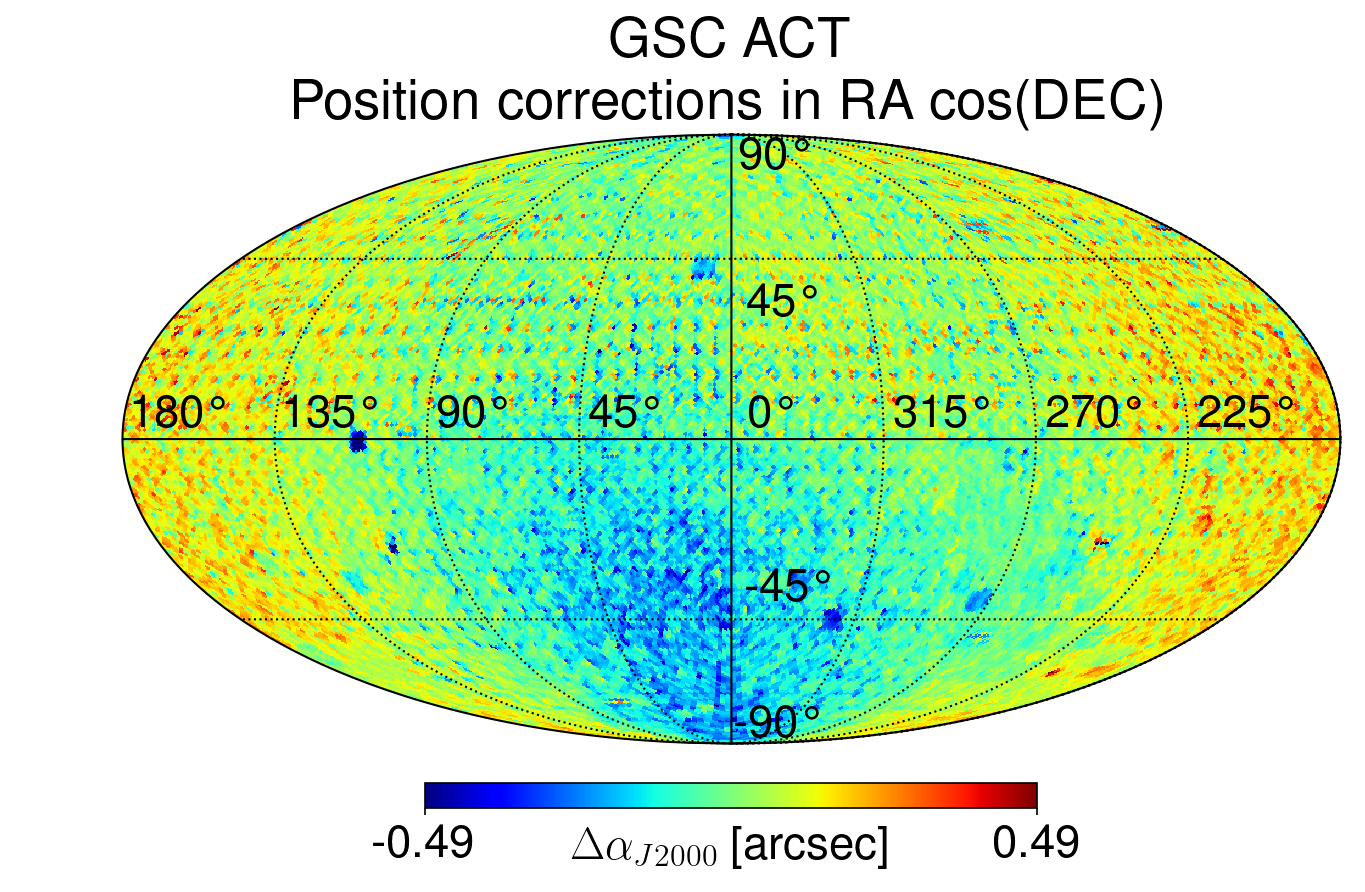} &
 \includegraphics[width=0.5\linewidth]{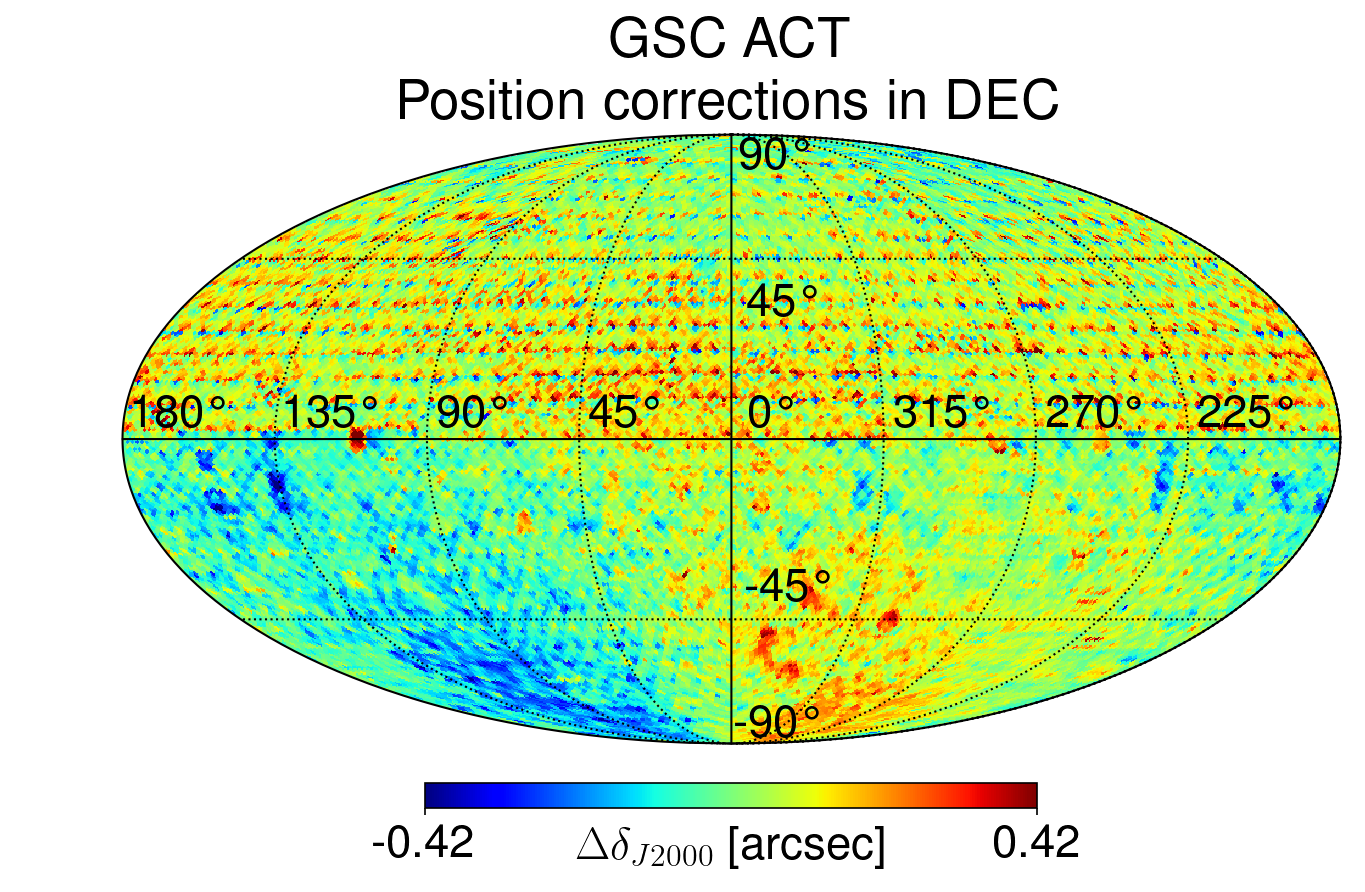} \\
 \end{tabular}
 \caption{Same as Figure \ref{fig:2mass} for the GSC 1.1 catalog (top panels), the GSC 1.2 catalog
 (center panels) and GSC ACT (bottom panels).  \label{fig:gsc}}
 \end{figure}

\begin{figure}
 \begin{tabular}{ll}
 \includegraphics[width=0.5\linewidth]{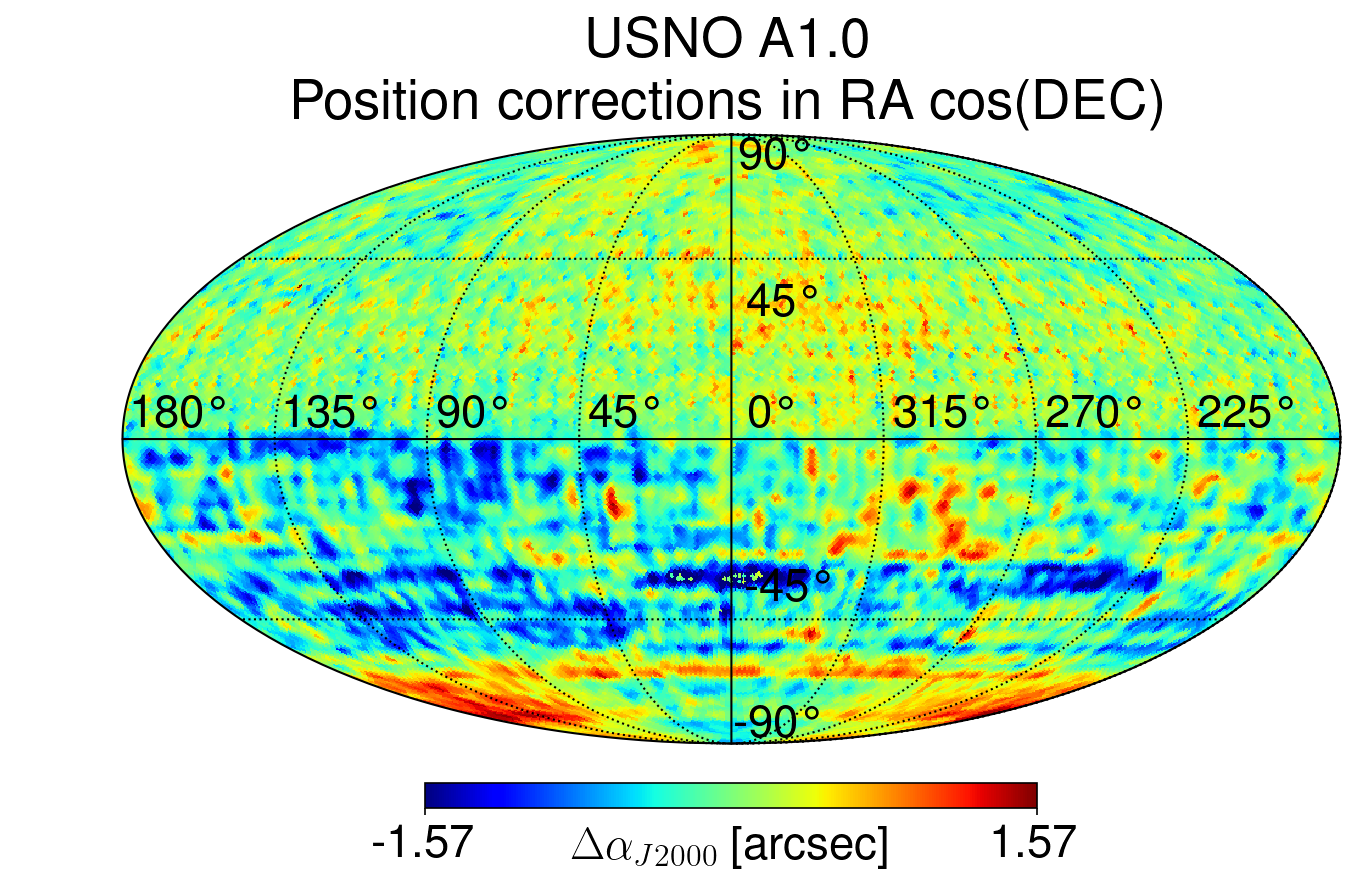} &
 \includegraphics[width=0.5\linewidth]{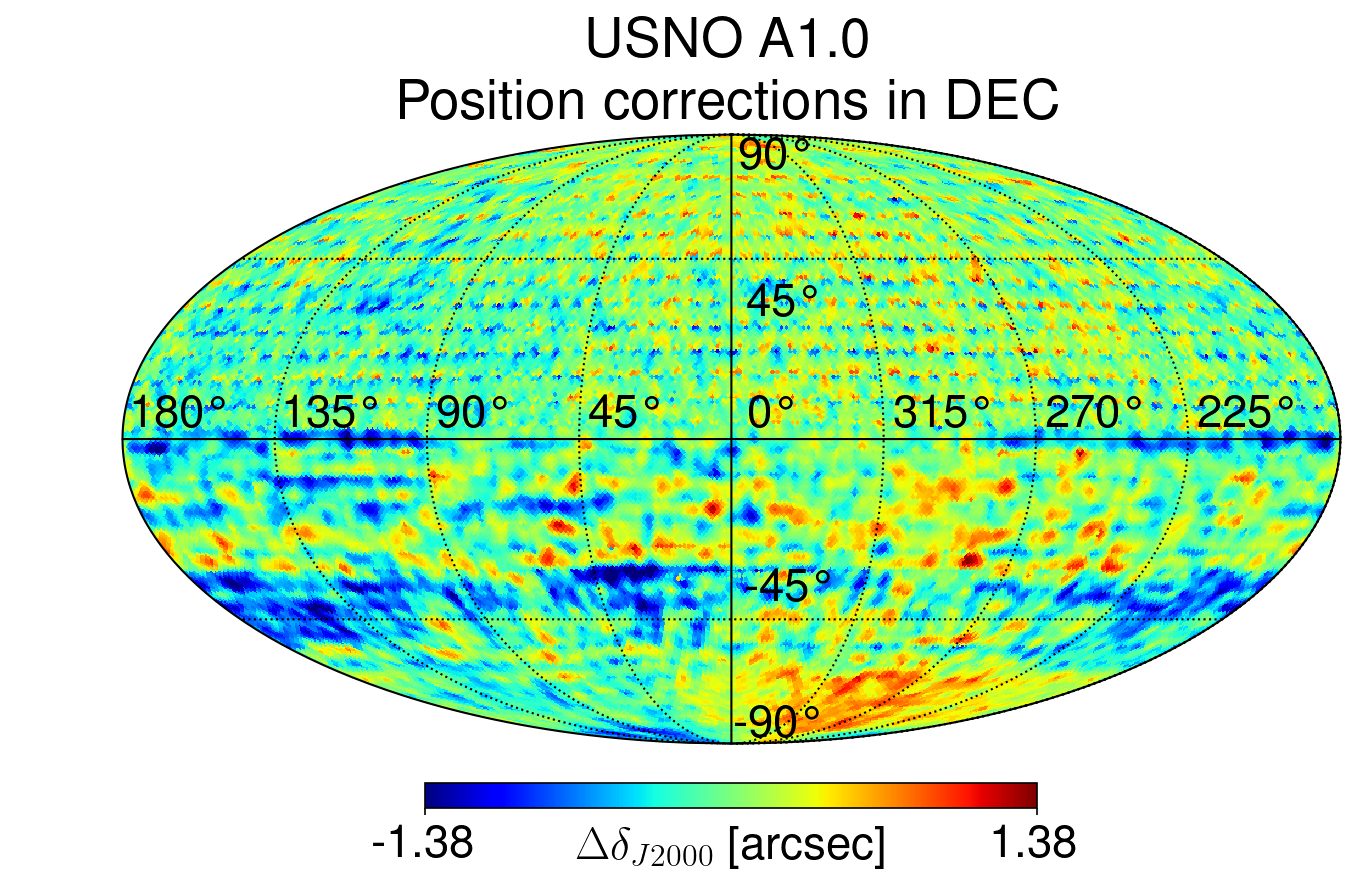} \\
 \includegraphics[width=0.5\linewidth]{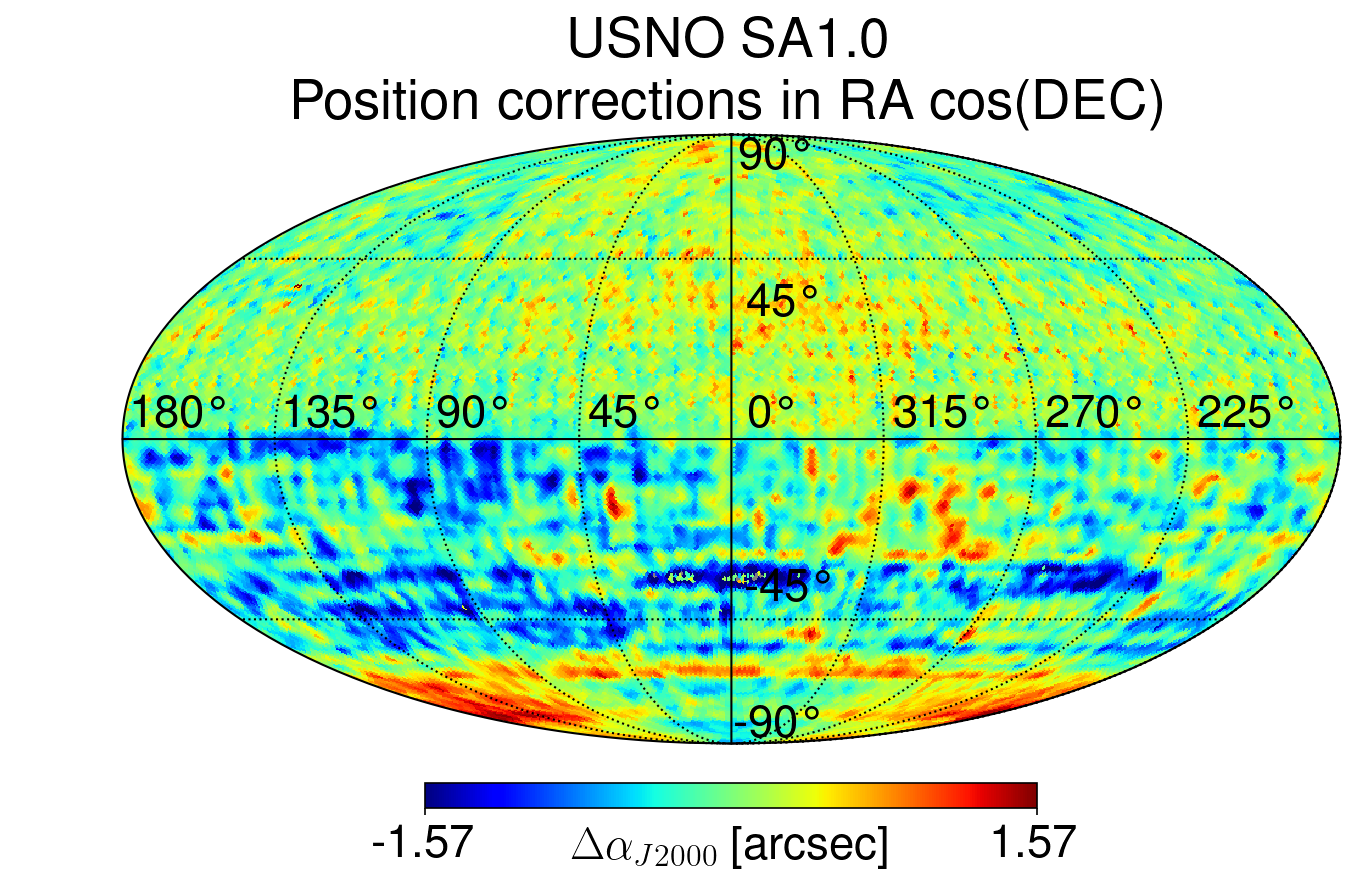} &
 \includegraphics[width=0.5\linewidth]{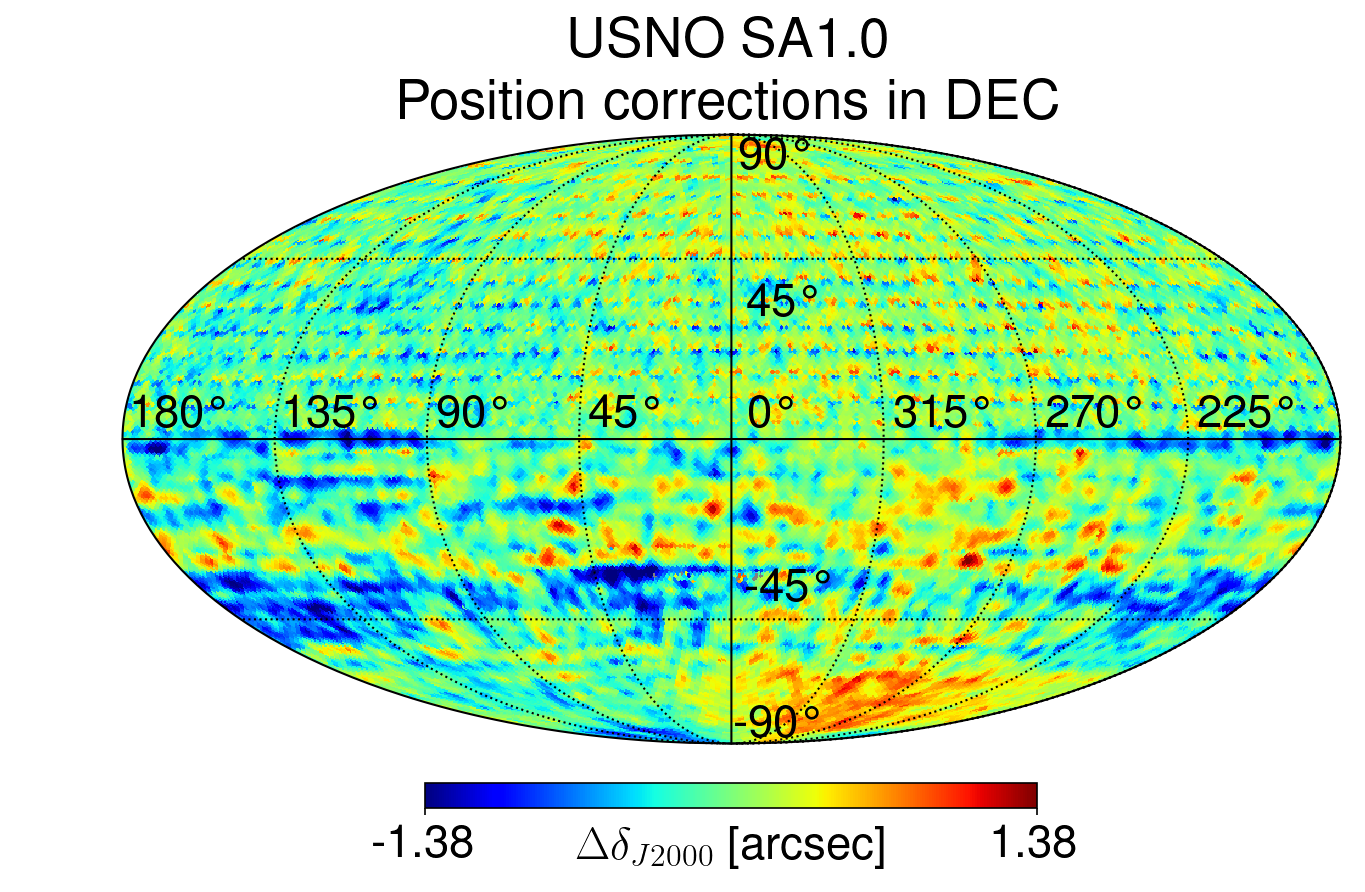} \\
 \includegraphics[width=0.5\linewidth]{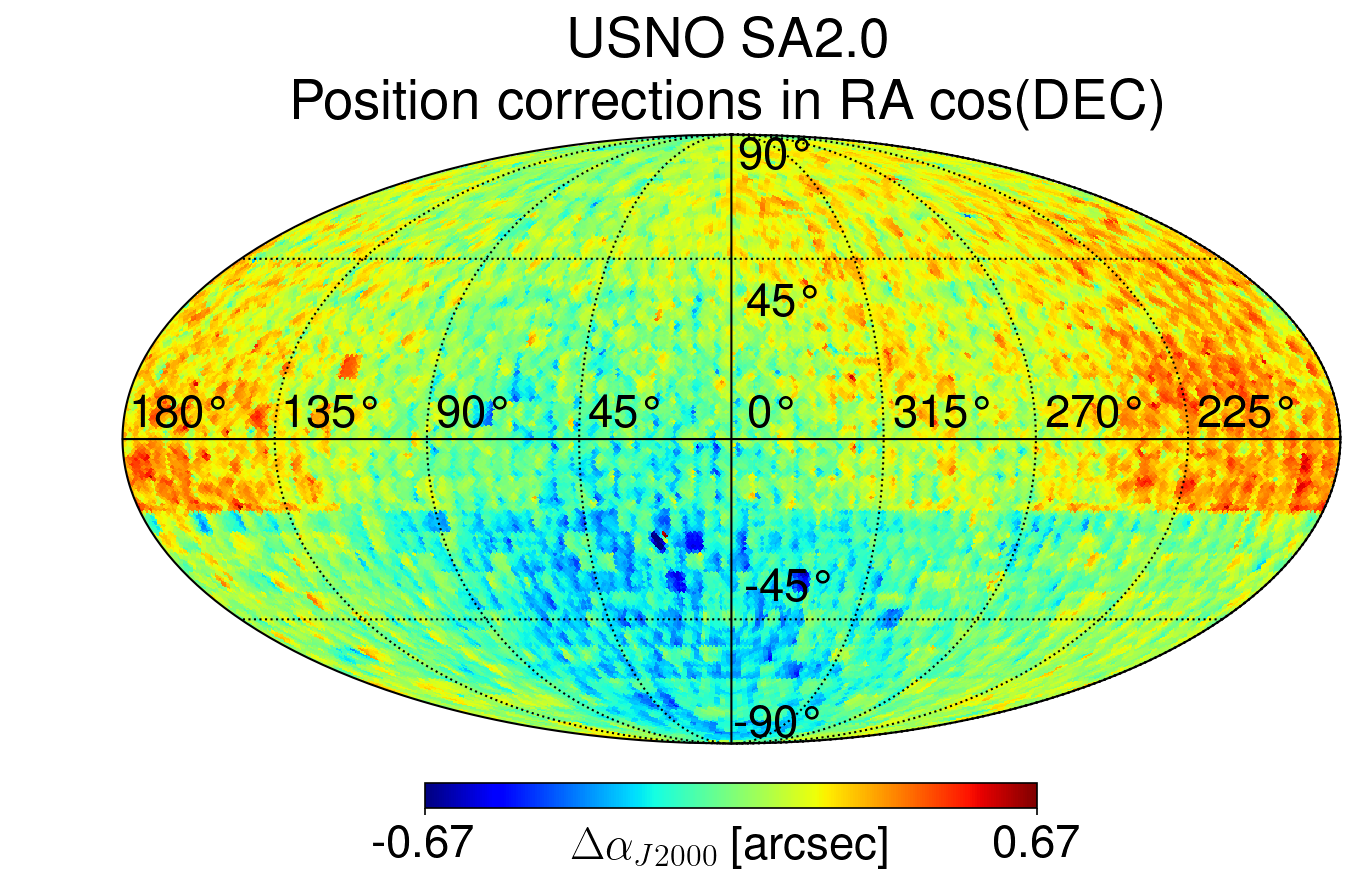} &
 \includegraphics[width=0.5\linewidth]{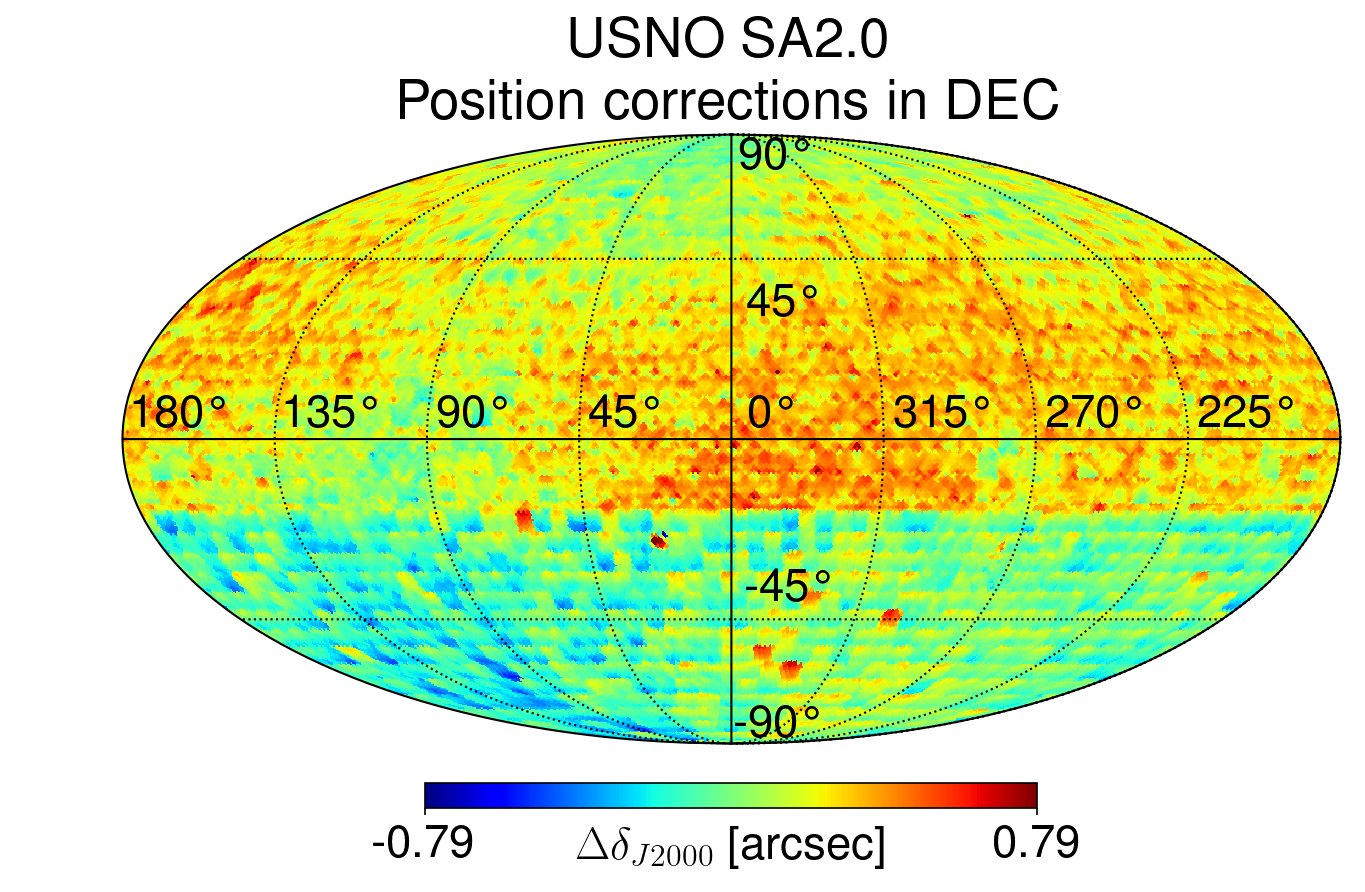} \\
 \end{tabular}
 \caption{Same as Figure \ref{fig:2mass} for the USNO A1.0 (top panels), the USNO SA1.0,
 (center panels) and the USNO SA2.0 catalogs (bottom panels).  \label{fig:usnoa}}
 \end{figure}

\begin{figure}
 \begin{tabular}{ll}
 \includegraphics[width=0.5\linewidth]{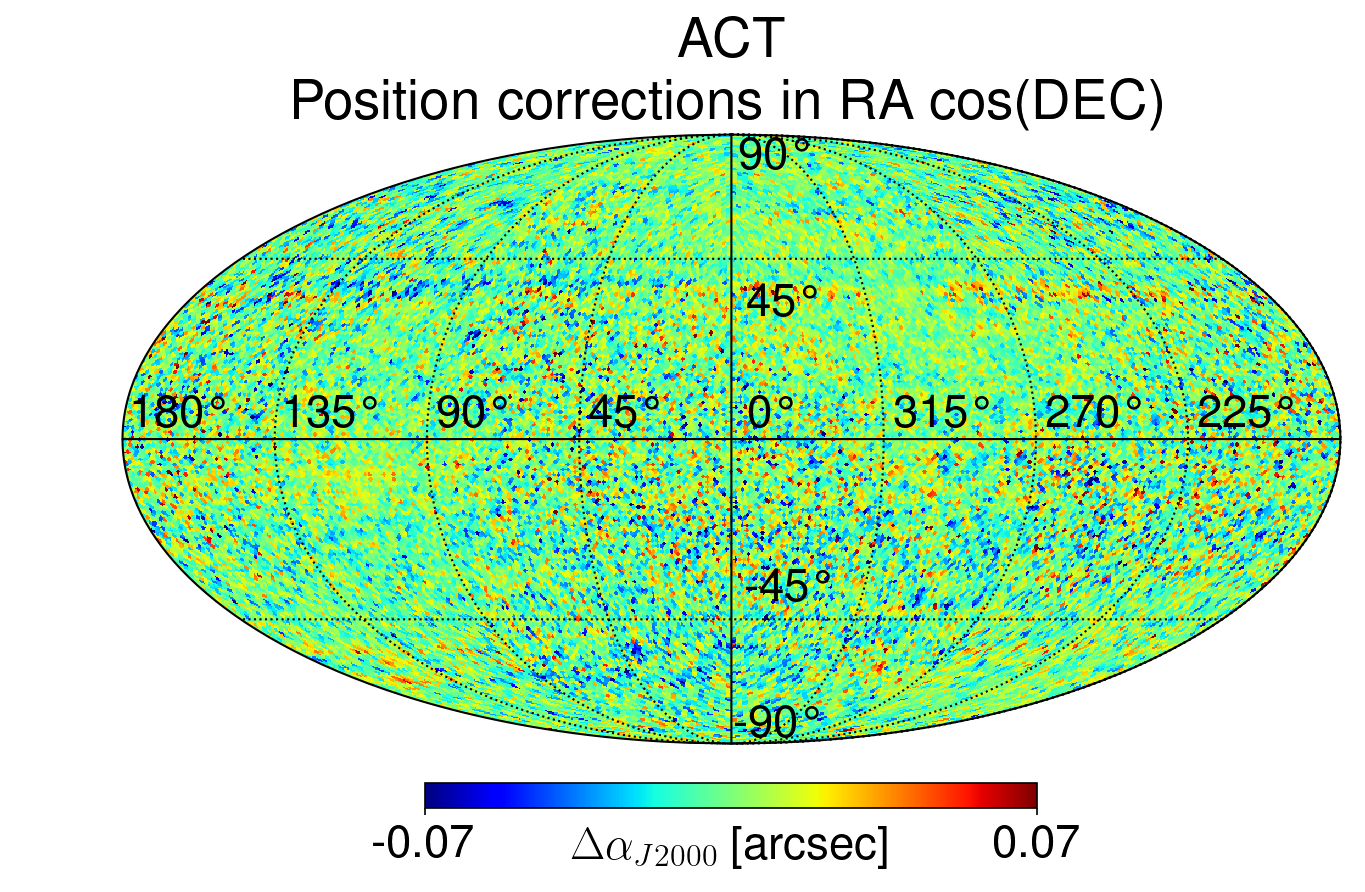} &
 \includegraphics[width=0.5\linewidth]{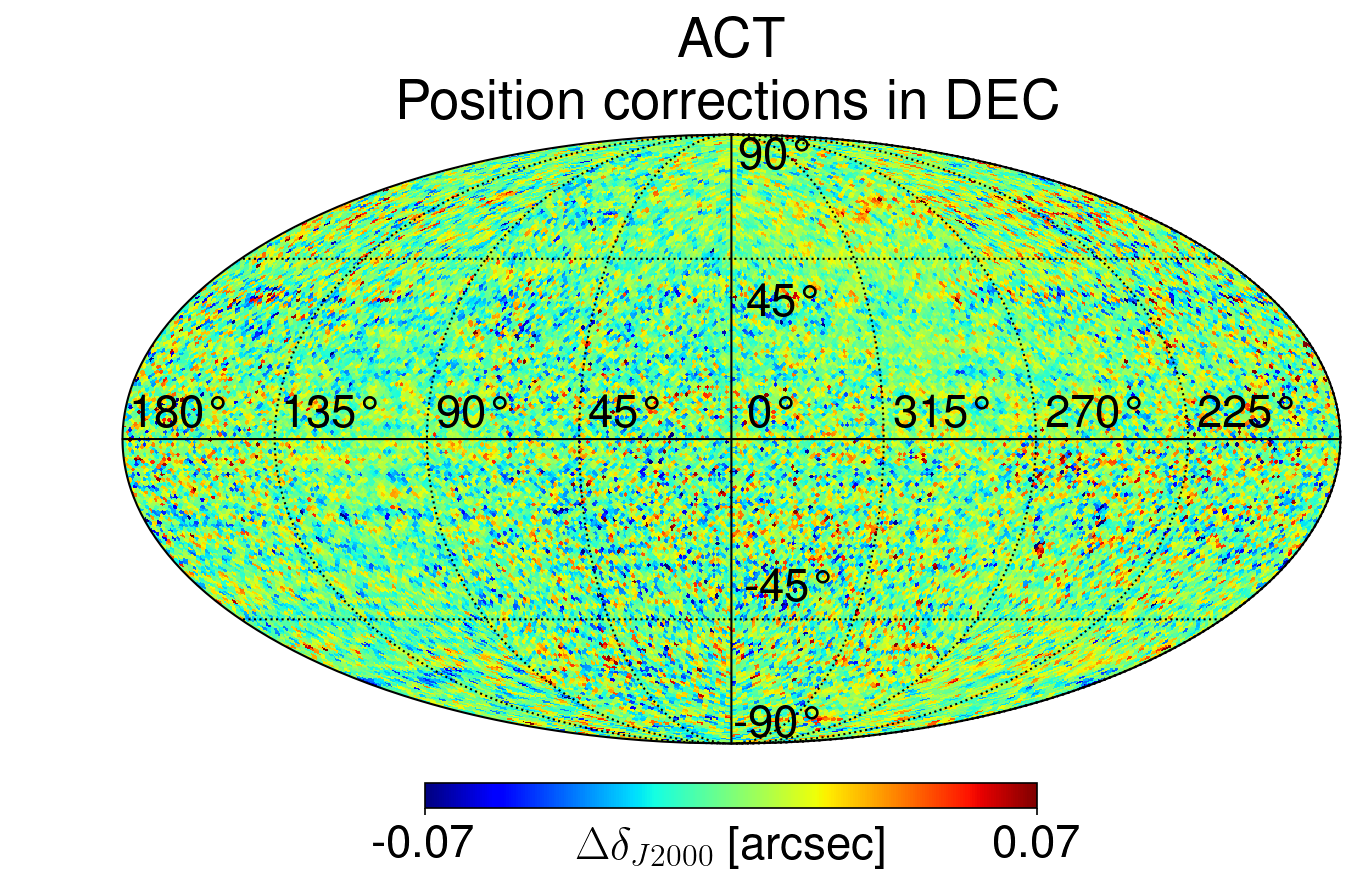} \\
 \includegraphics[width=0.5\linewidth]{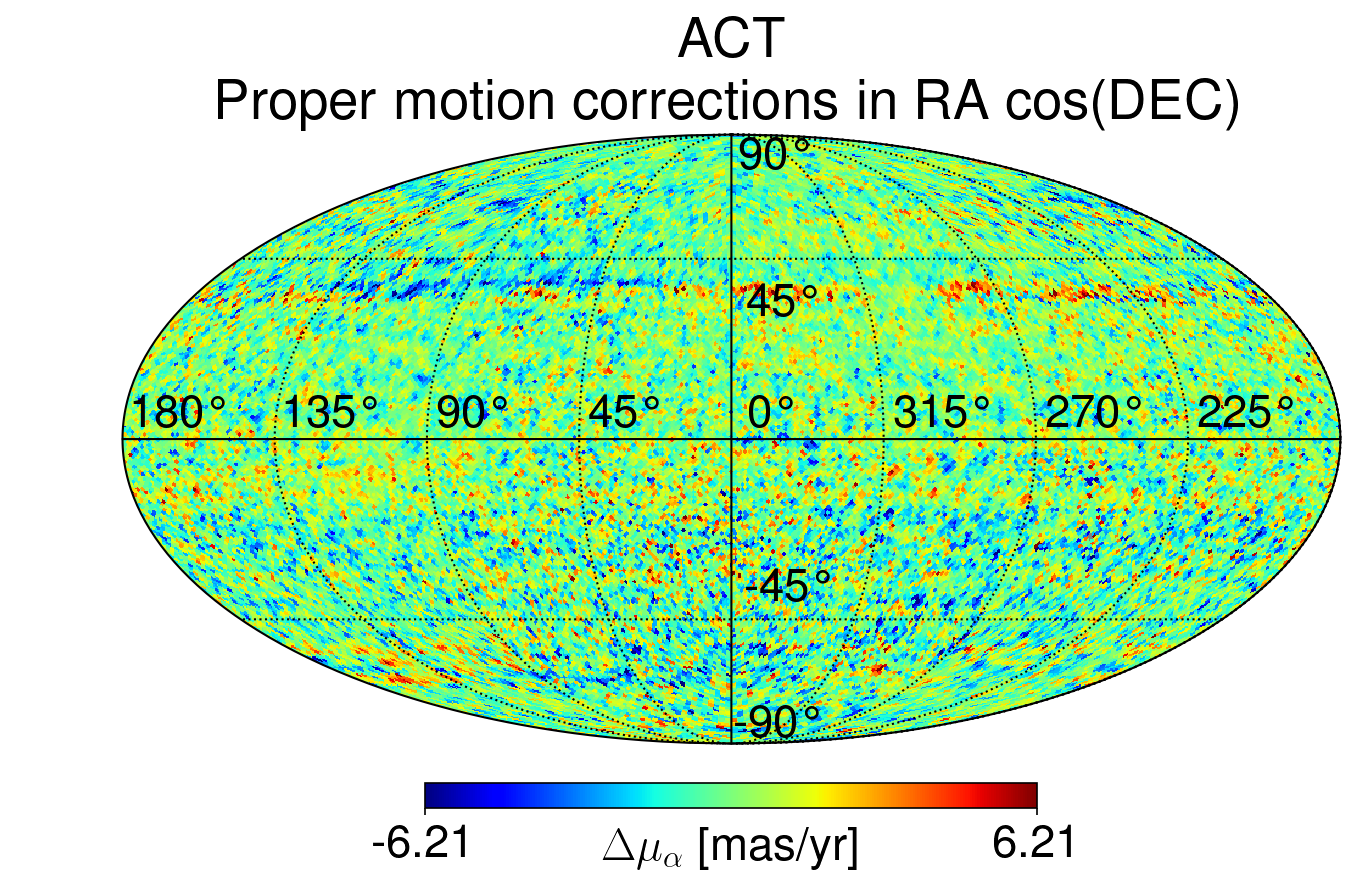} &
 \includegraphics[width=0.5\linewidth]{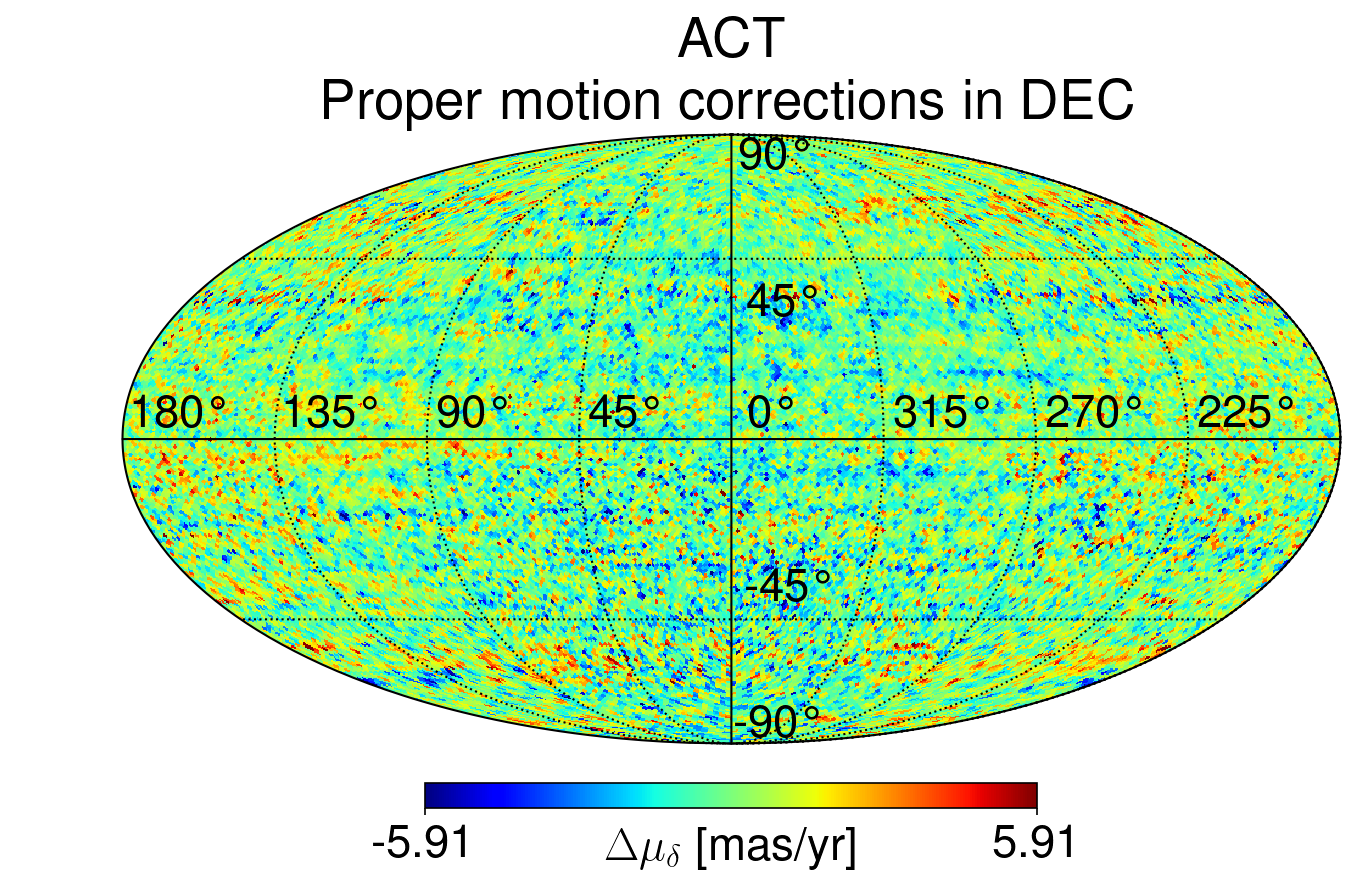}
 \end{tabular}
 \caption{Corrections in stellar positions and proper motion for the ACT catalog. \label{fig:act}}
 \end{figure}

\begin{figure}
 \begin{tabular}{ll}
 \includegraphics[width=0.5\linewidth]{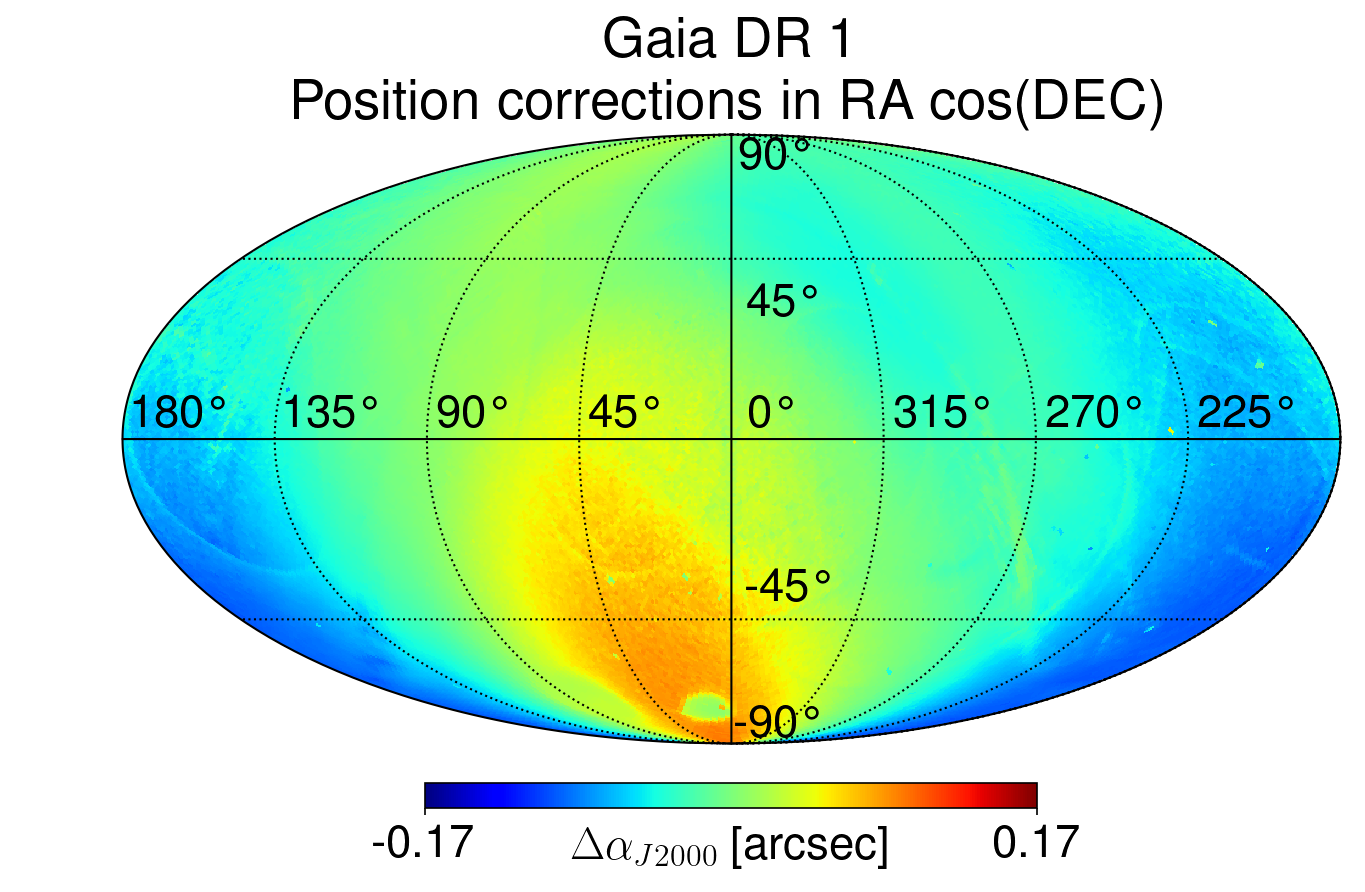} &
 \includegraphics[width=0.5\linewidth]{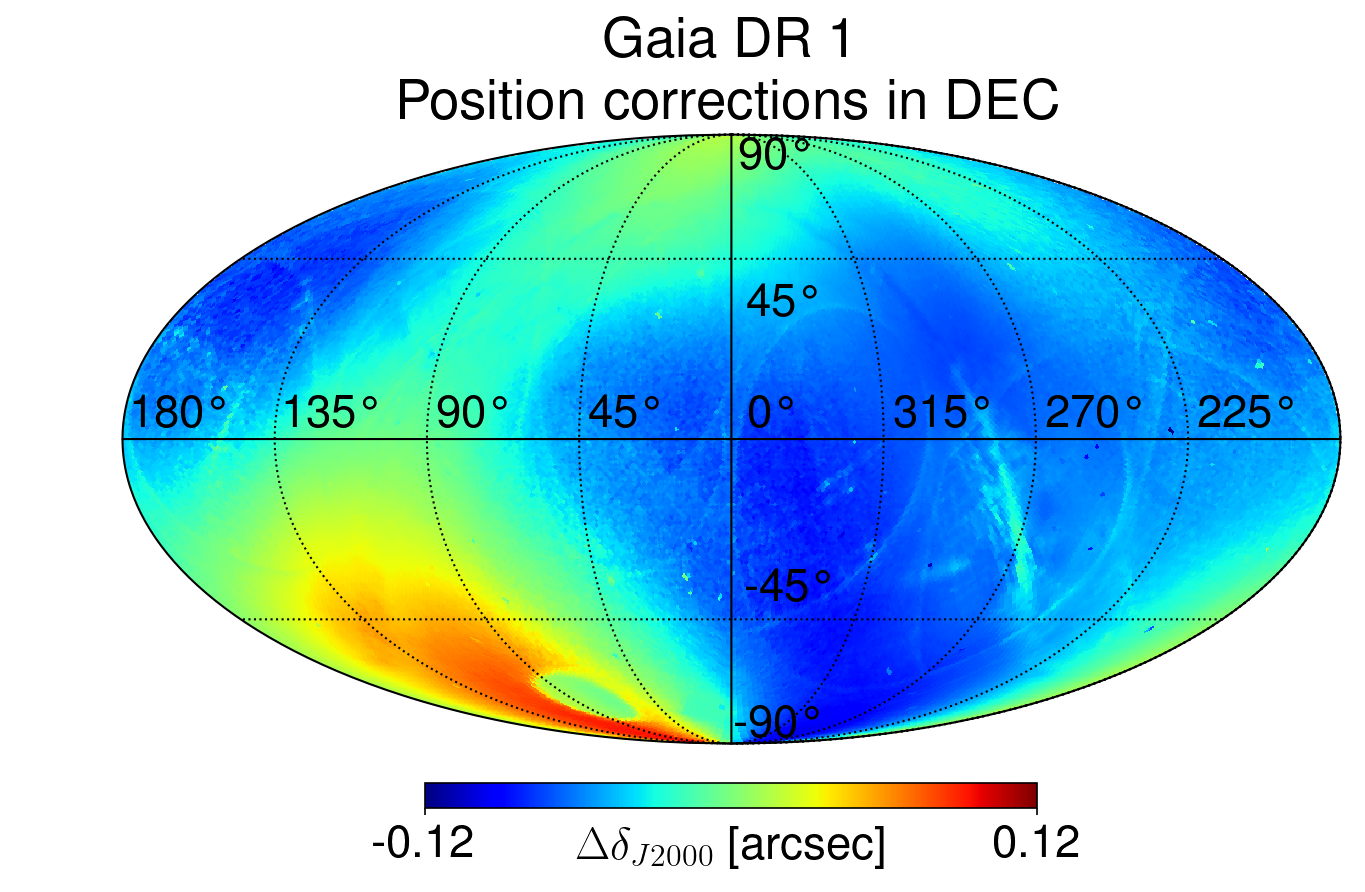} \\
 \includegraphics[width=0.5\linewidth]{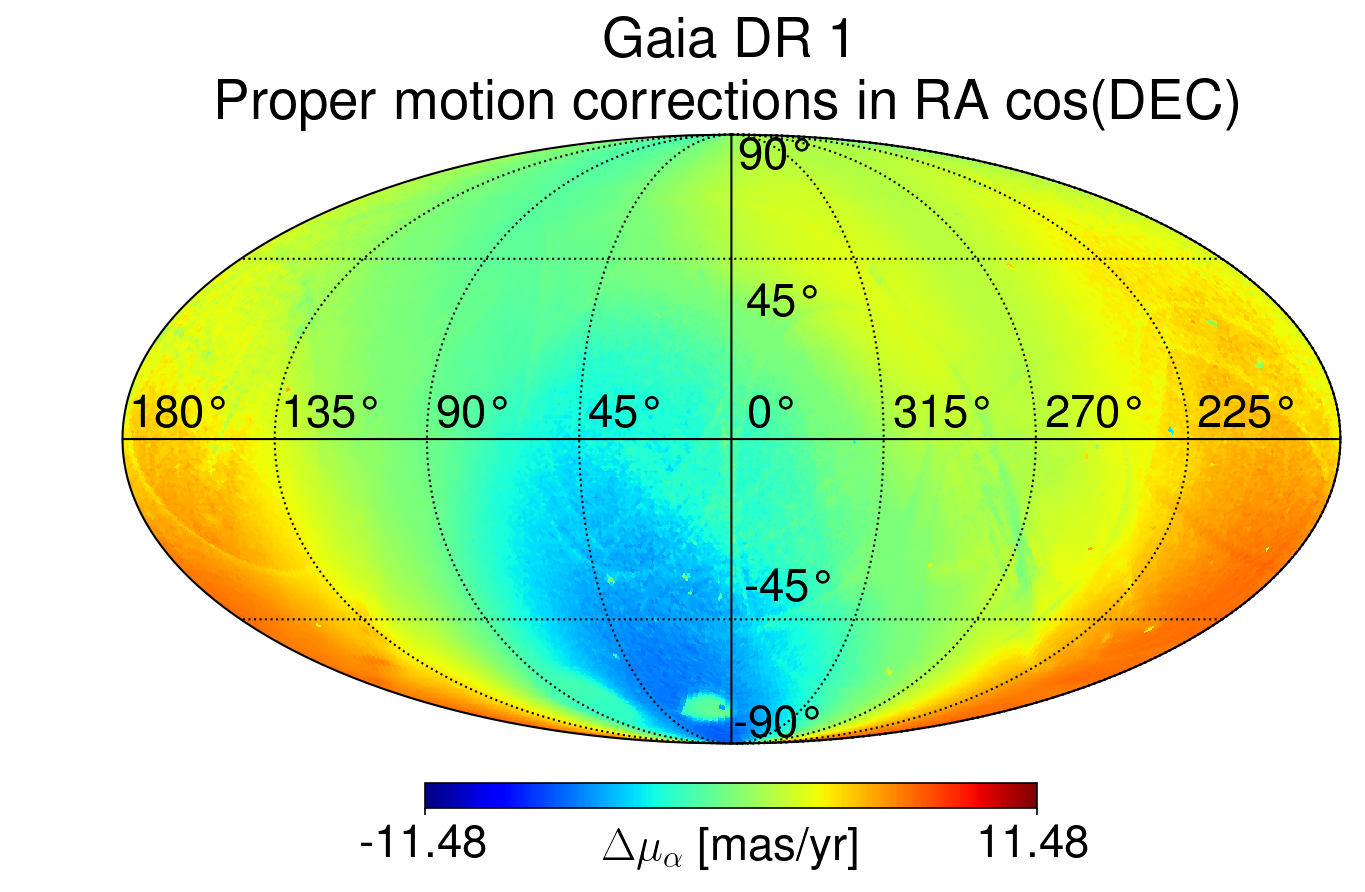} &
 \includegraphics[width=0.5\linewidth]{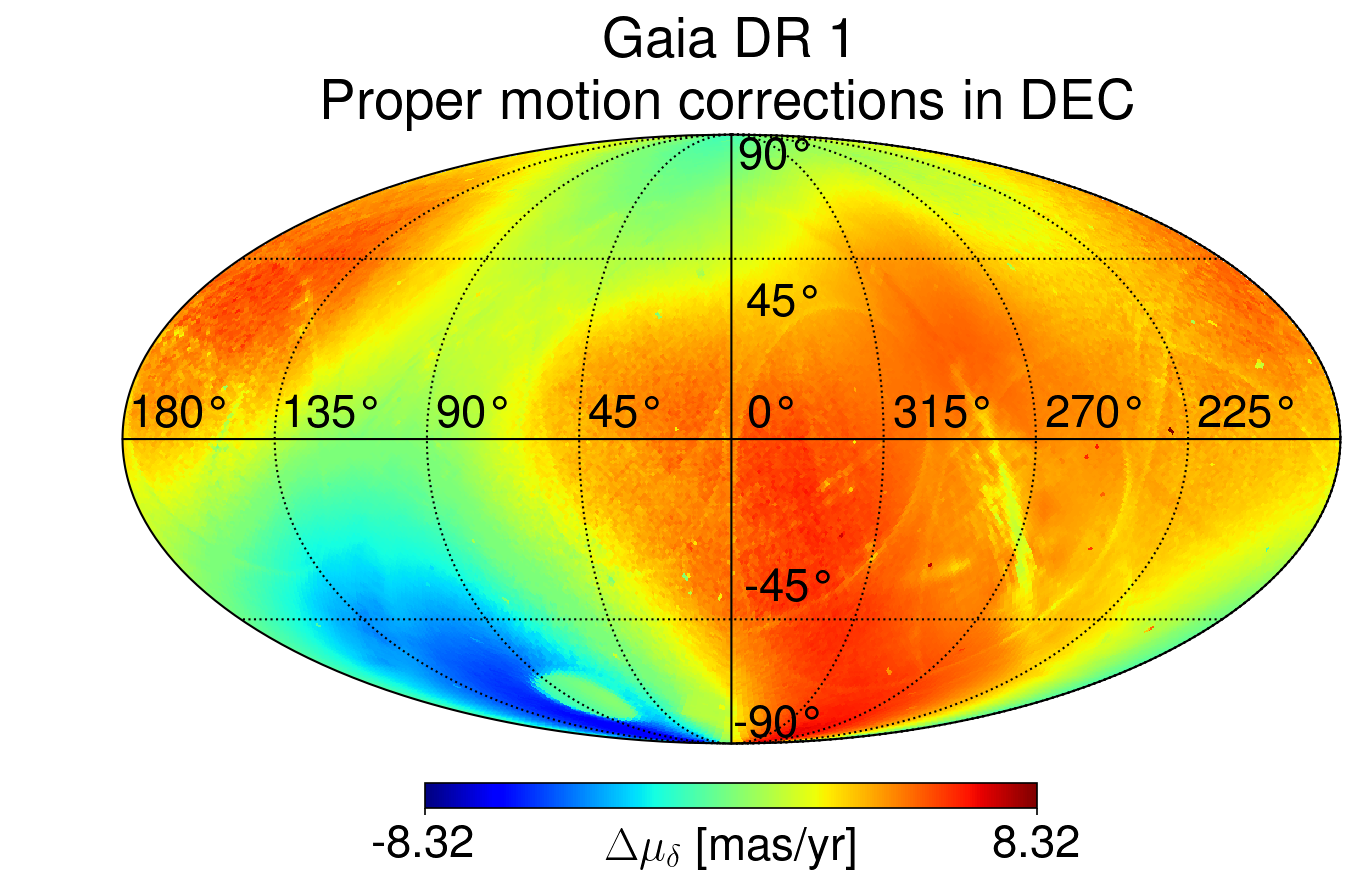}
 \end{tabular}
 \caption{Corrections in stellar positions and proper motion for the Gaia DR 1 catalog at epoch J2000. \label{fig:gaiadr1}}
 \end{figure}

\begin{figure}
 \begin{tabular}{ll}
 \includegraphics[width=0.5\linewidth]{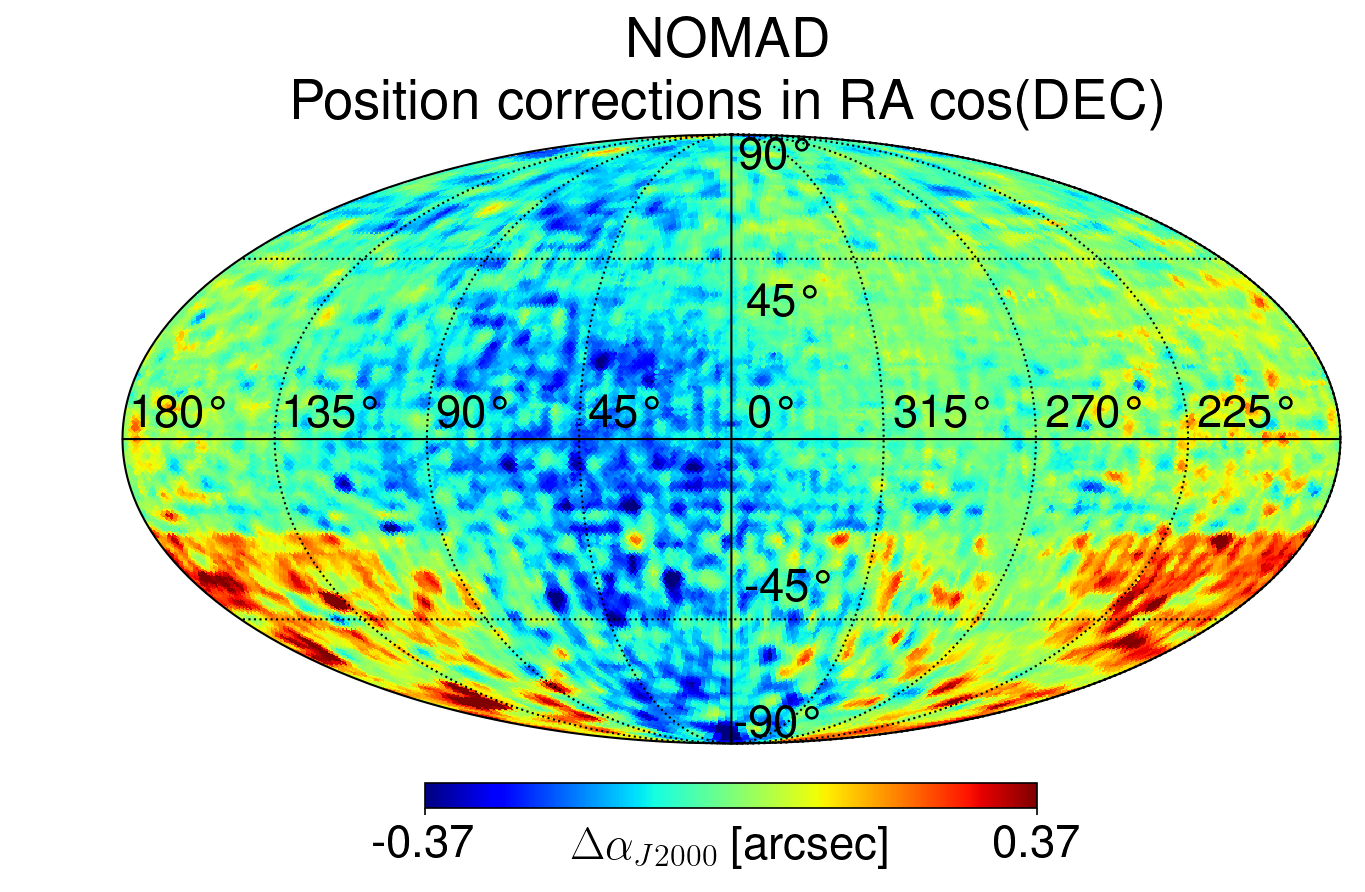} &
 \includegraphics[width=0.5\linewidth]{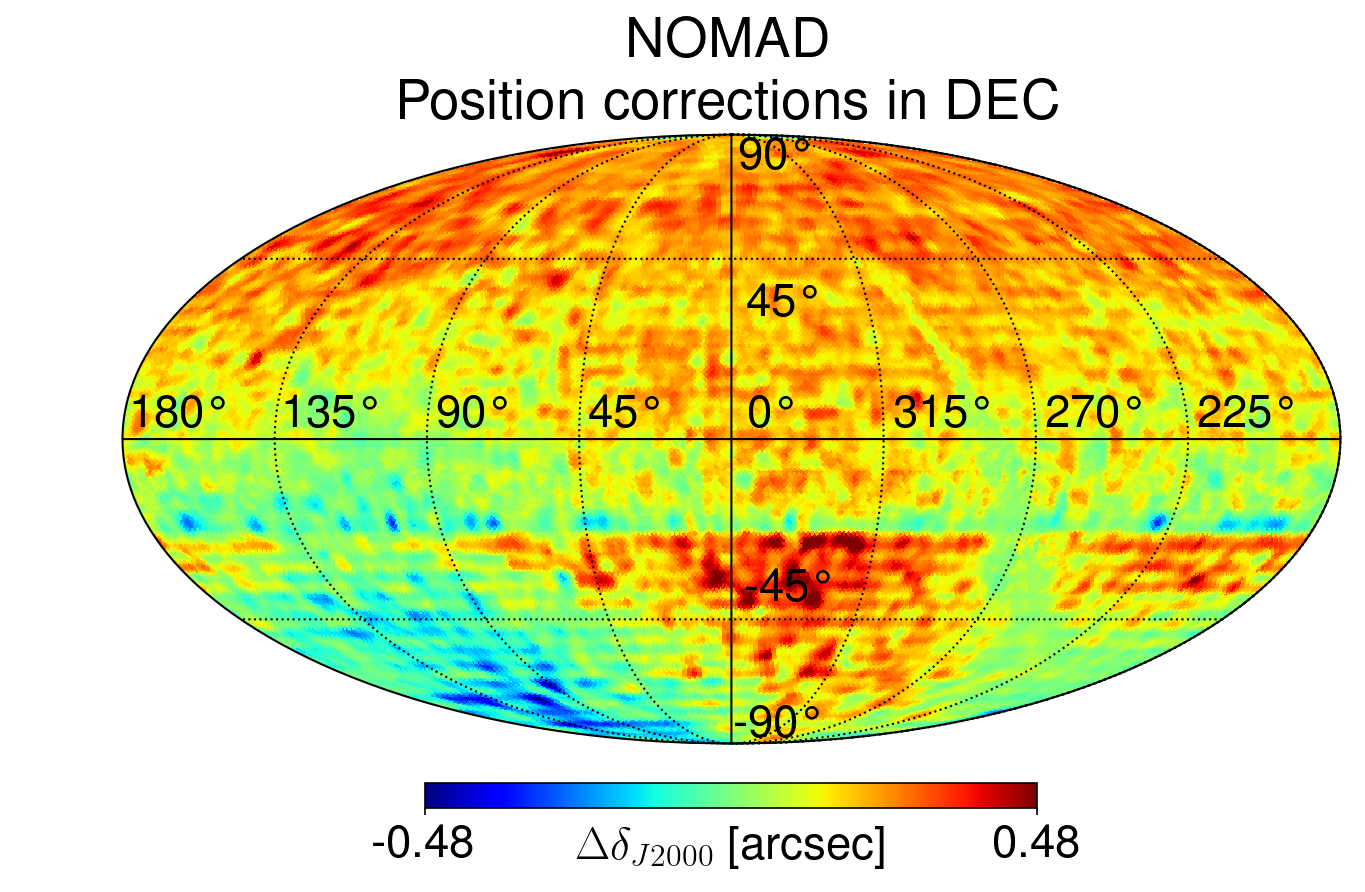} \\
 \includegraphics[width=0.5\linewidth]{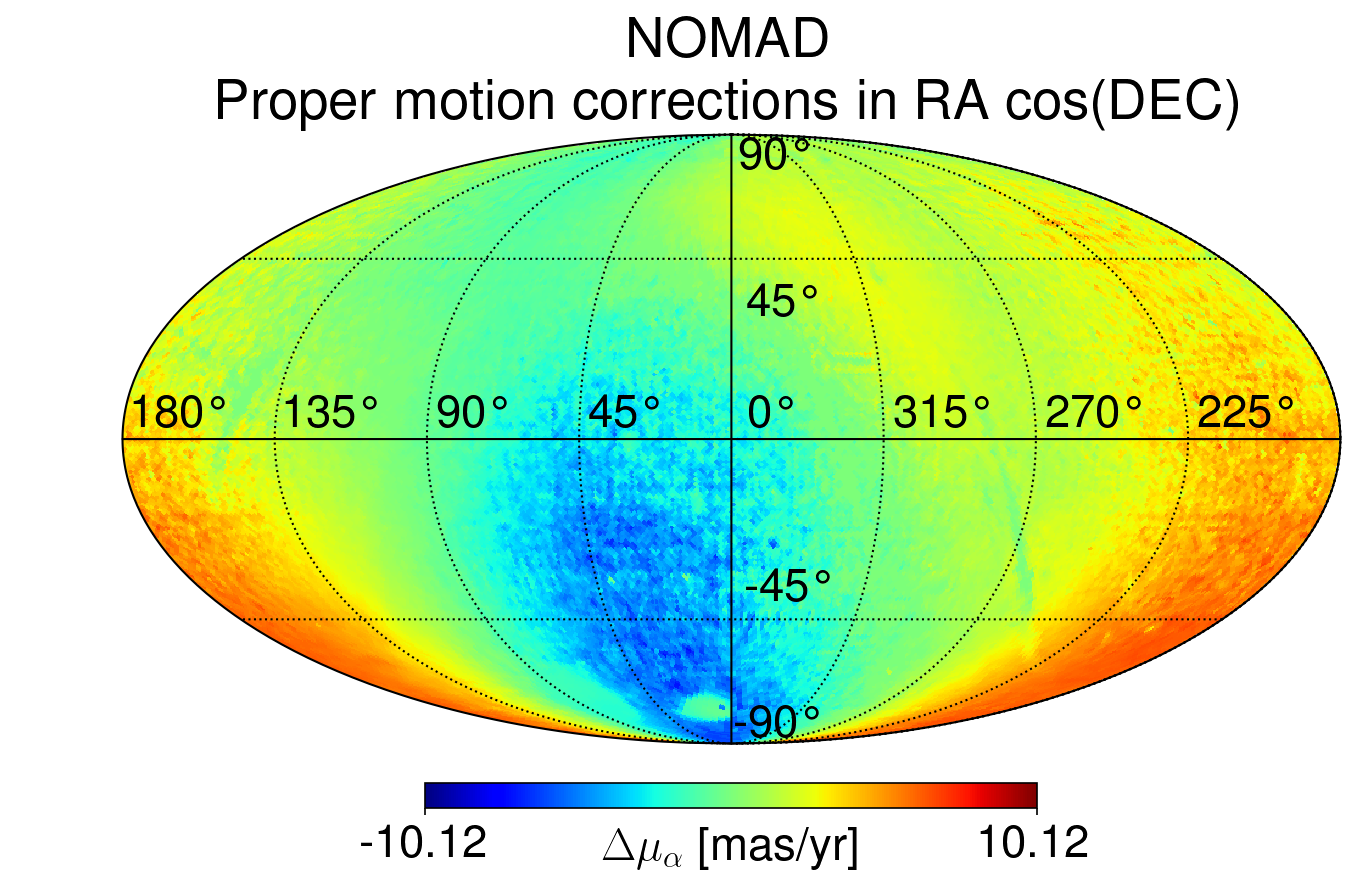} &
 \includegraphics[width=0.5\linewidth]{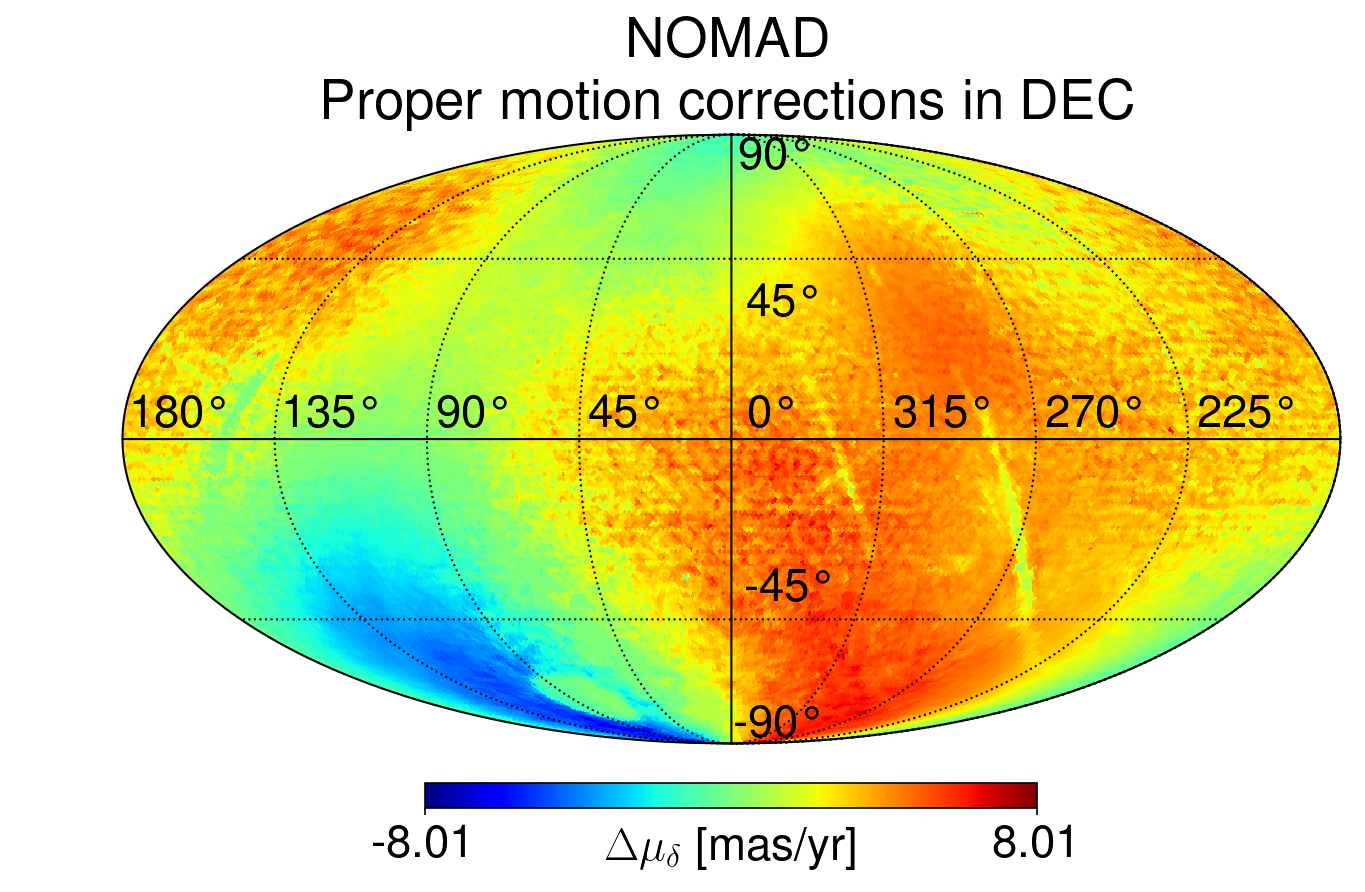}
 \end{tabular}
 \caption{Corrections in stellar positions and proper motion for the NOMAD catalog. \label{fig:nomad}}
 \end{figure}

\begin{figure}
 \begin{tabular}{ll}
 \includegraphics[width=0.5\linewidth]{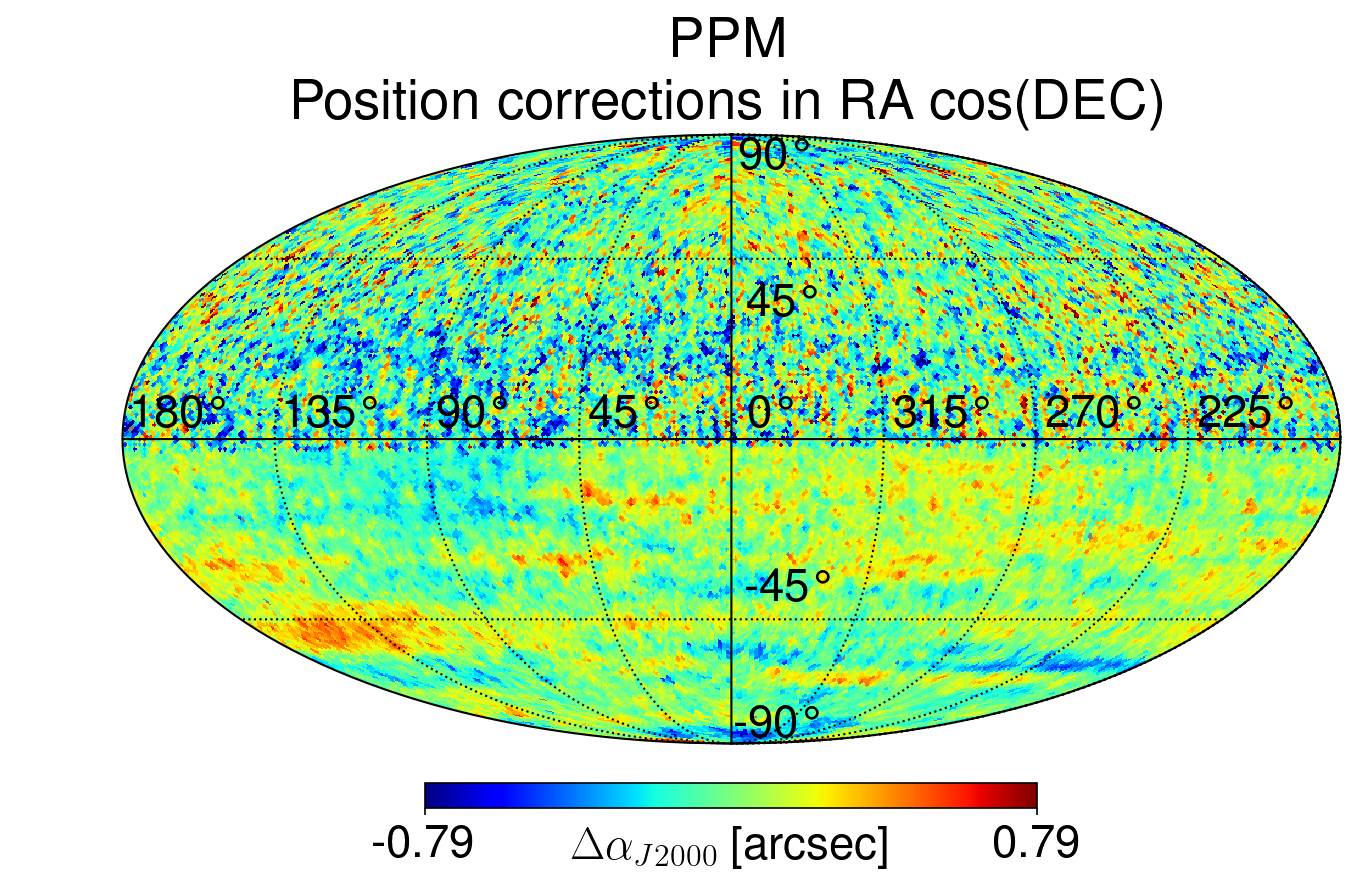} &
 \includegraphics[width=0.5\linewidth]{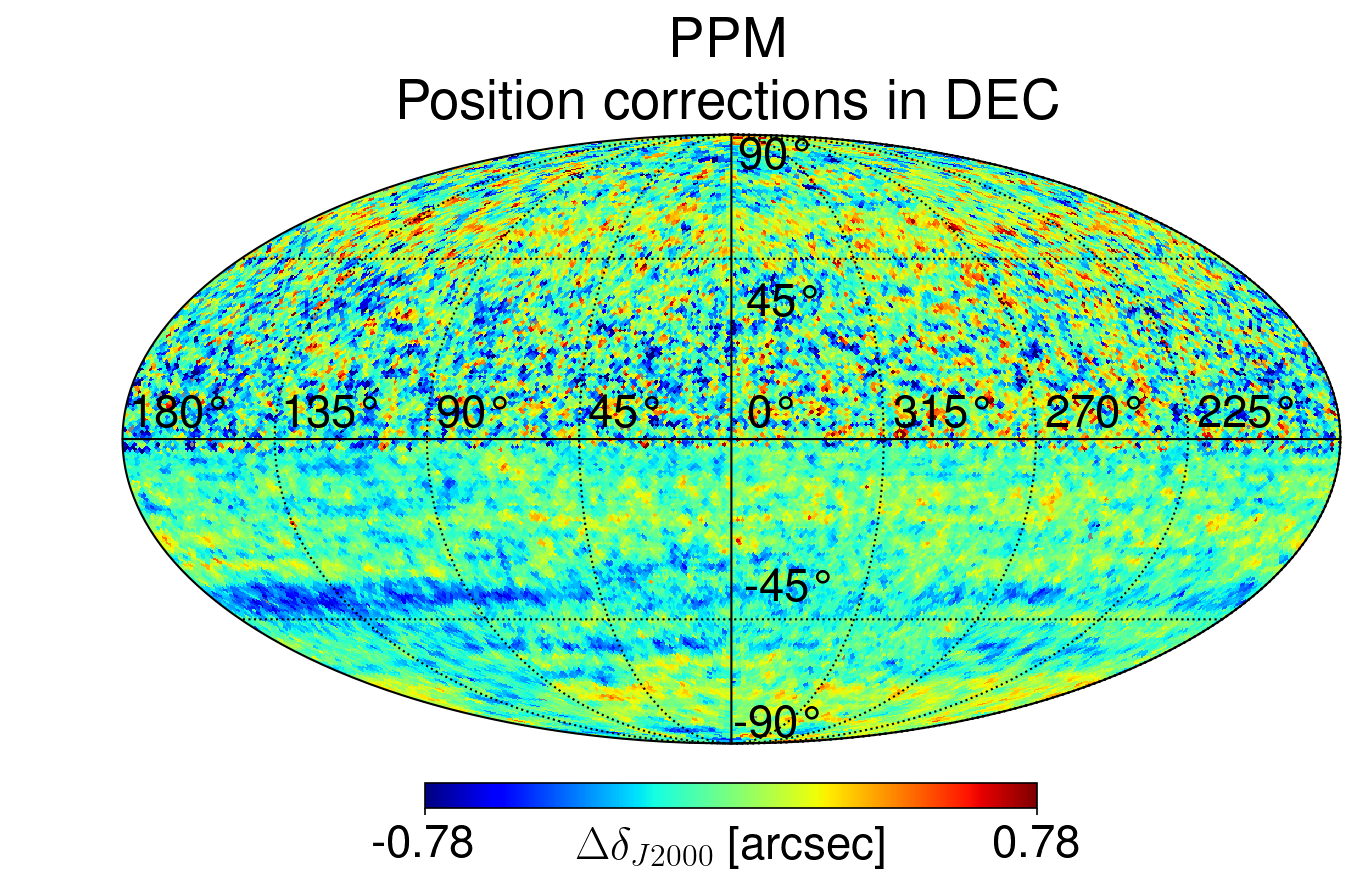} \\
 \includegraphics[width=0.5\linewidth]{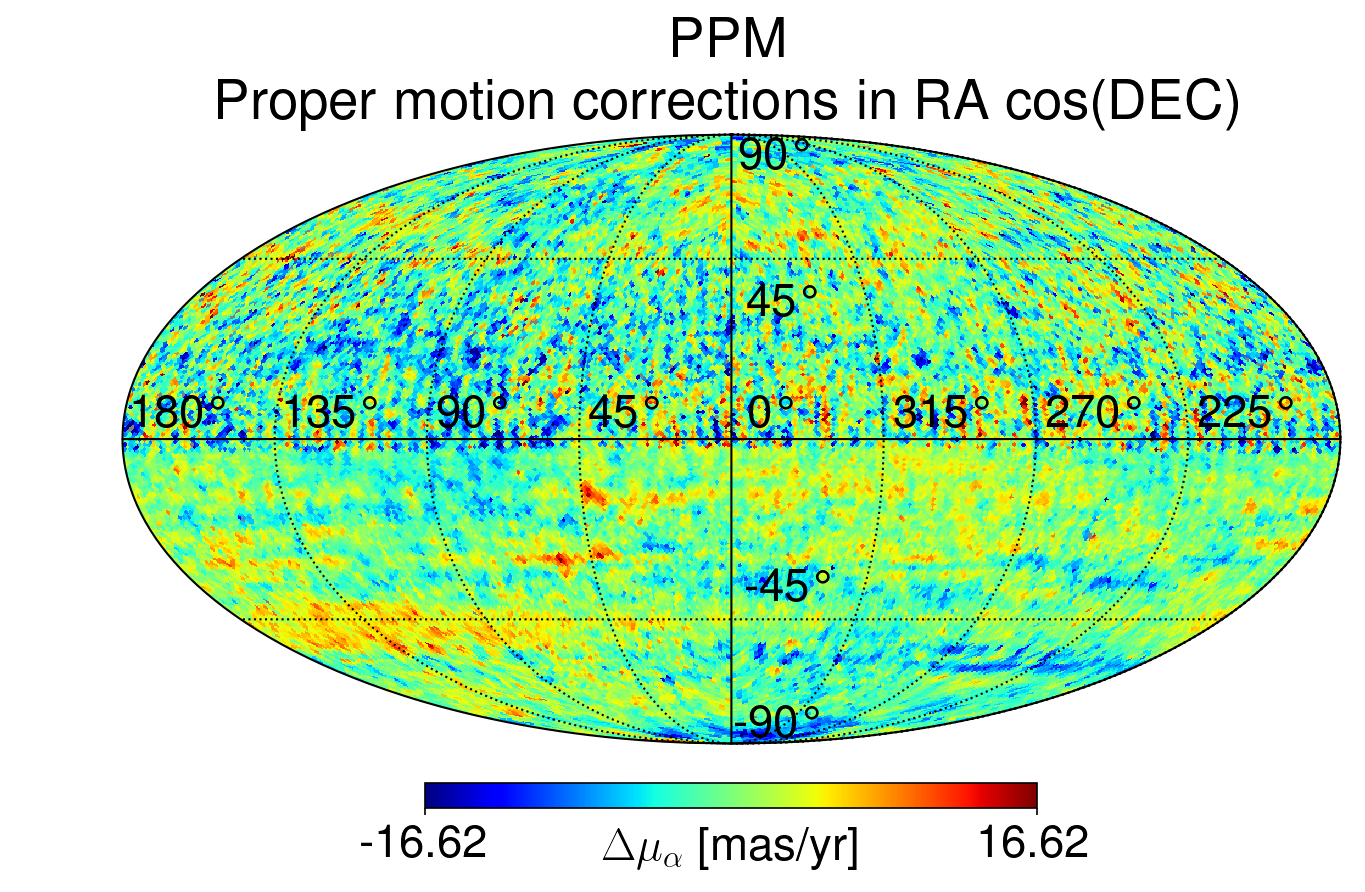} &
 \includegraphics[width=0.5\linewidth]{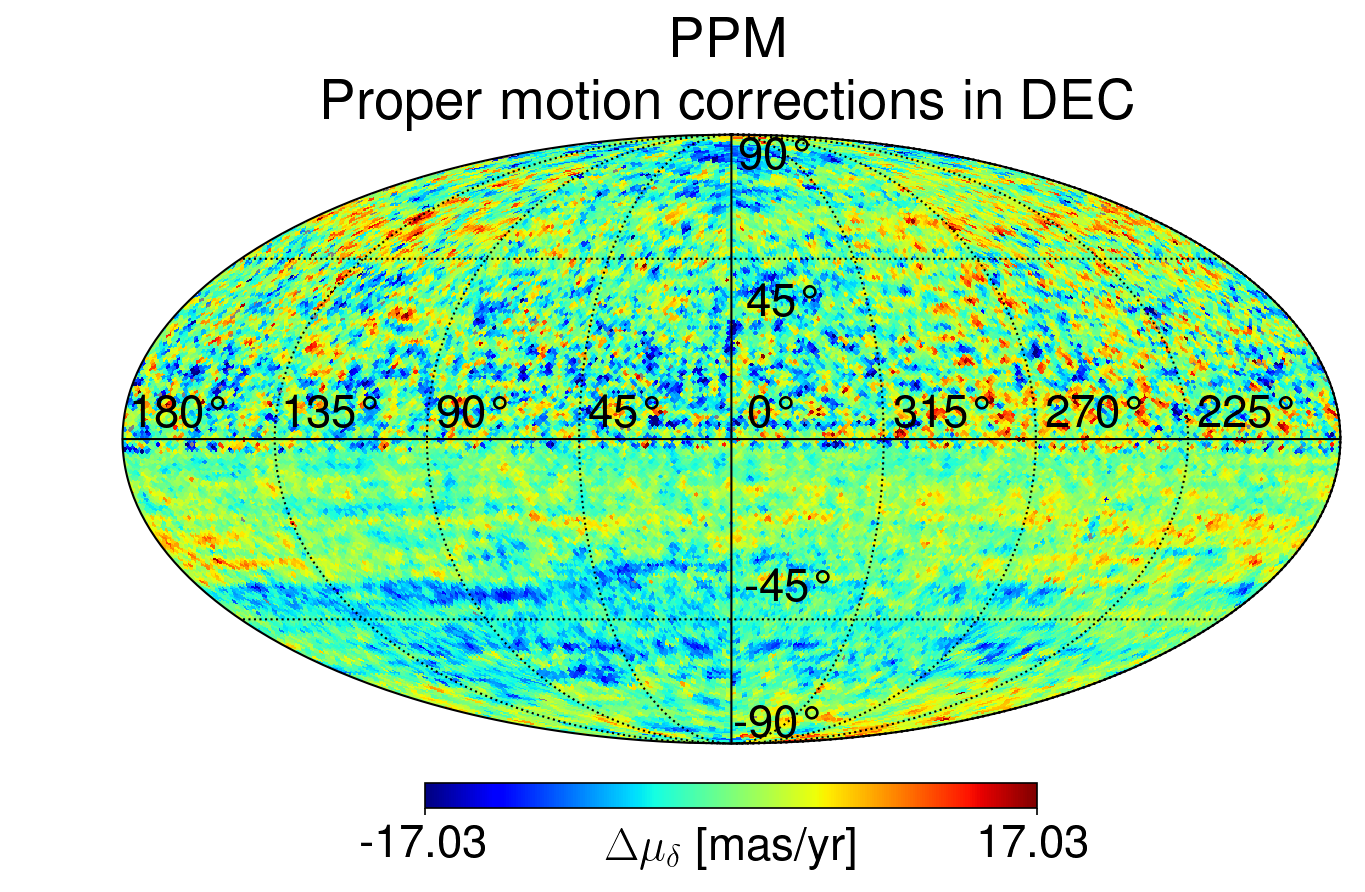}
 \end{tabular}
 \caption{Corrections in stellar positions and proper motion for the PPM catalog. \label{fig:ppm}}
 \end{figure}

\begin{figure}
 \begin{tabular}{ll}
 \includegraphics[width=0.5\linewidth]{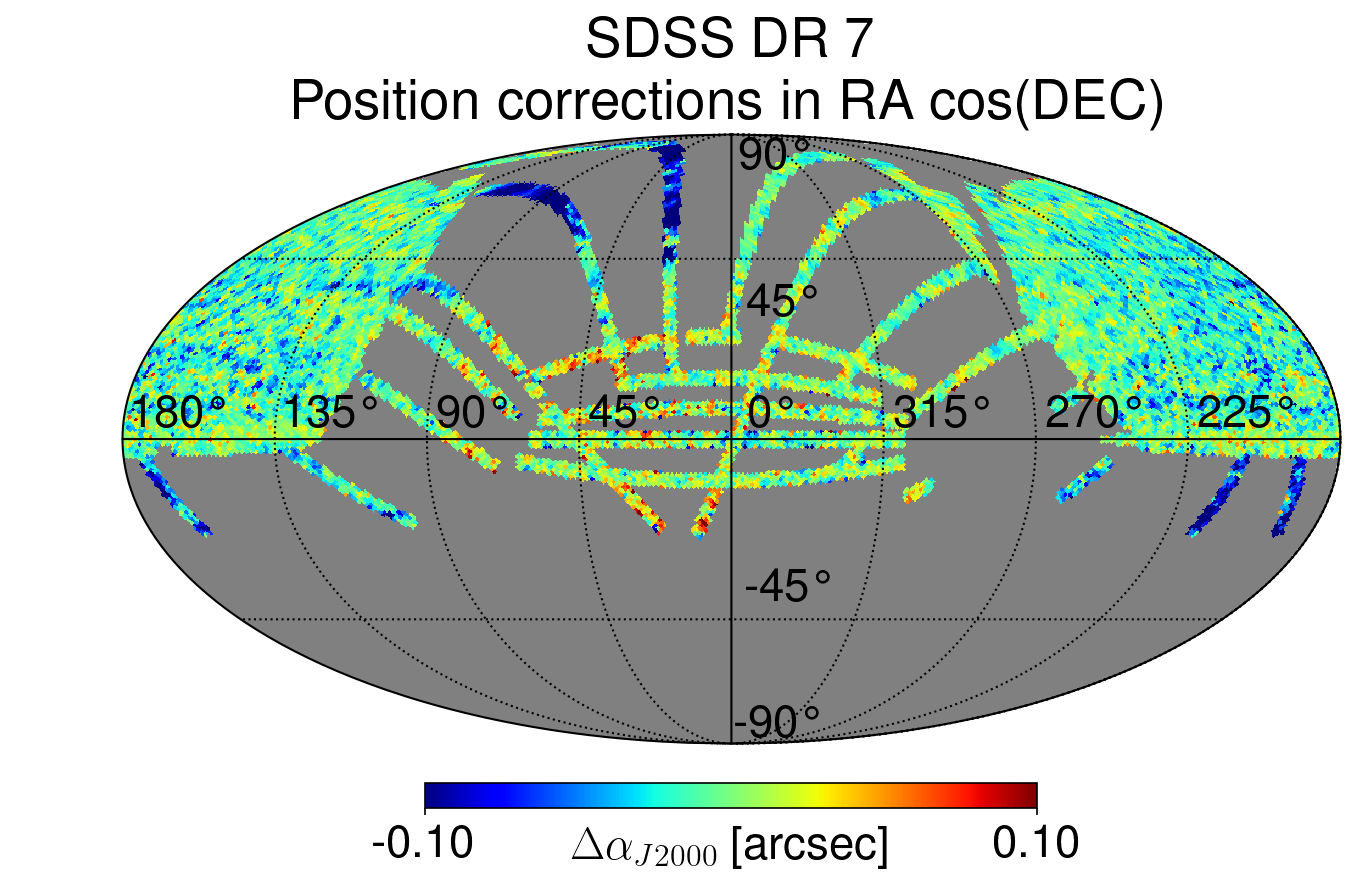} &
 \includegraphics[width=0.5\linewidth]{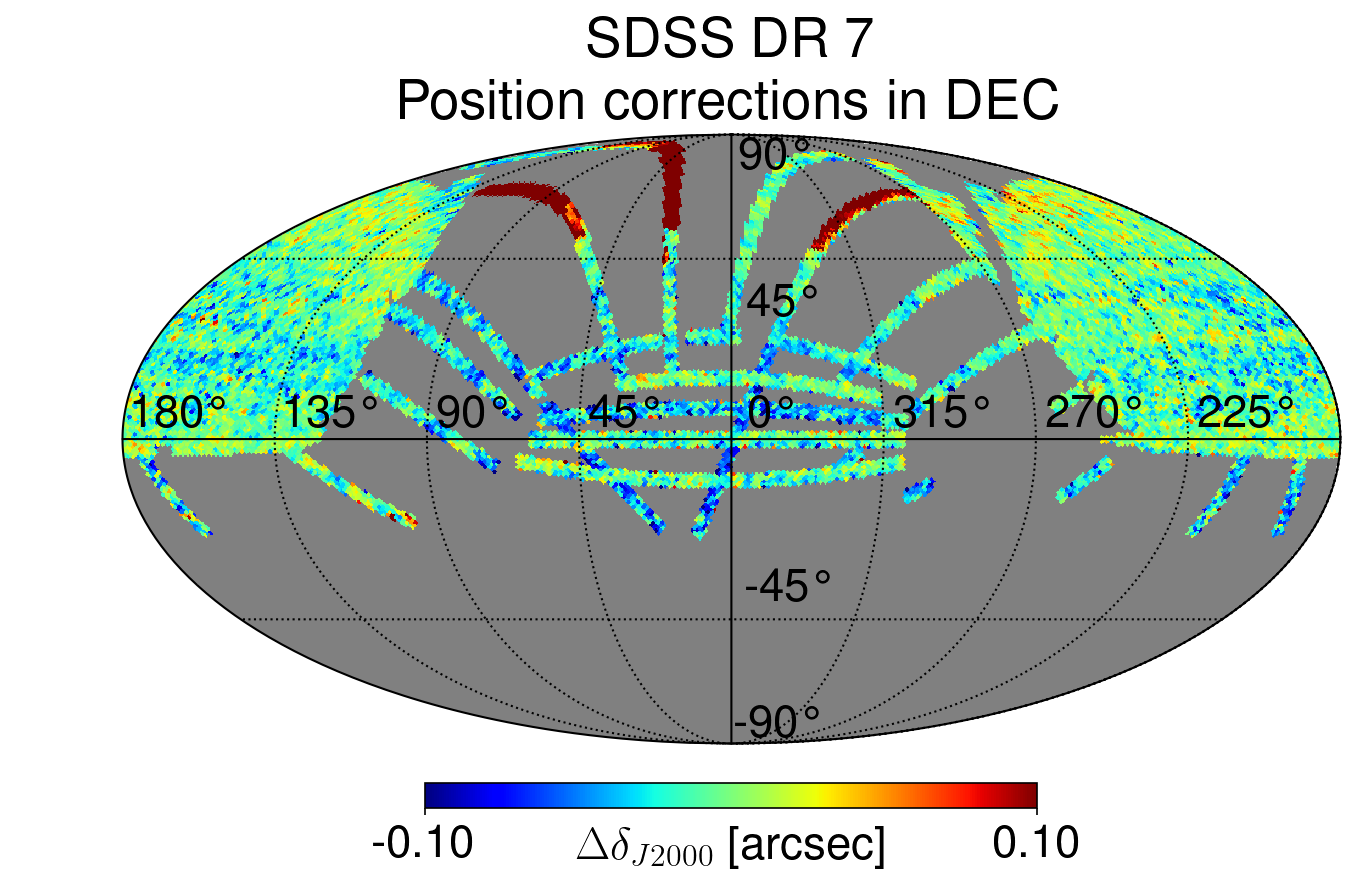} \\
 \includegraphics[width=0.5\linewidth]{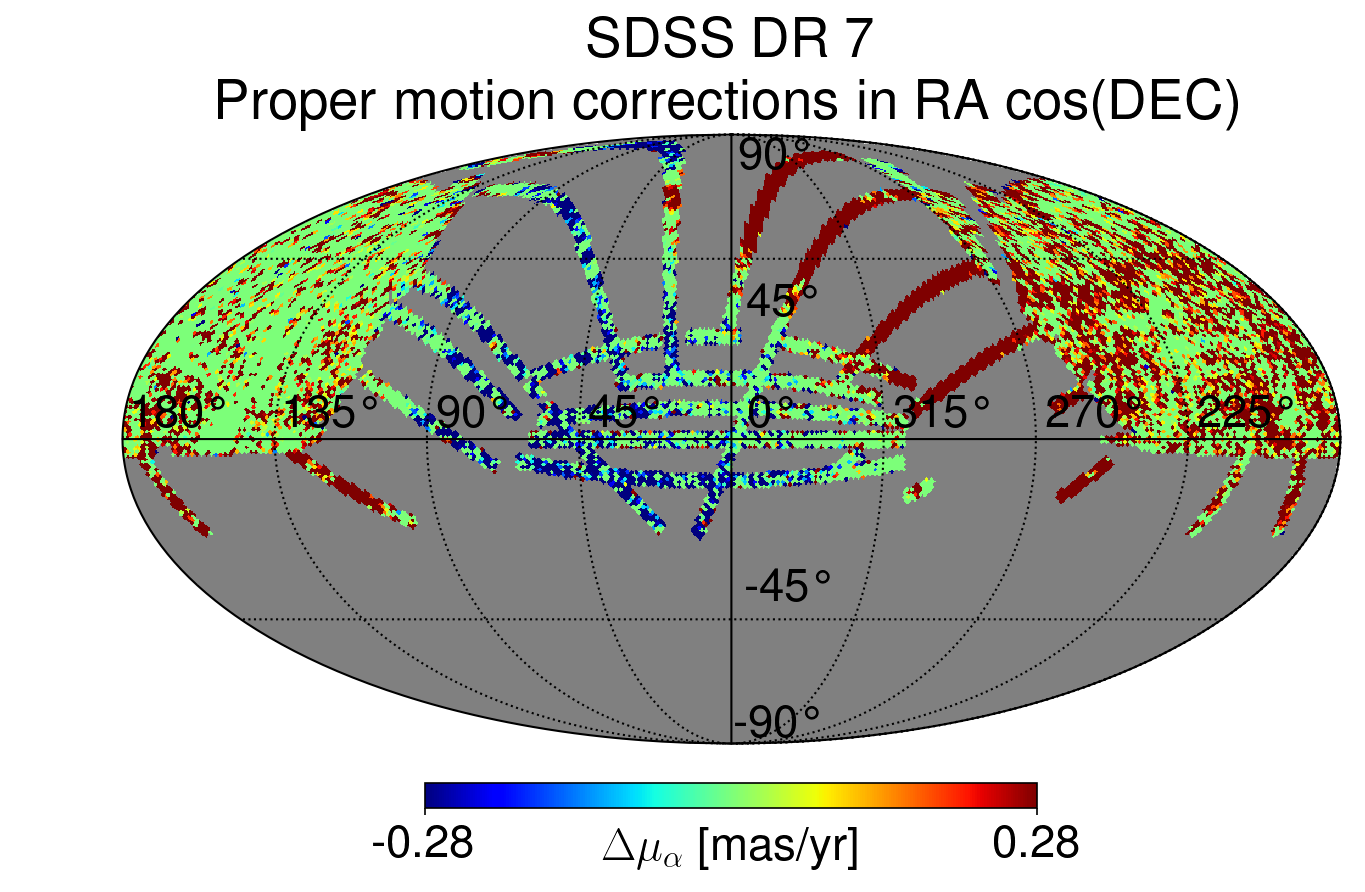} &
 \includegraphics[width=0.5\linewidth]{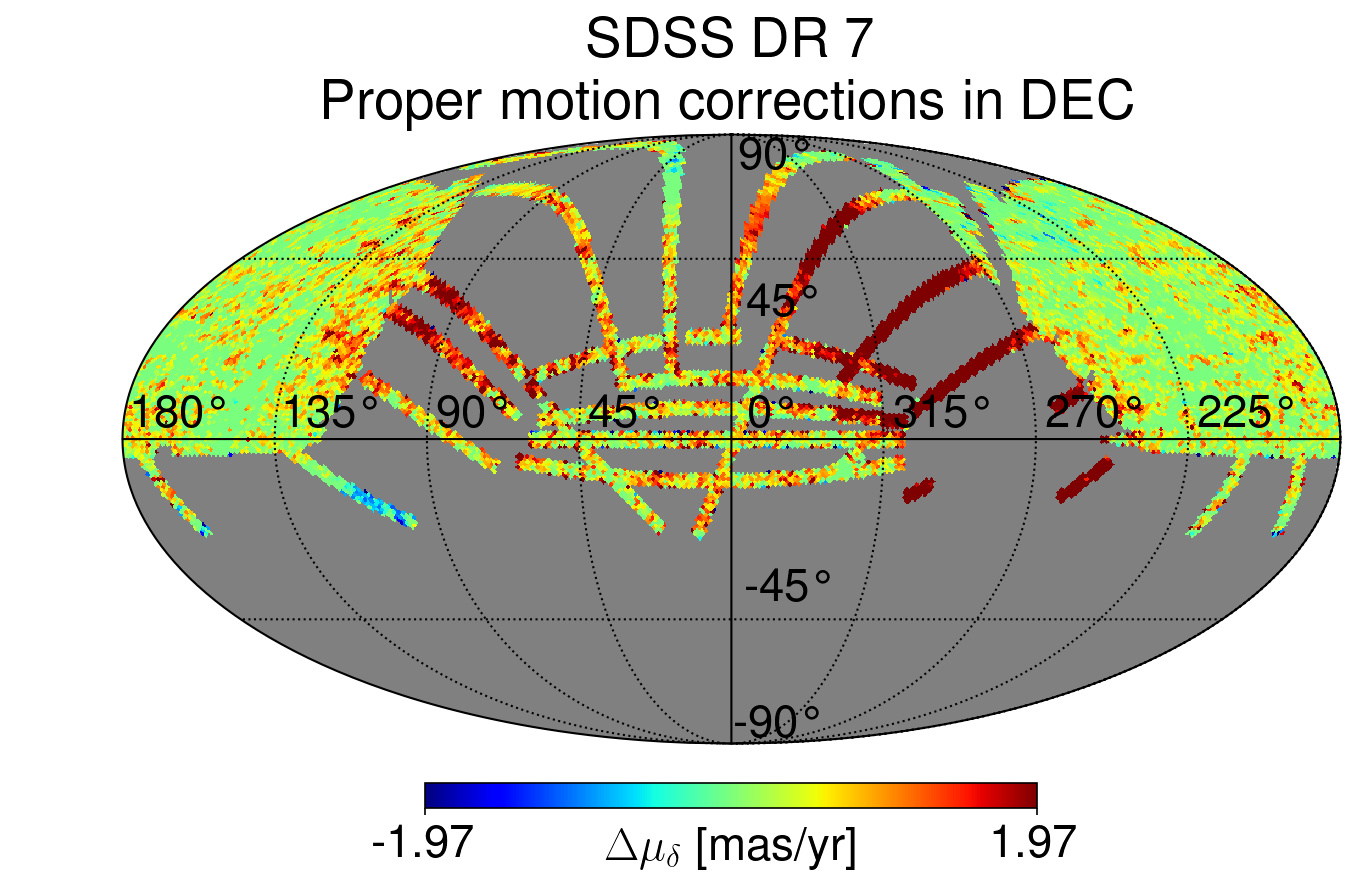}
 \end{tabular}
 \caption{Corrections in stellar positions and proper motion for the SDSS DR 7 catalog. \label{fig:sdss7}}
 \end{figure}

\begin{figure}
 \begin{tabular}{ll}
 \includegraphics[width=0.5\linewidth]{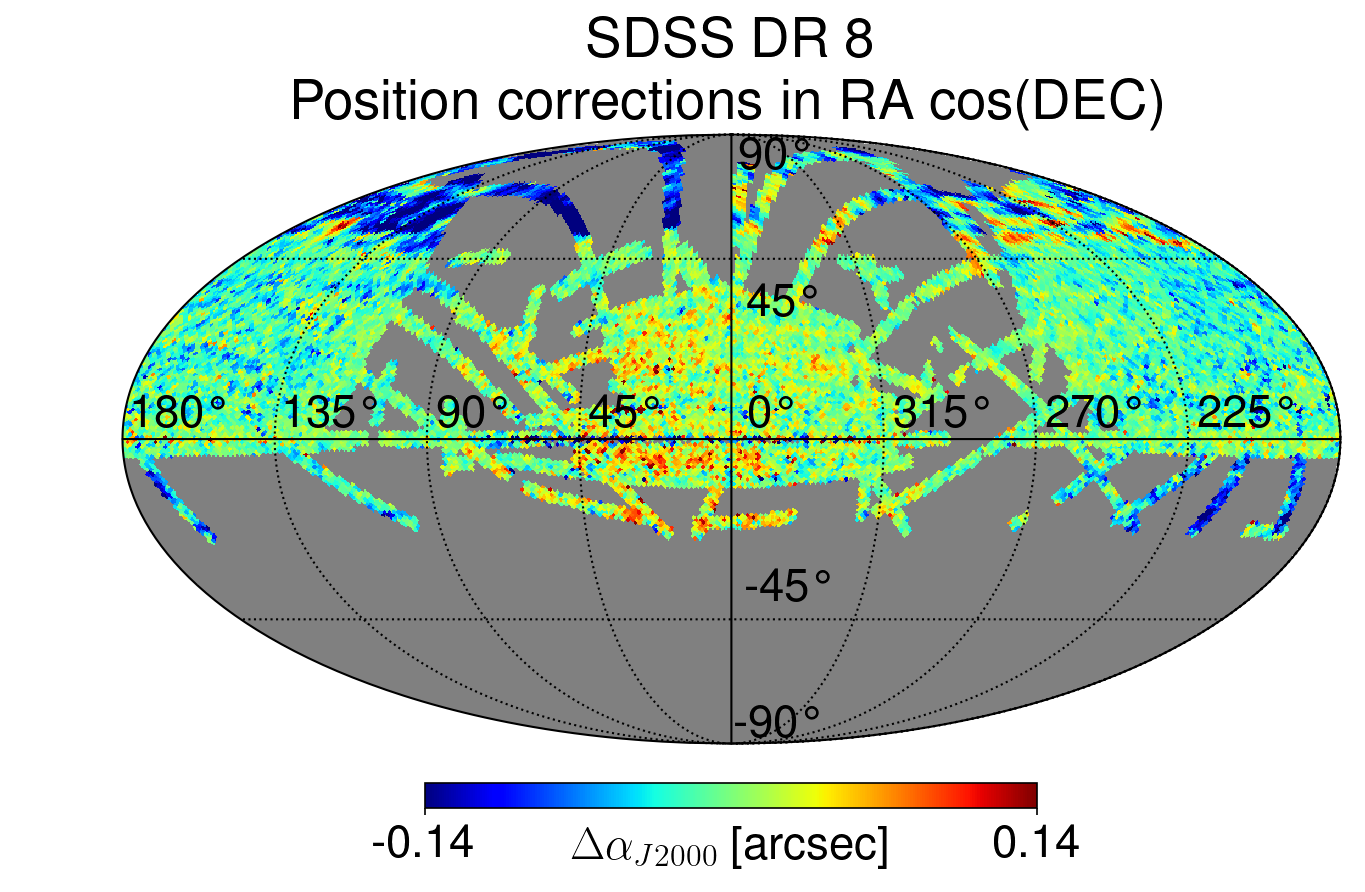} &
 \includegraphics[width=0.5\linewidth]{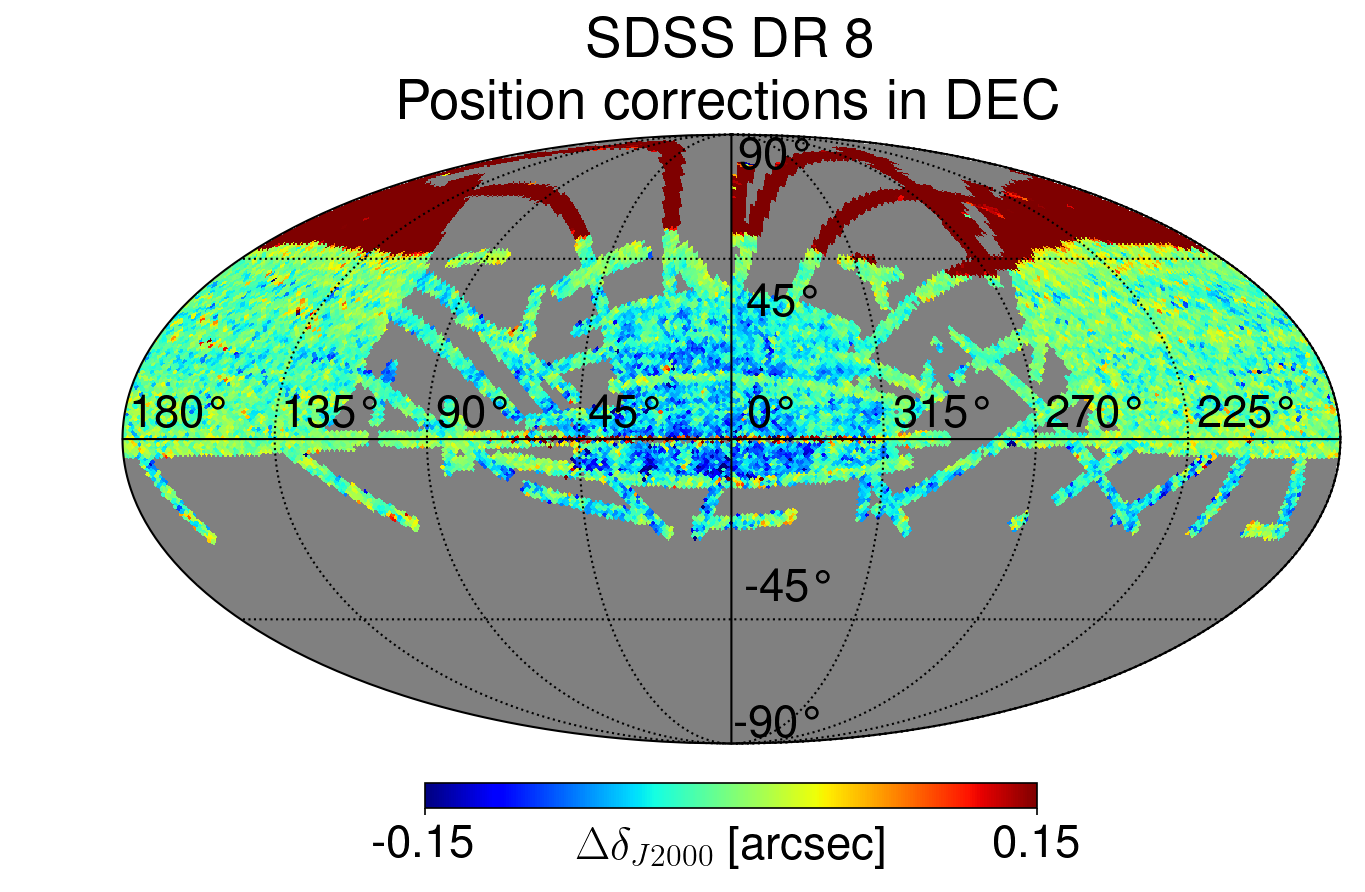} \\
 \includegraphics[width=0.5\linewidth]{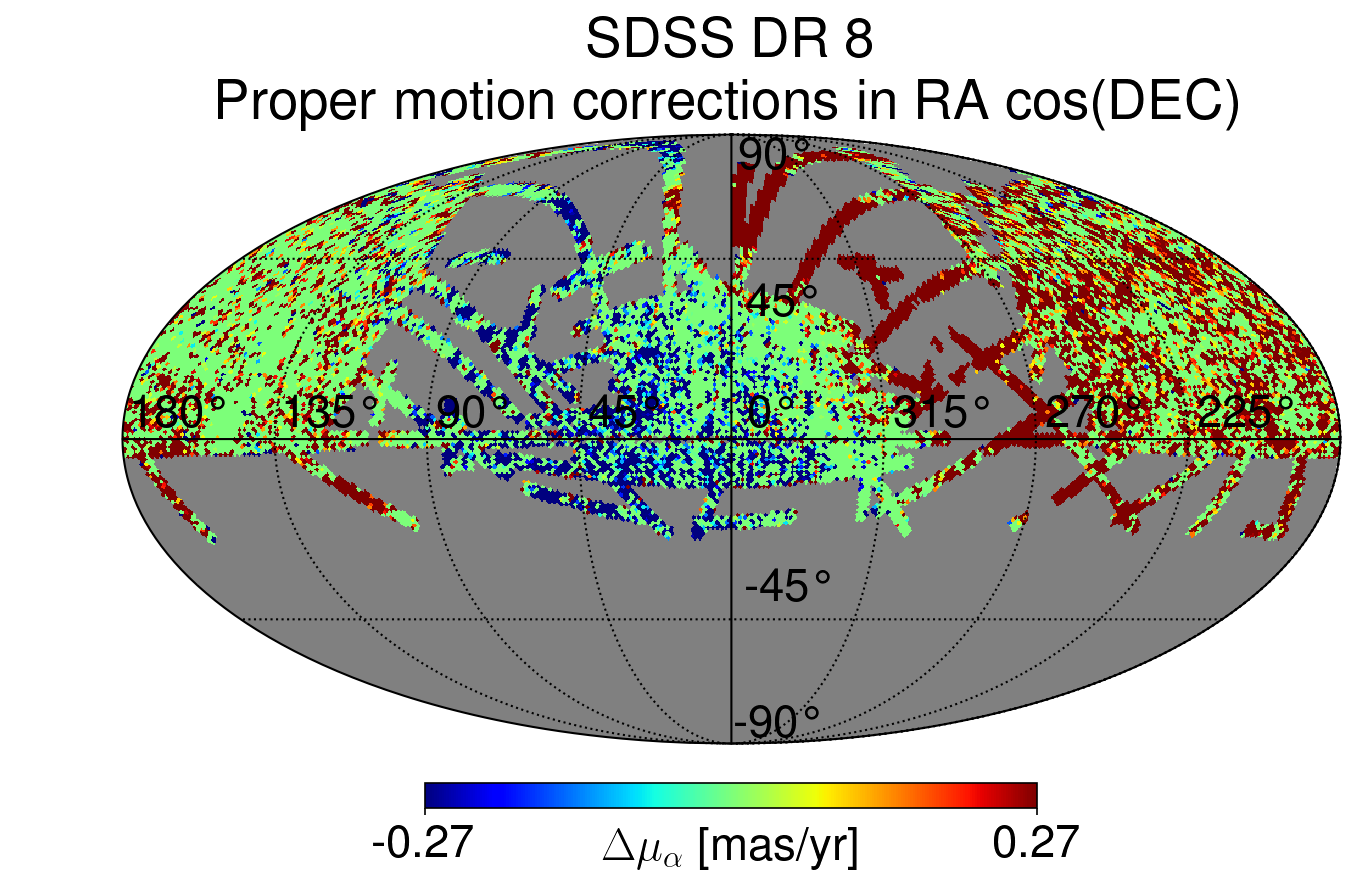} &
 \includegraphics[width=0.5\linewidth]{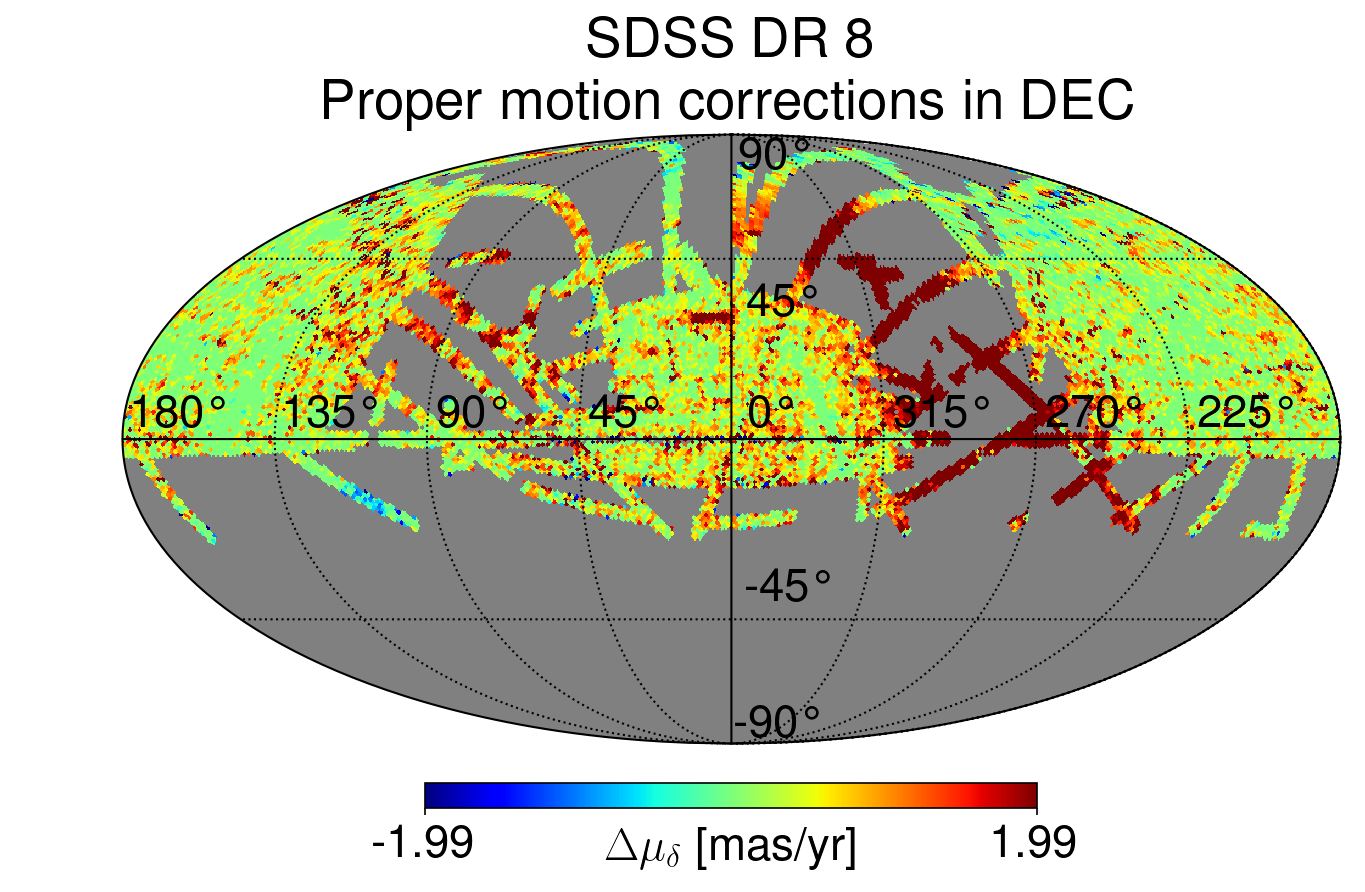}
 \end{tabular}
 \caption{Corrections in stellar positions and proper motion for the SDSS DR 8 catalog. \label{fig:sdss8}}
 \end{figure}

\begin{figure}
 \begin{tabular}{ll}
 \includegraphics[width=0.5\linewidth]{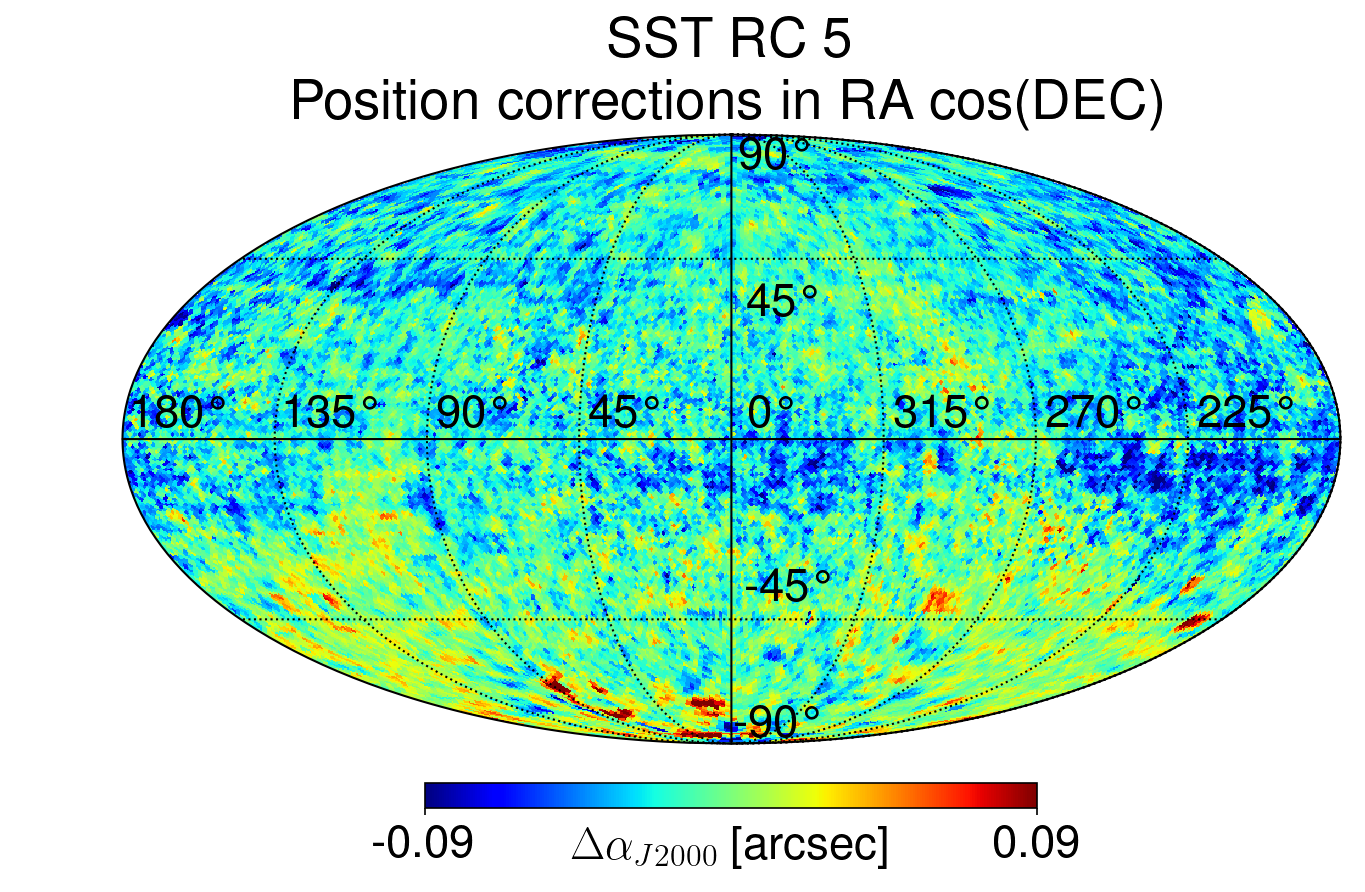} &
 \includegraphics[width=0.5\linewidth]{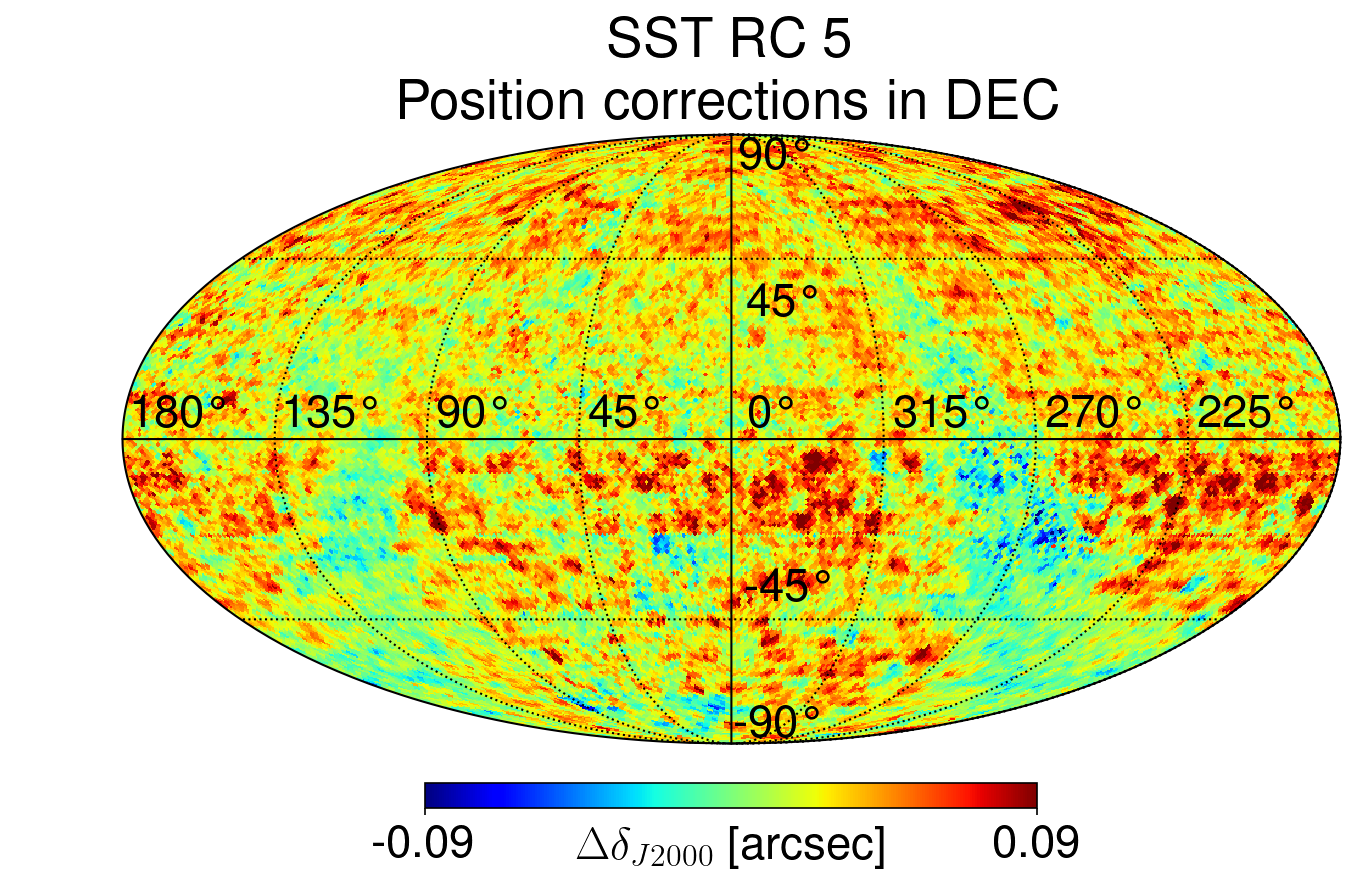} \\
 \includegraphics[width=0.5\linewidth]{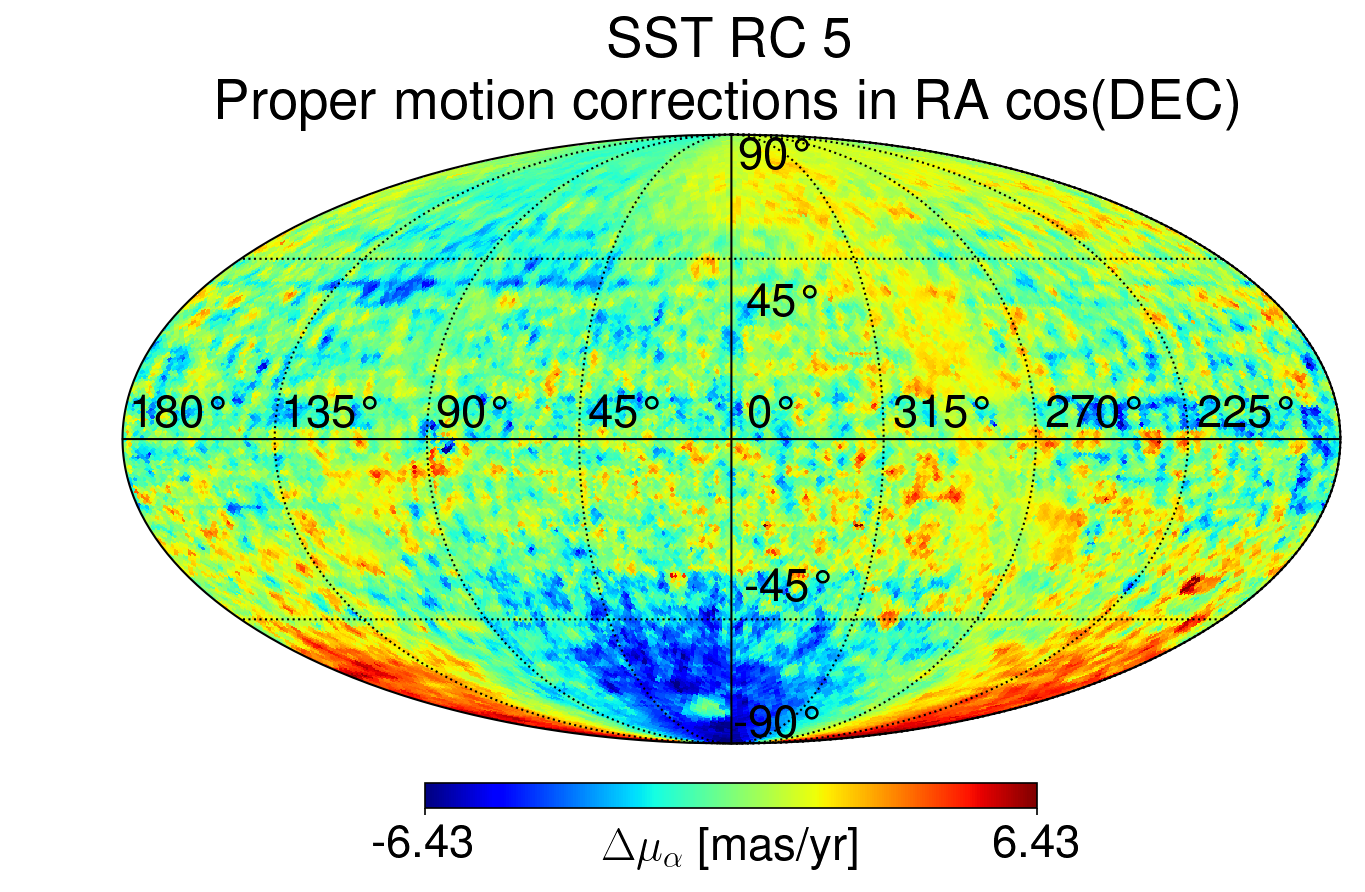} &
 \includegraphics[width=0.5\linewidth]{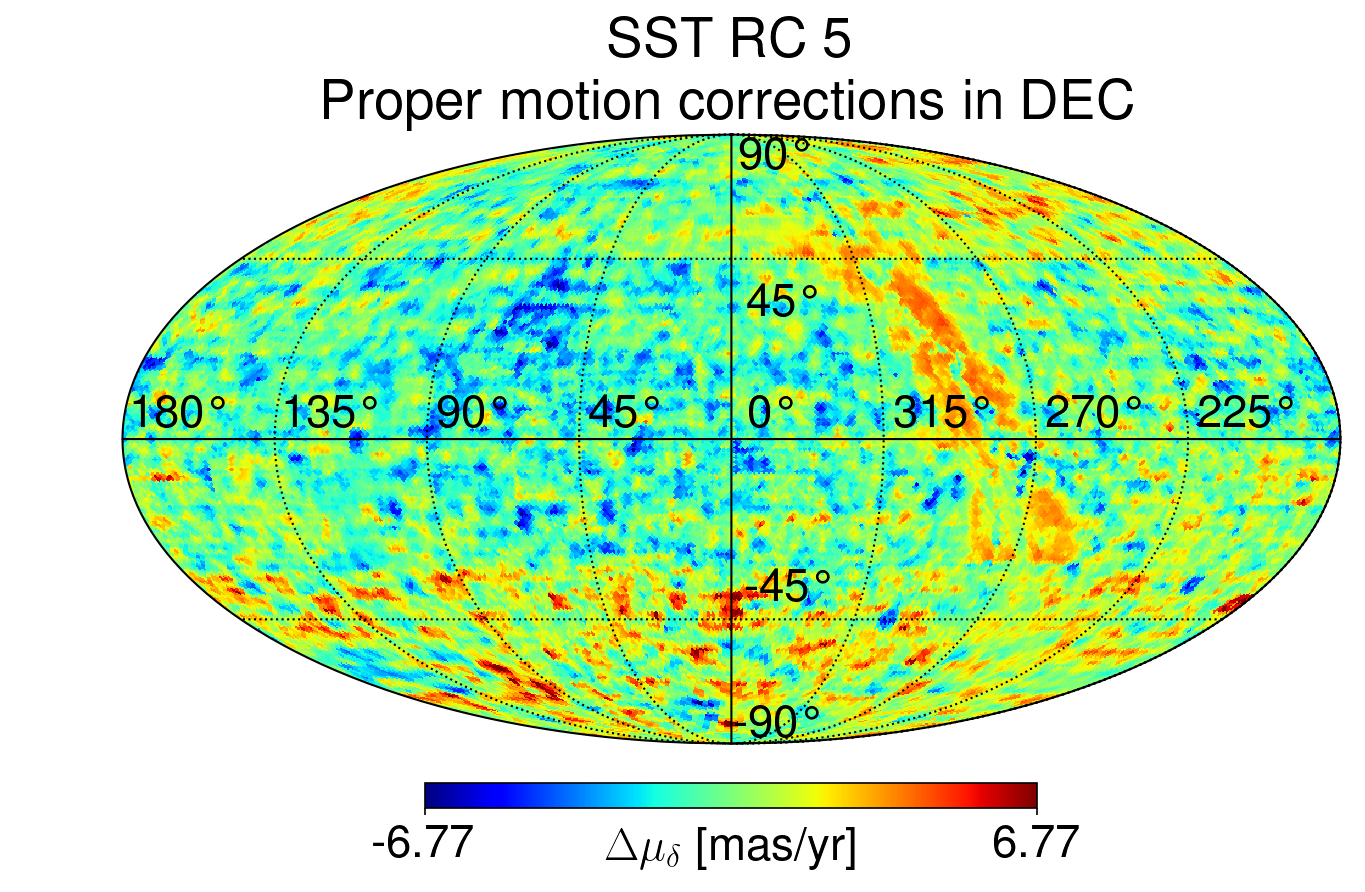}
 \end{tabular}
 \caption{Corrections in stellar positions and proper motion for the SST RC 5 catalog. \label{fig:sstrc5}}
 \end{figure}

\begin{figure}
 \begin{tabular}{ll}
 \includegraphics[width=0.5\linewidth]{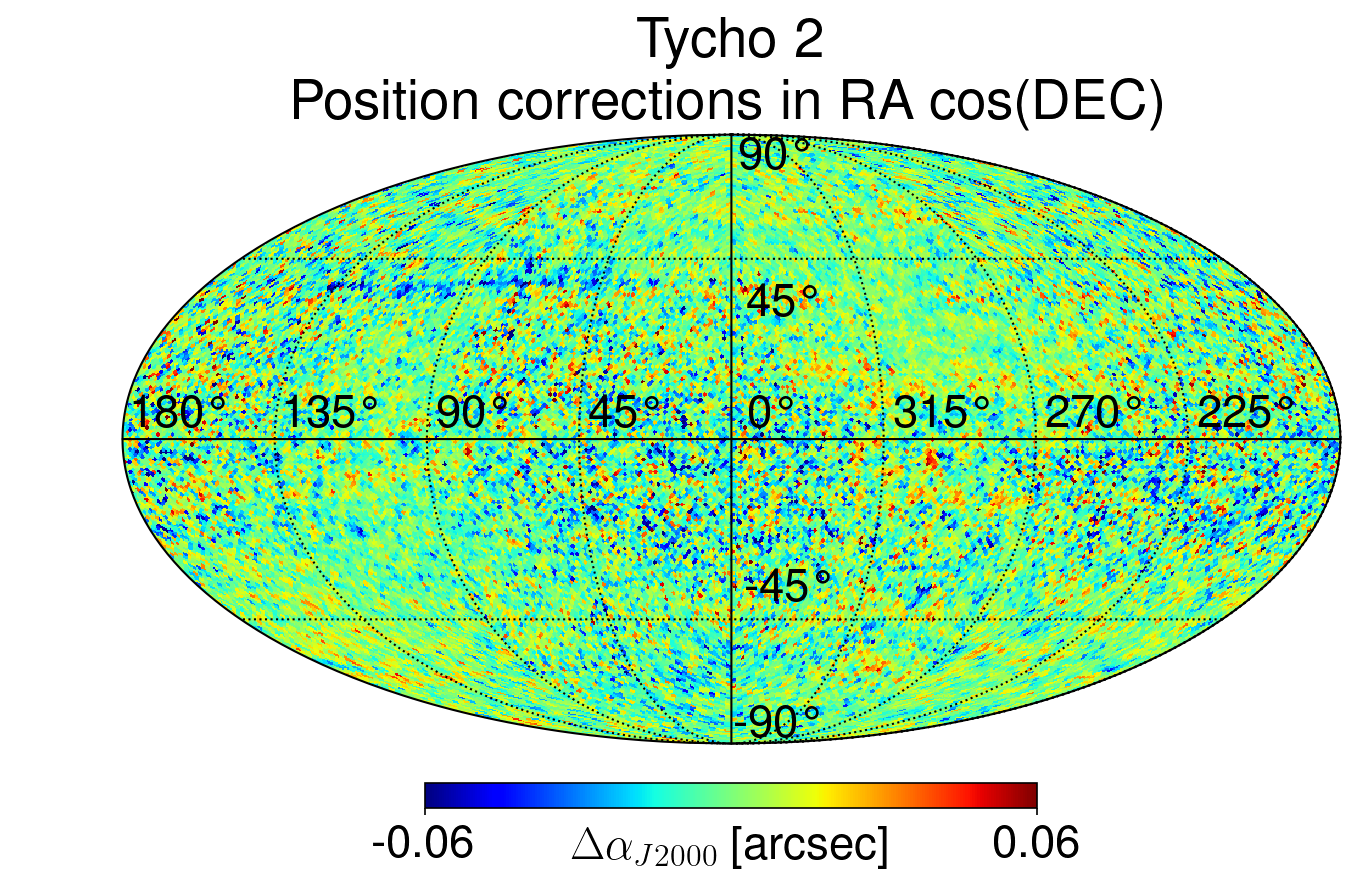} &
 \includegraphics[width=0.5\linewidth]{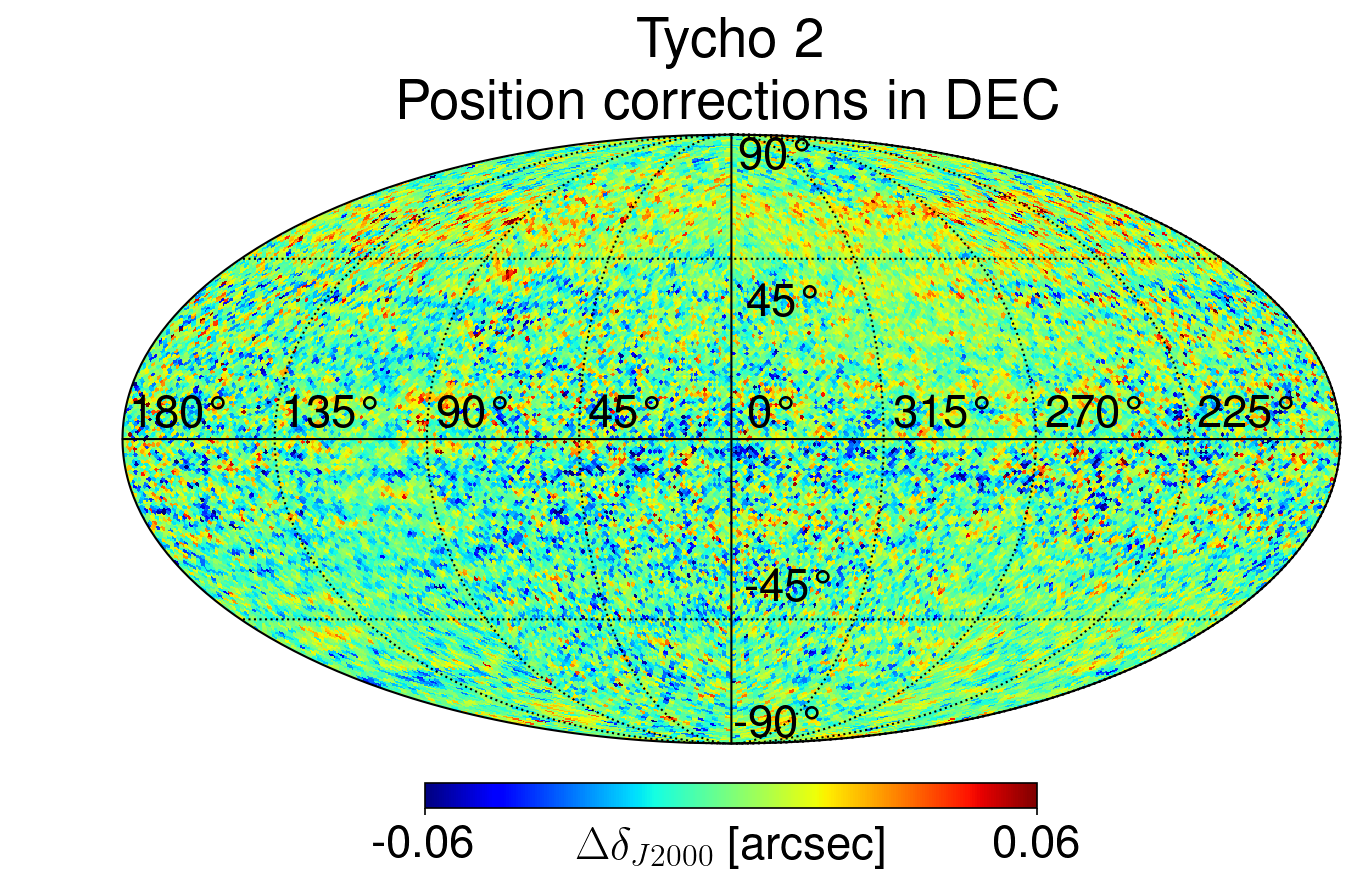} \\
 \includegraphics[width=0.5\linewidth]{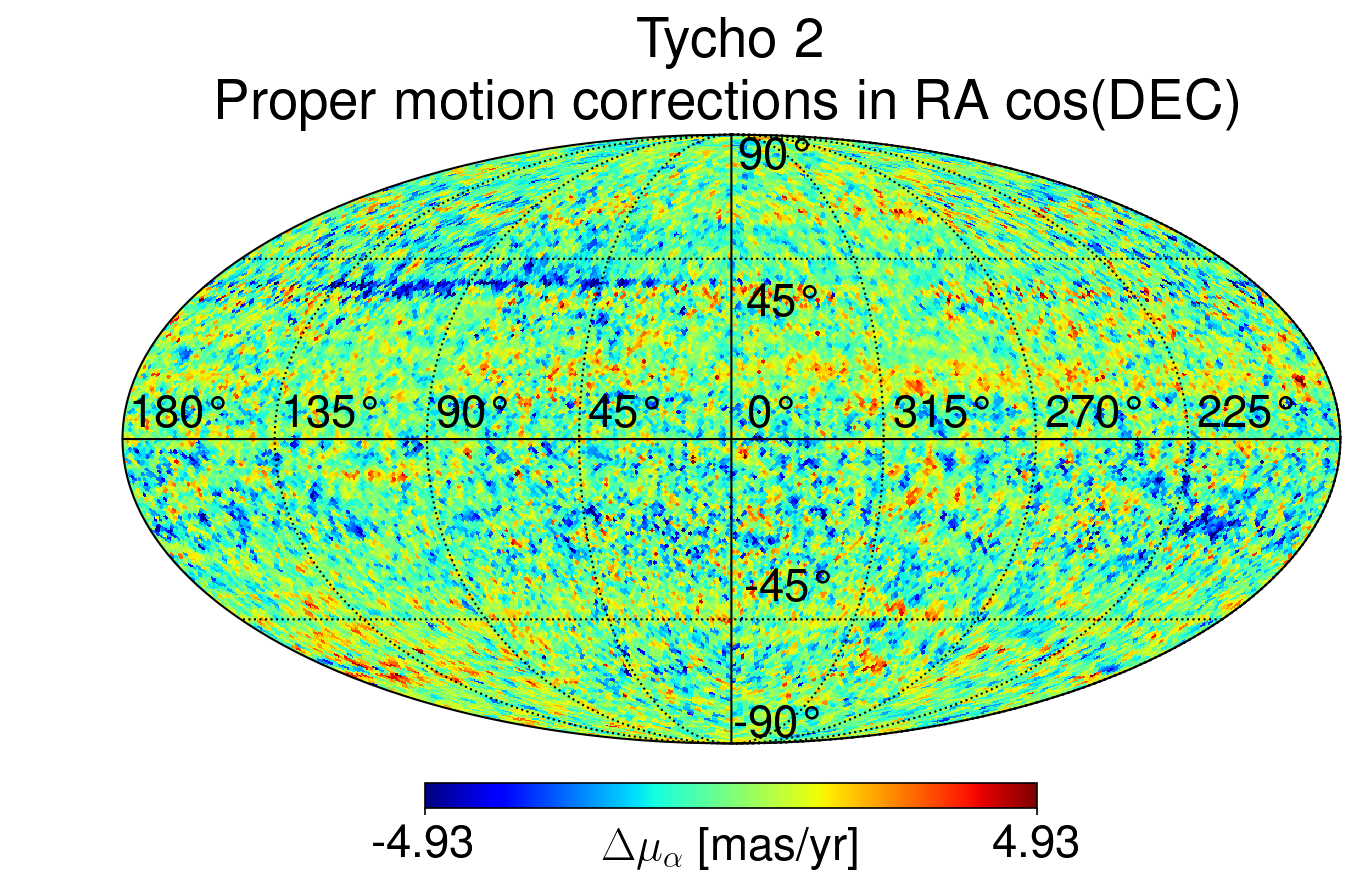} &
 \includegraphics[width=0.5\linewidth]{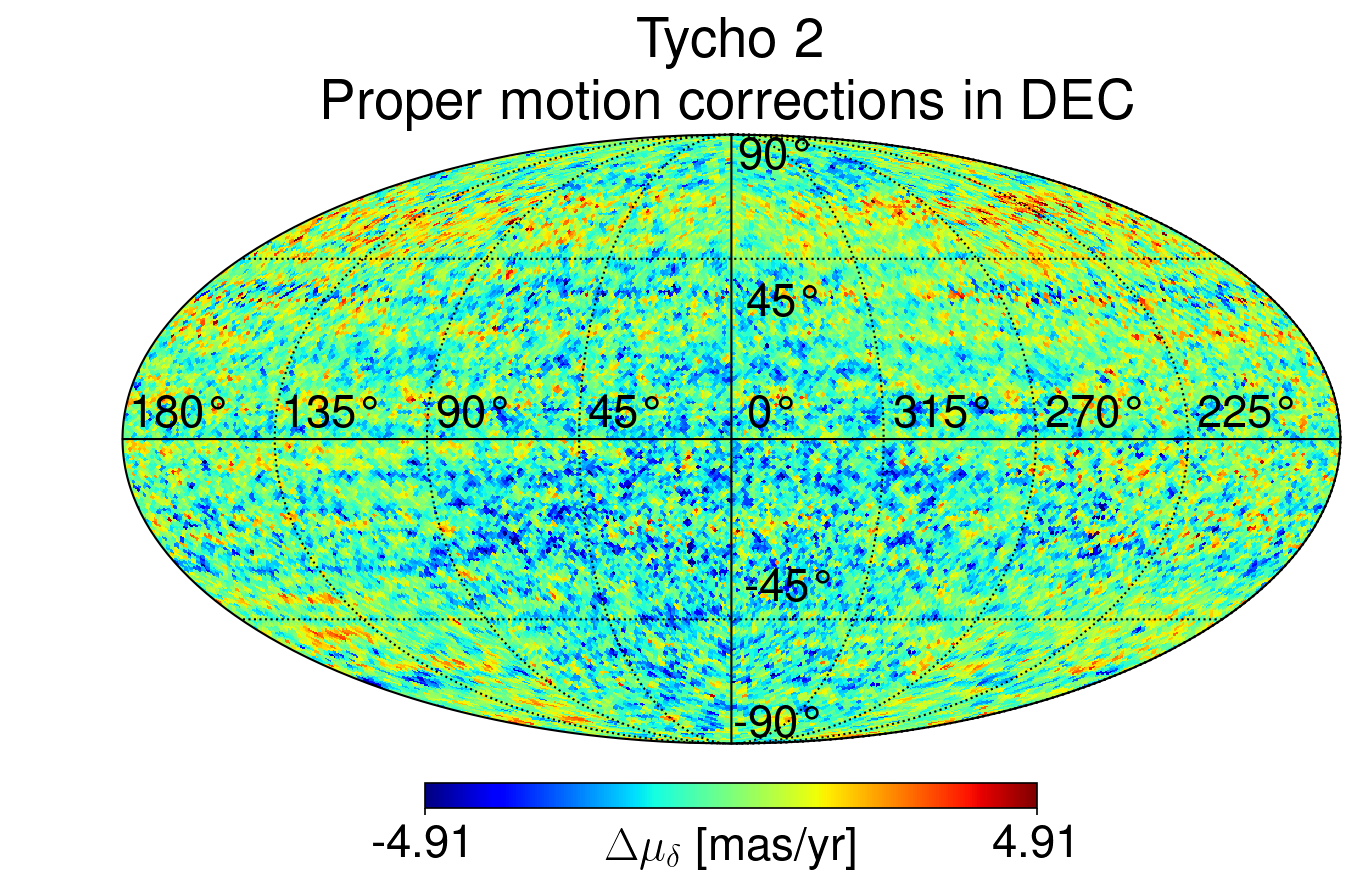}
 \end{tabular}
 \caption{Corrections in stellar positions and proper motion for the Tycho 2 catalog. \label{fig:tycho2}}
 \end{figure}

\begin{figure}
 \begin{tabular}{ll}
 \includegraphics[width=0.5\linewidth]{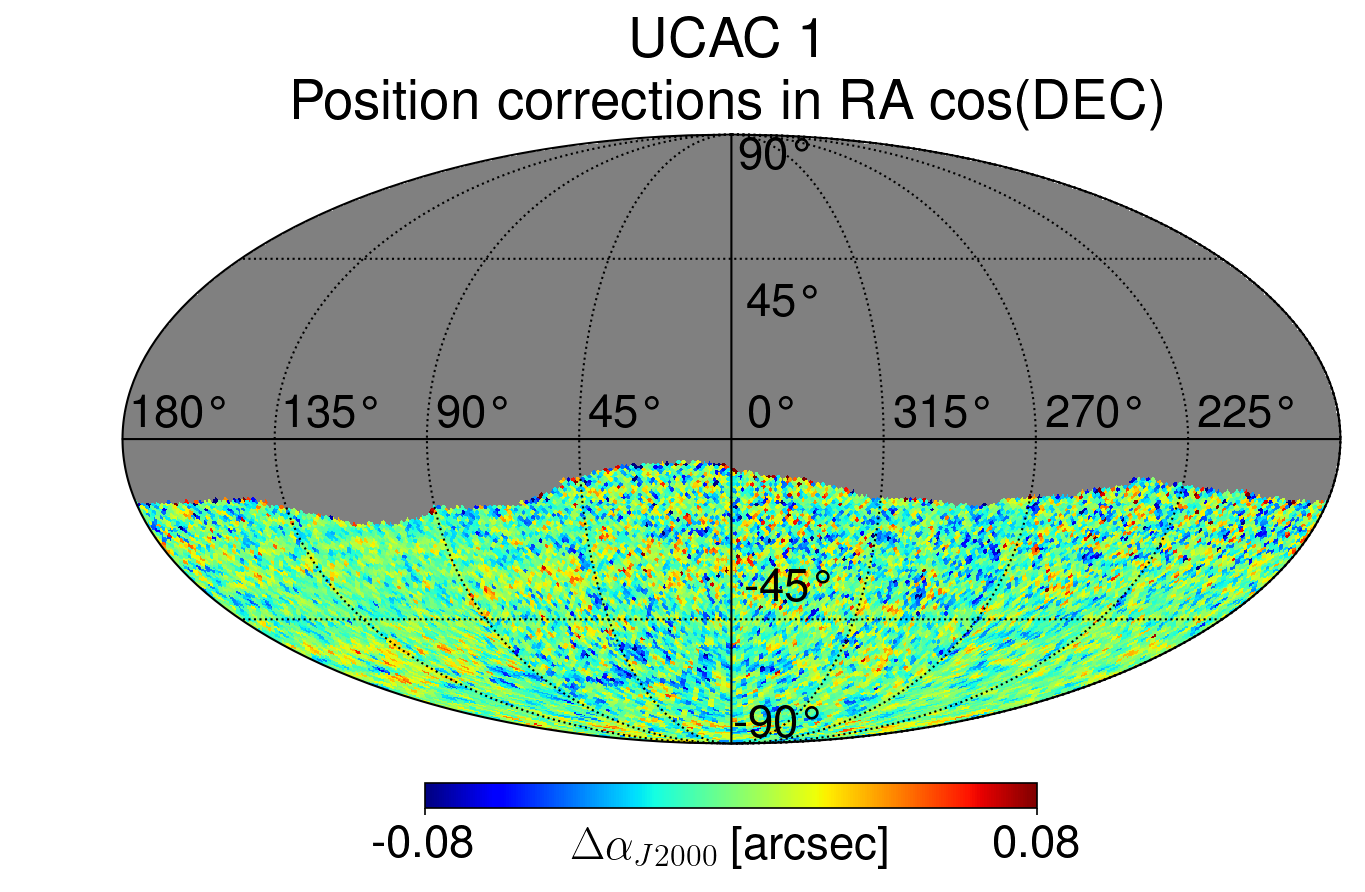} &
 \includegraphics[width=0.5\linewidth]{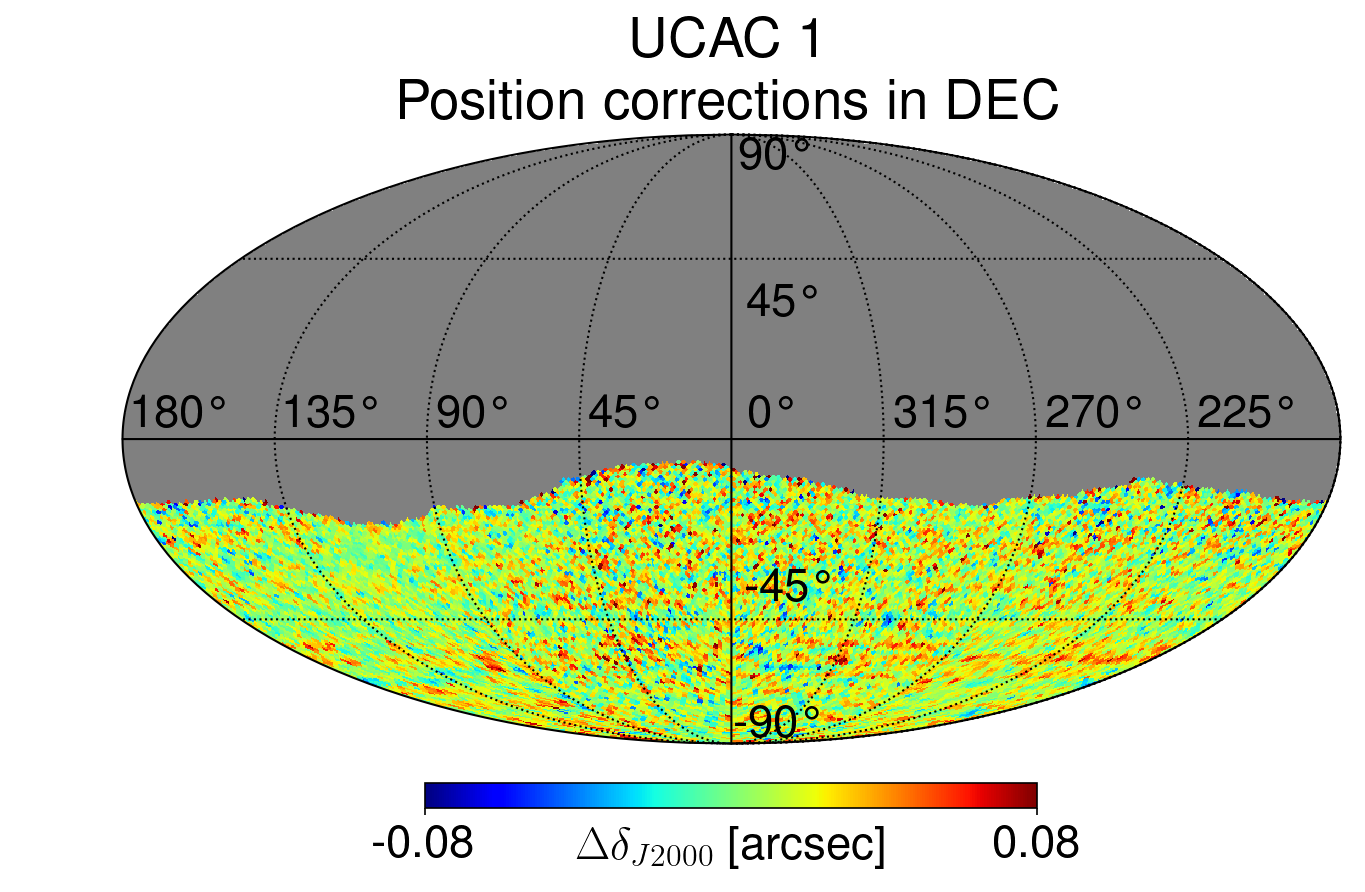} \\
 \includegraphics[width=0.5\linewidth]{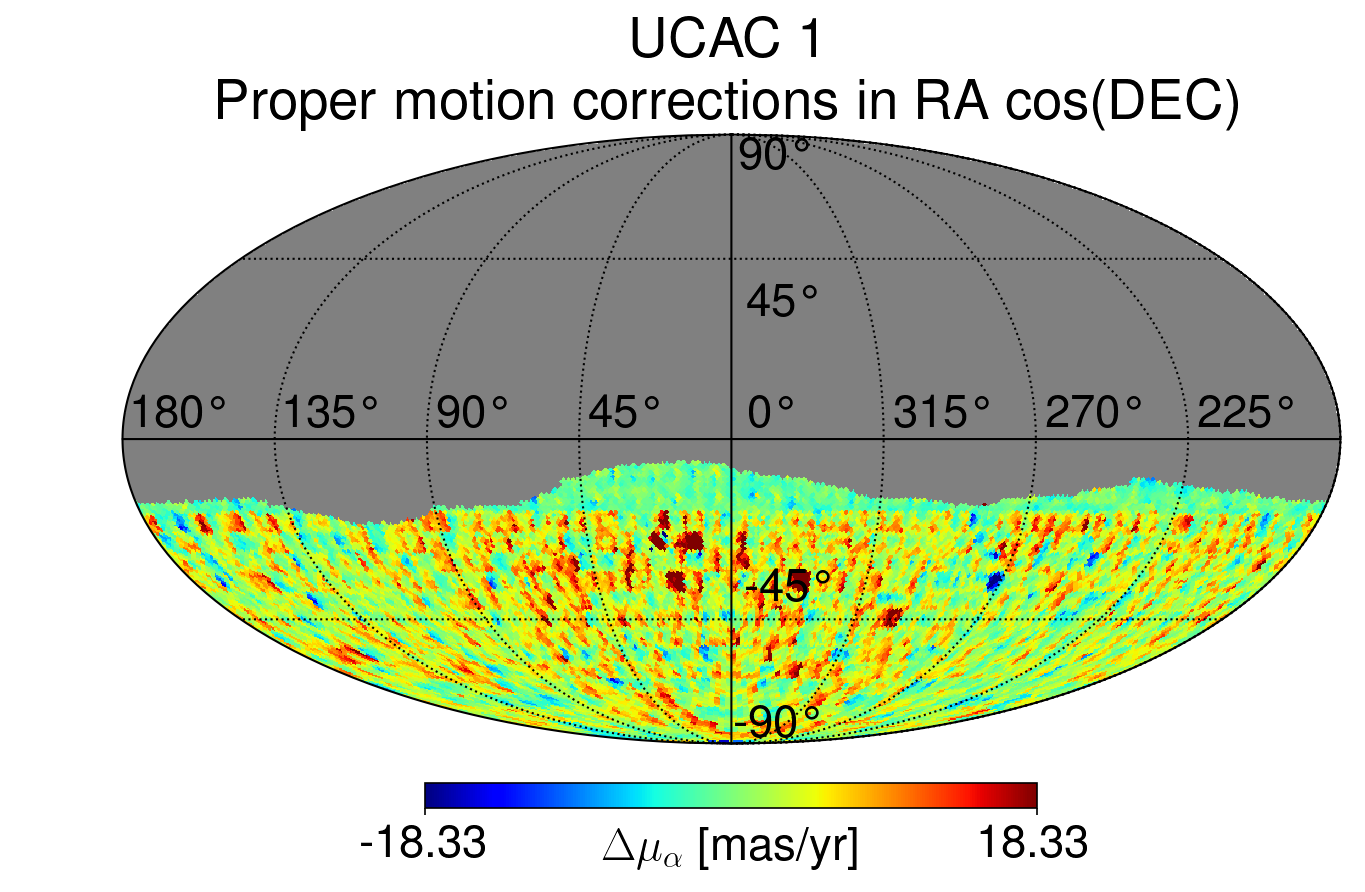} &
 \includegraphics[width=0.5\linewidth]{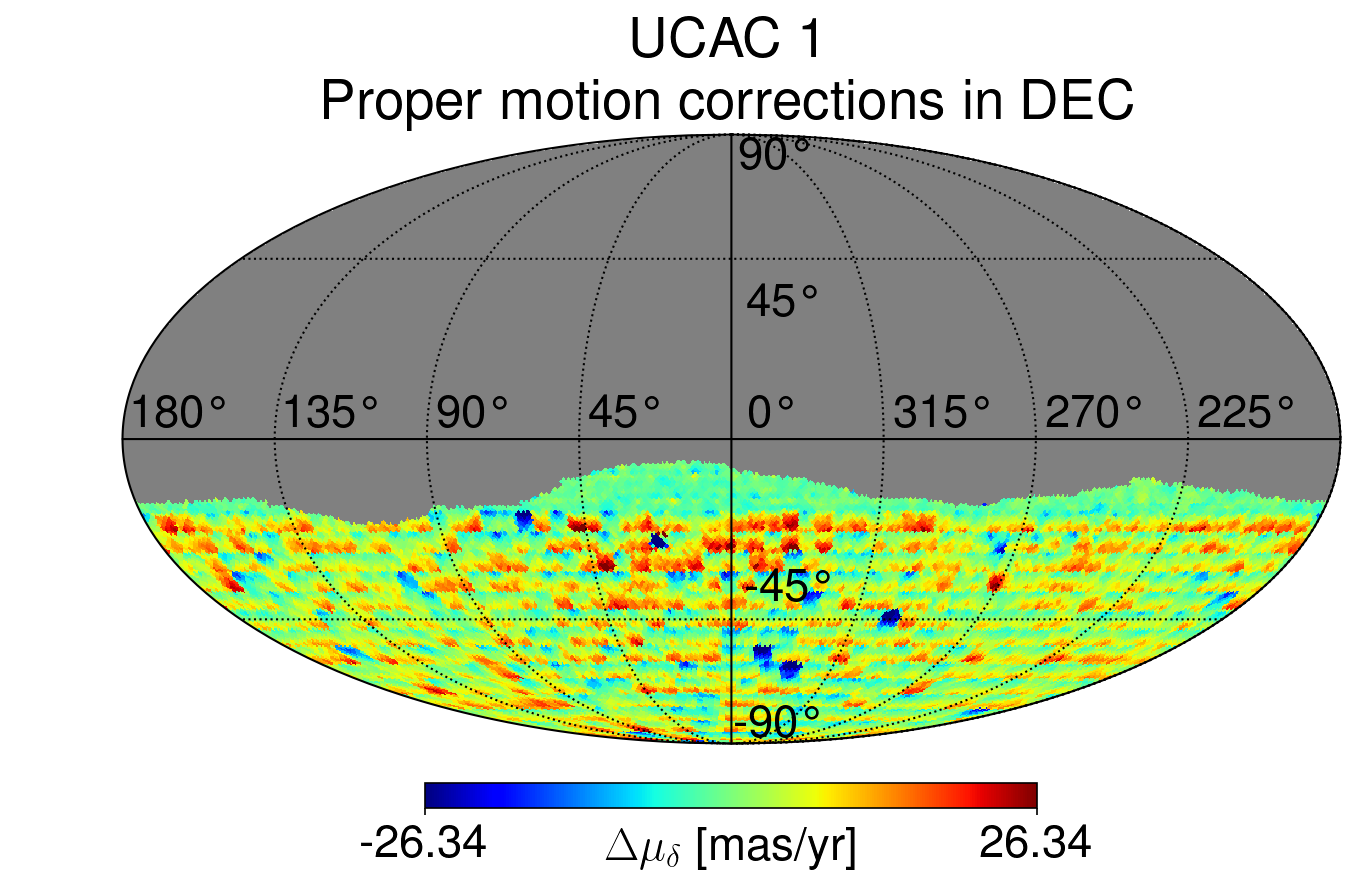}
 \end{tabular}
 \caption{Corrections in stellar positions and proper motion for the UCAC 1 catalog. \label{fig:ucac1}}
 \end{figure}

\begin{figure}
 \begin{tabular}{ll}
 \includegraphics[width=0.5\linewidth]{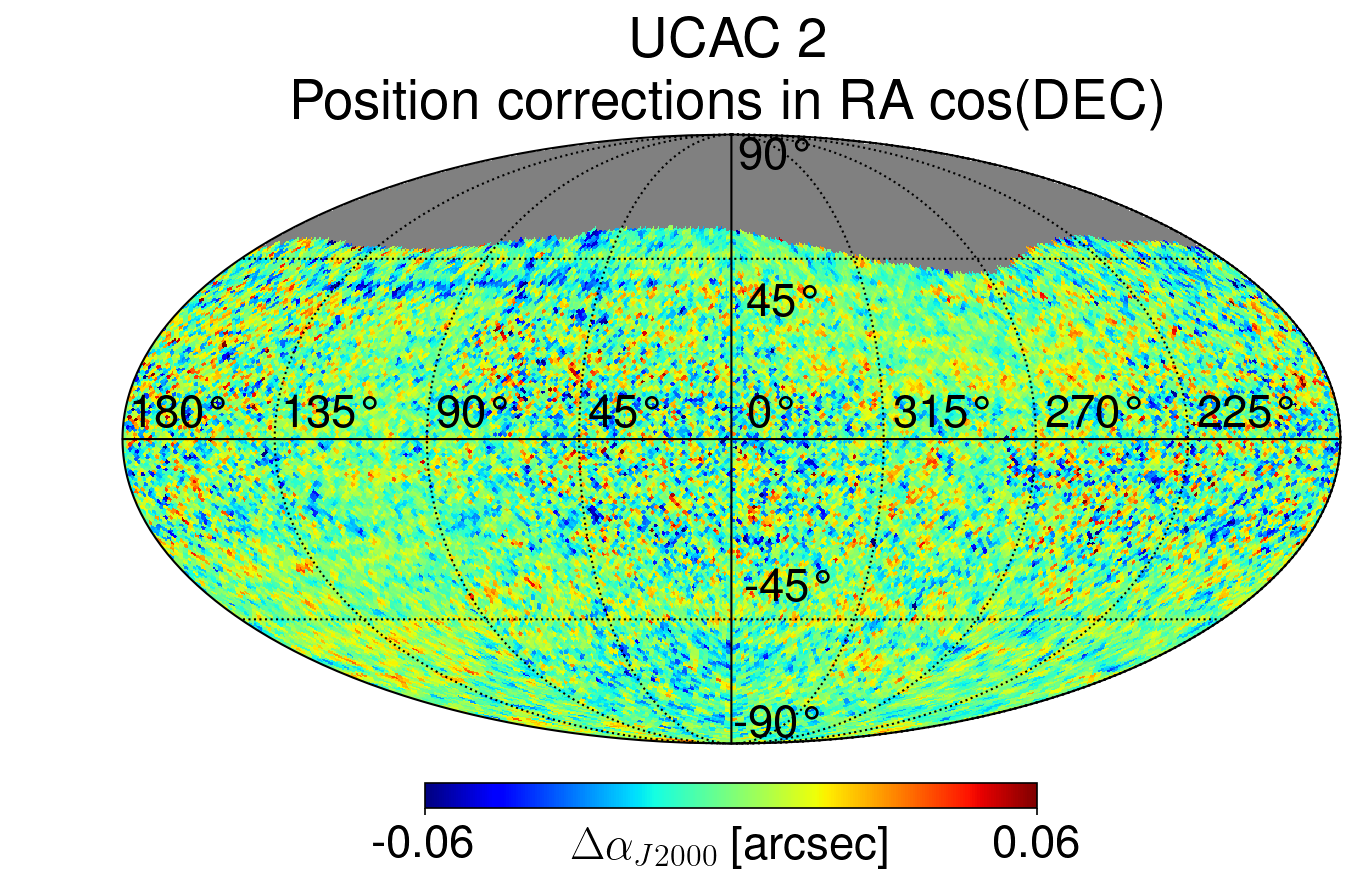} &
 \includegraphics[width=0.5\linewidth]{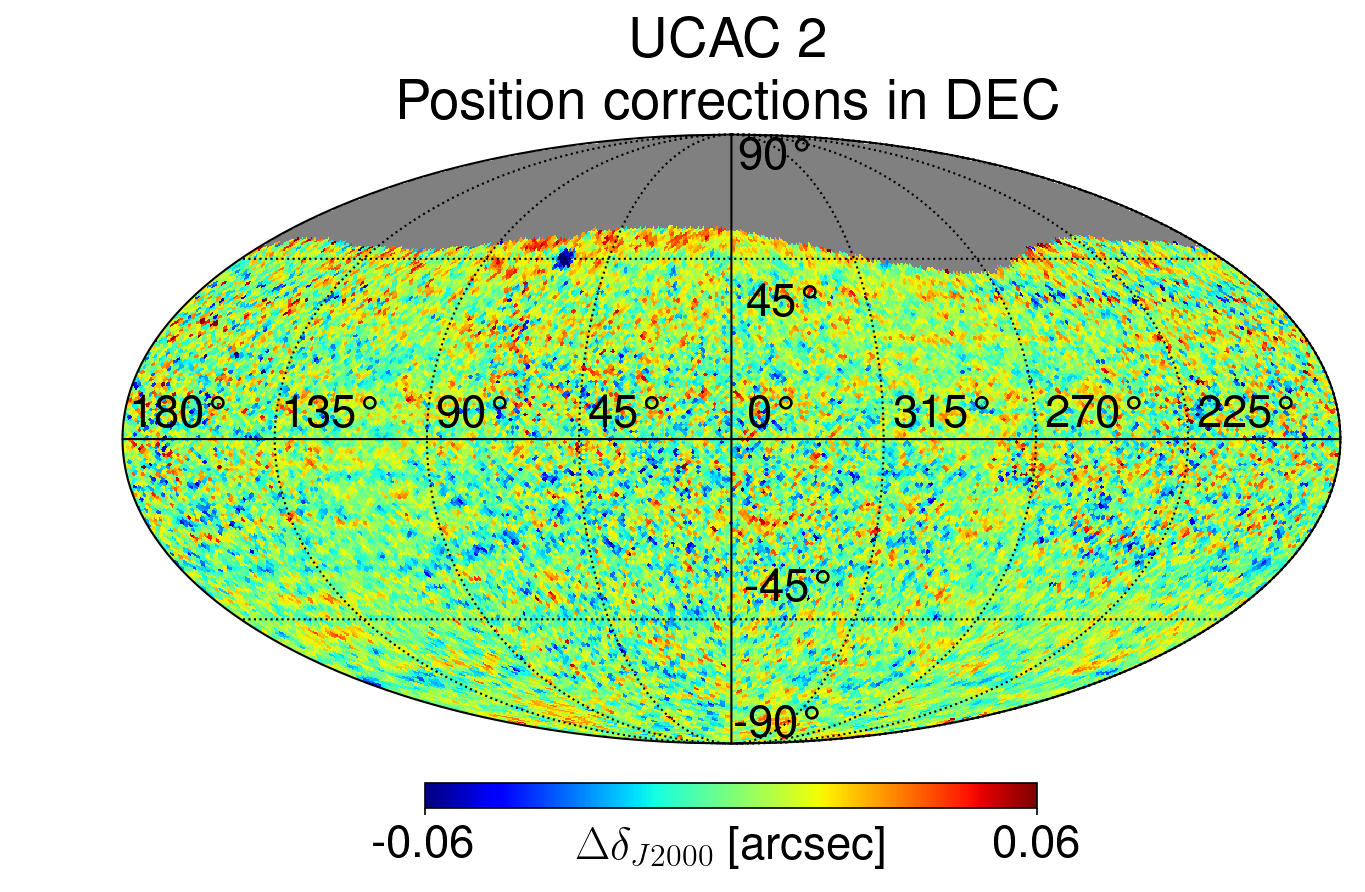} \\
 \includegraphics[width=0.5\linewidth]{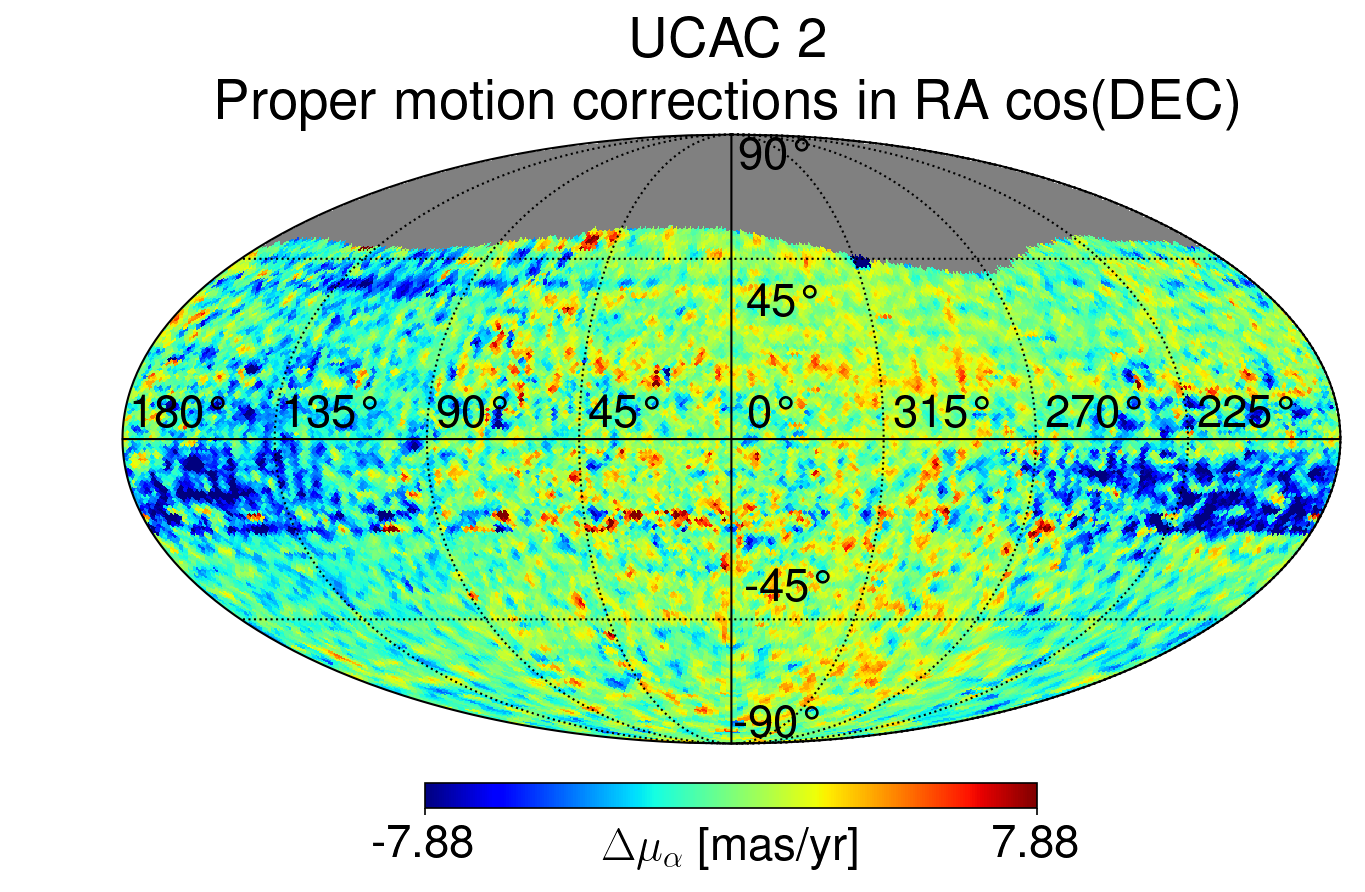} &
 \includegraphics[width=0.5\linewidth]{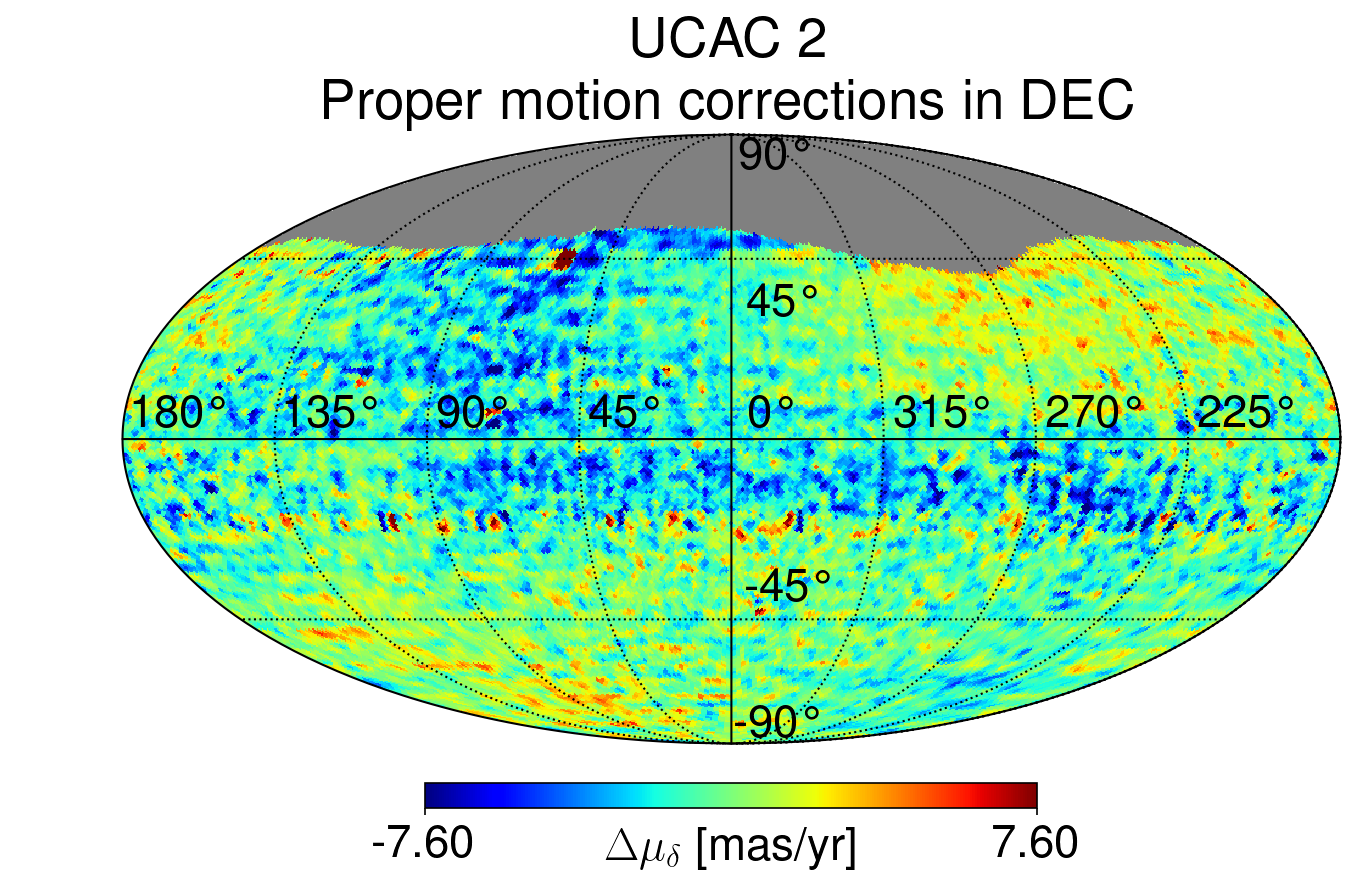}
 \end{tabular}
 \caption{Corrections in stellar positions and proper motion for  the UCAC 2 catalog. \label{fig:ucac2}}
 \end{figure}

\begin{figure}
 \begin{tabular}{ll}
 \includegraphics[width=0.5\linewidth]{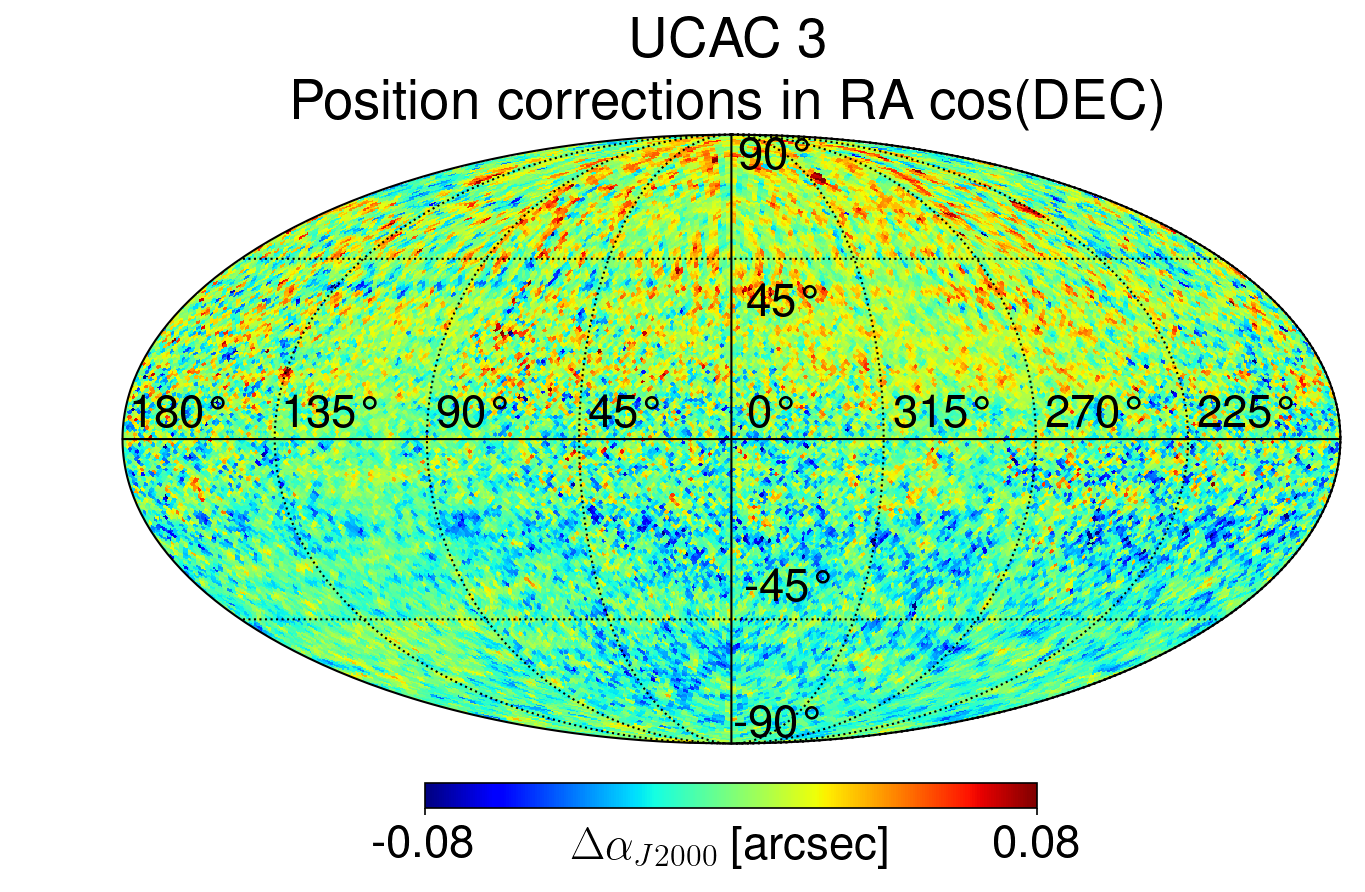} &
 \includegraphics[width=0.5\linewidth]{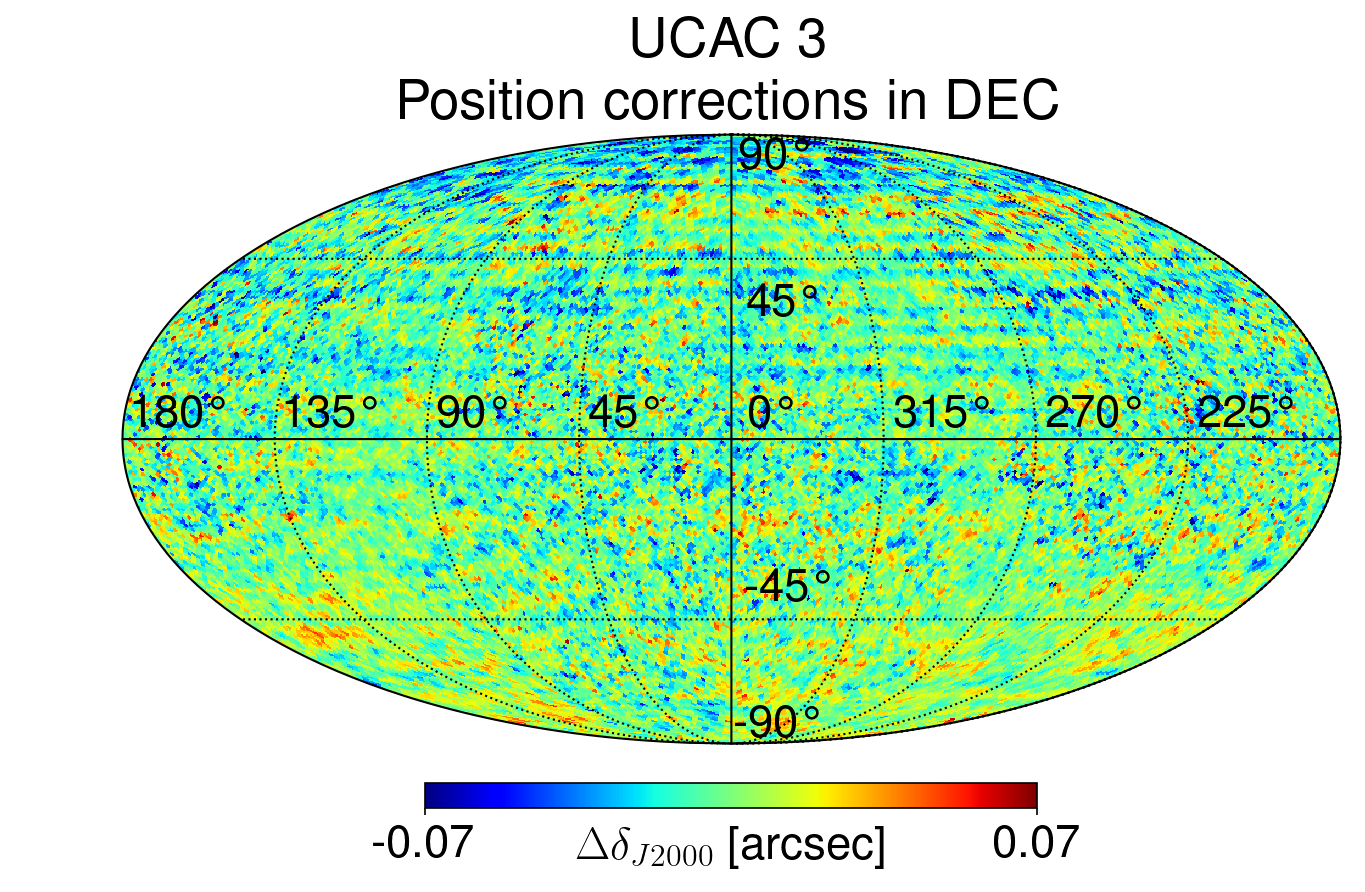} \\
 \includegraphics[width=0.5\linewidth]{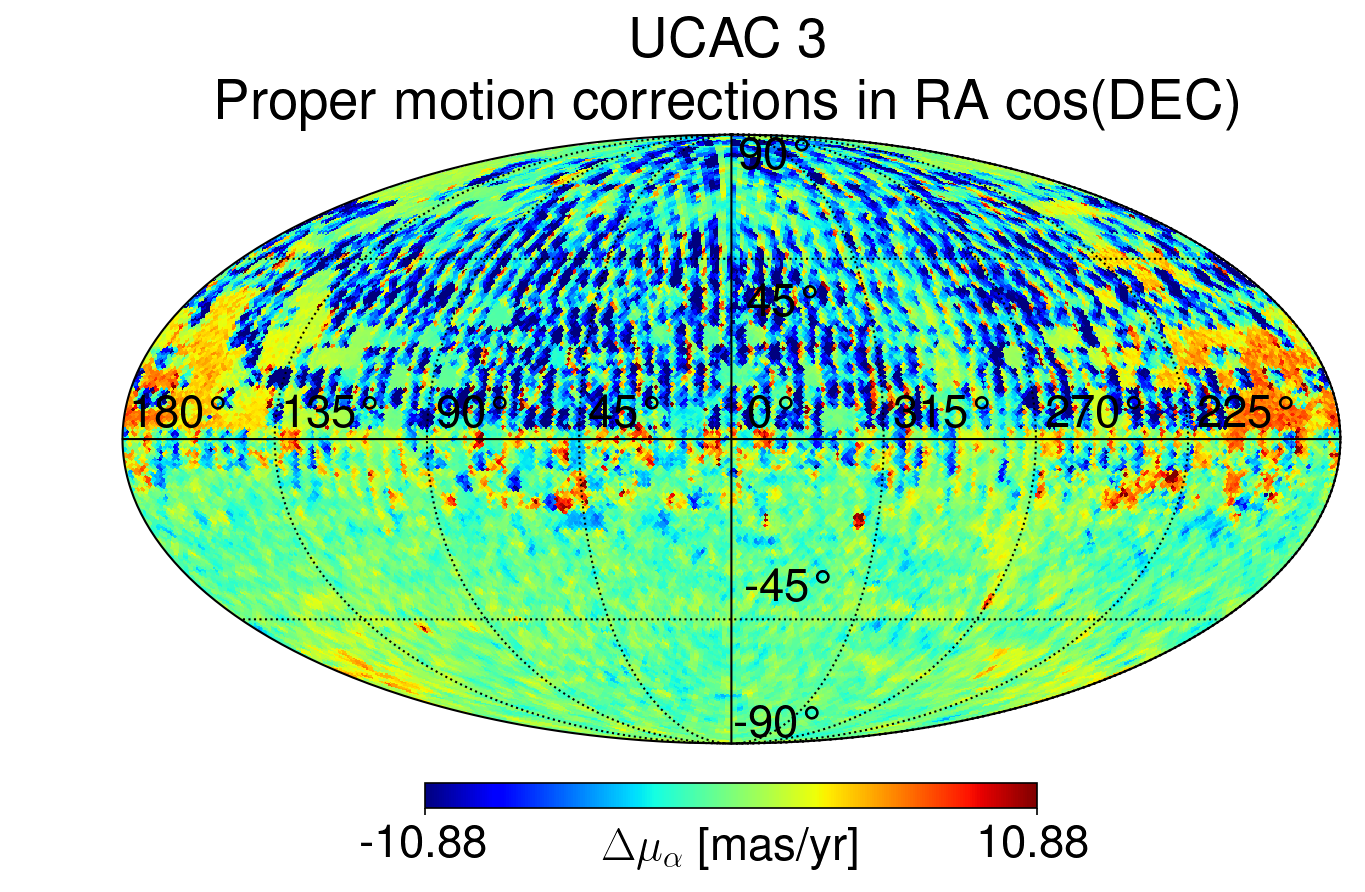} &
 \includegraphics[width=0.5\linewidth]{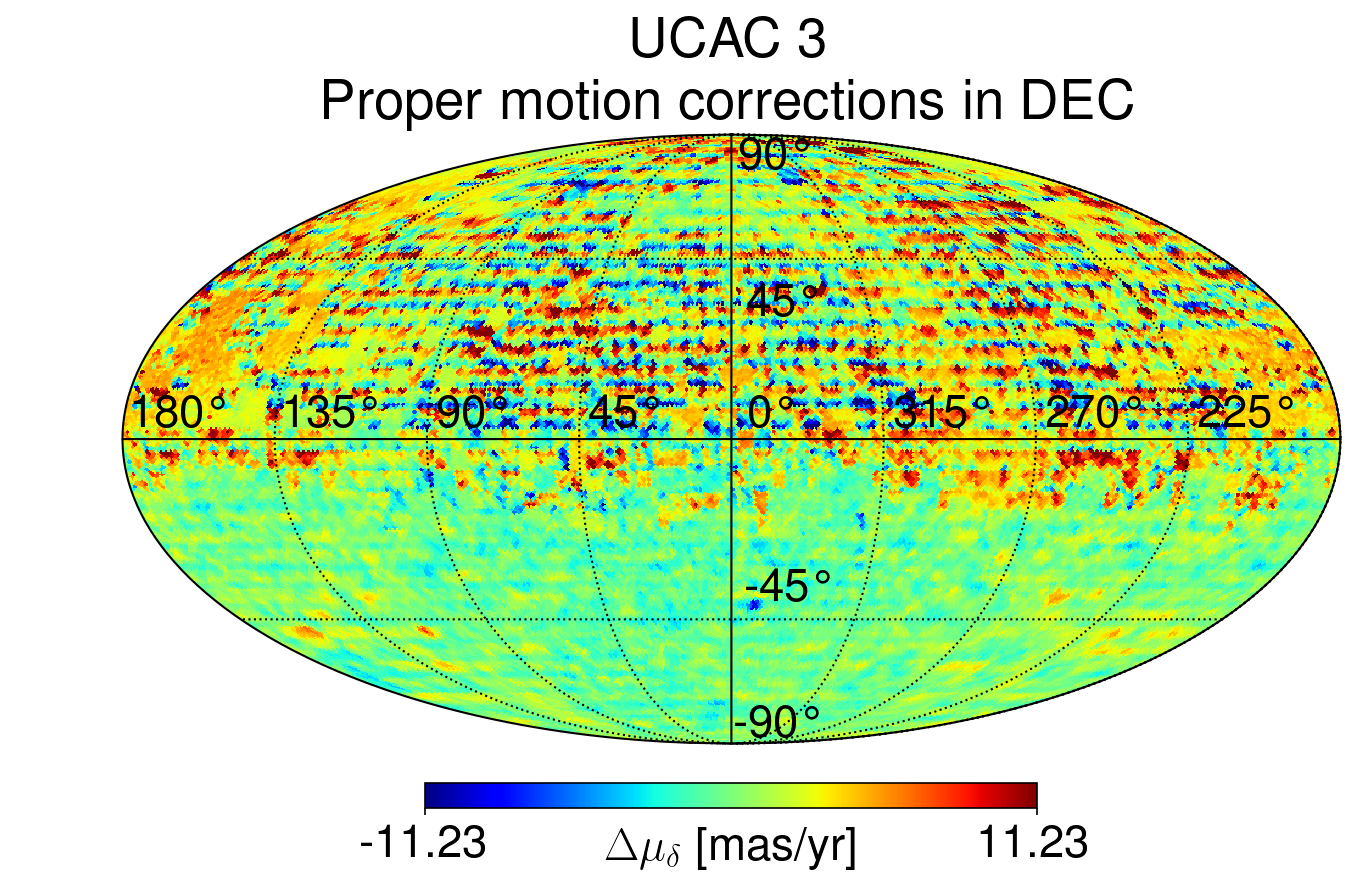}
 \end{tabular}
 \caption{Corrections in stellar positions and proper motion for the UCAC 3 catalog. \label{fig:ucac3}}
 \end{figure}

\begin{figure}
 \begin{tabular}{ll}
 \includegraphics[width=0.5\linewidth]{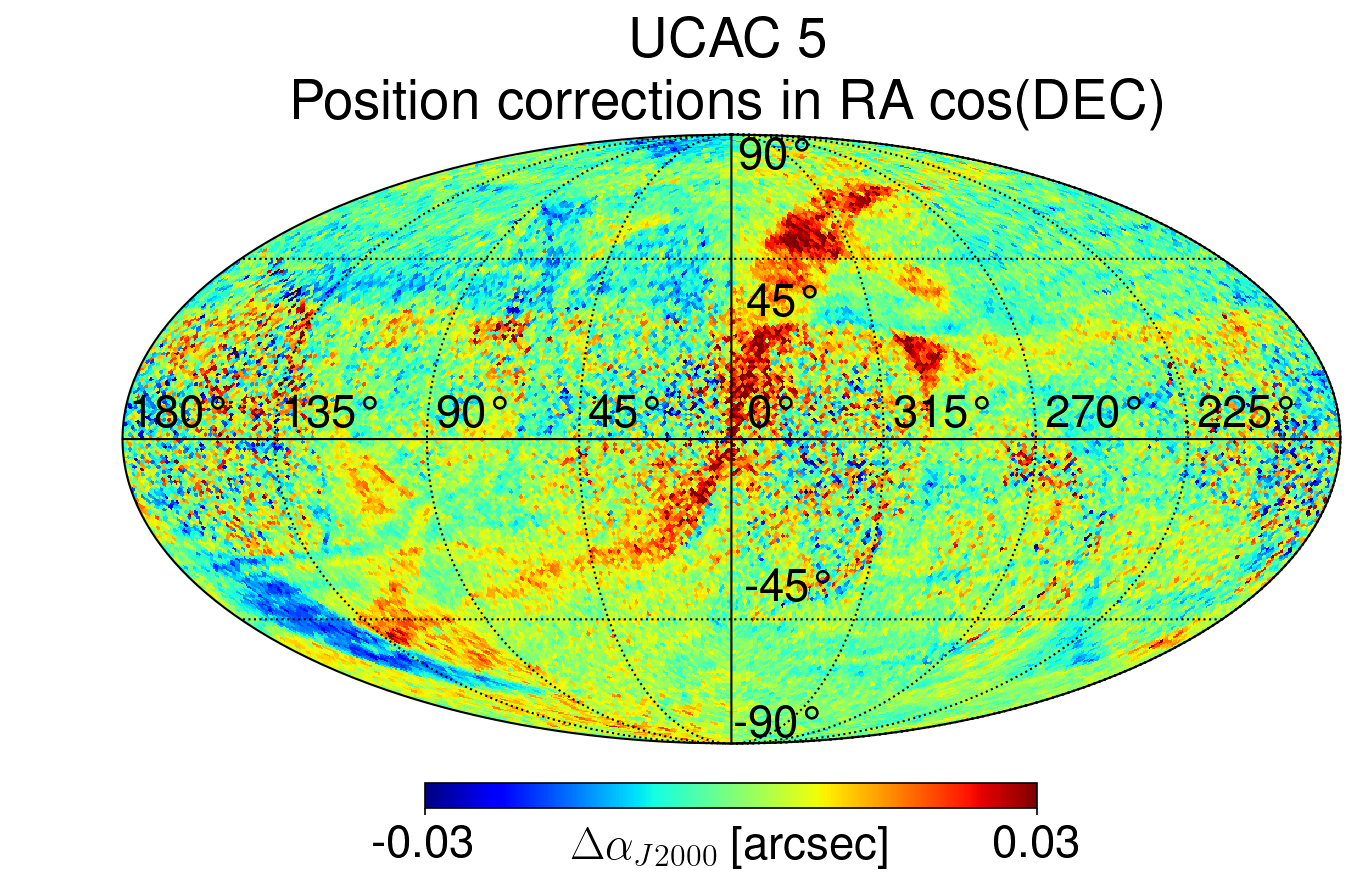} &
 \includegraphics[width=0.5\linewidth]{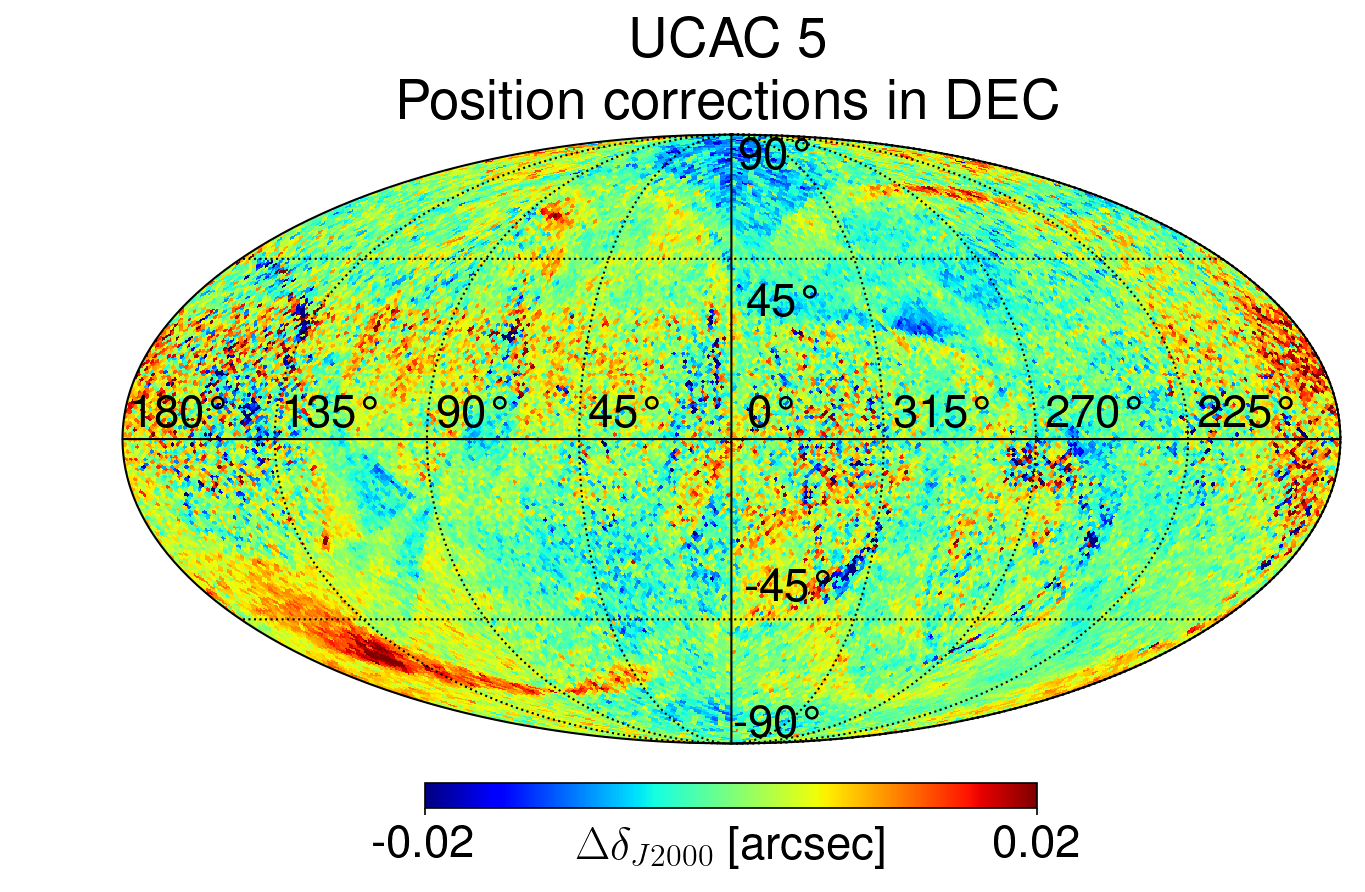} \\
 \includegraphics[width=0.5\linewidth]{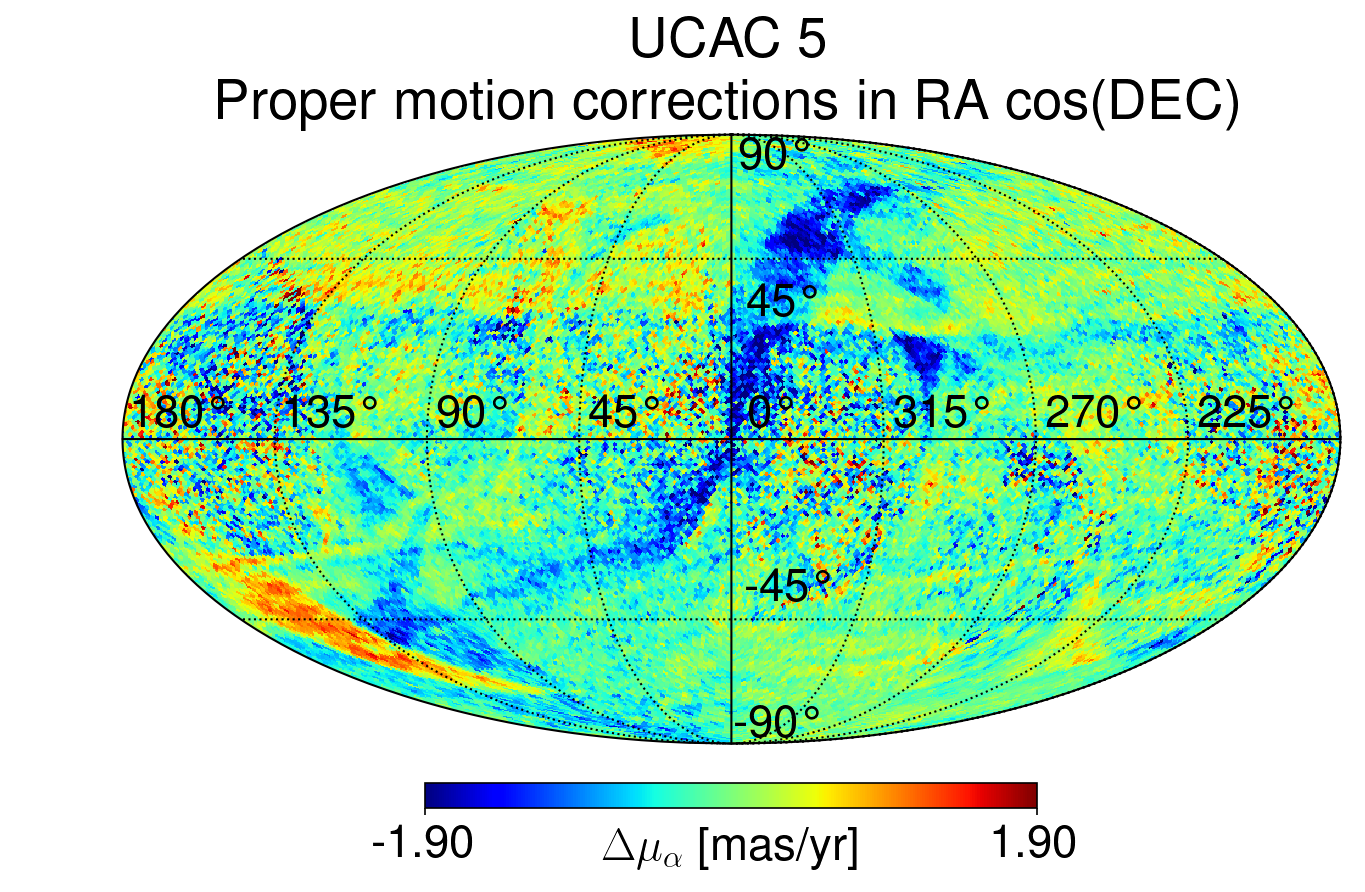} &
 \includegraphics[width=0.5\linewidth]{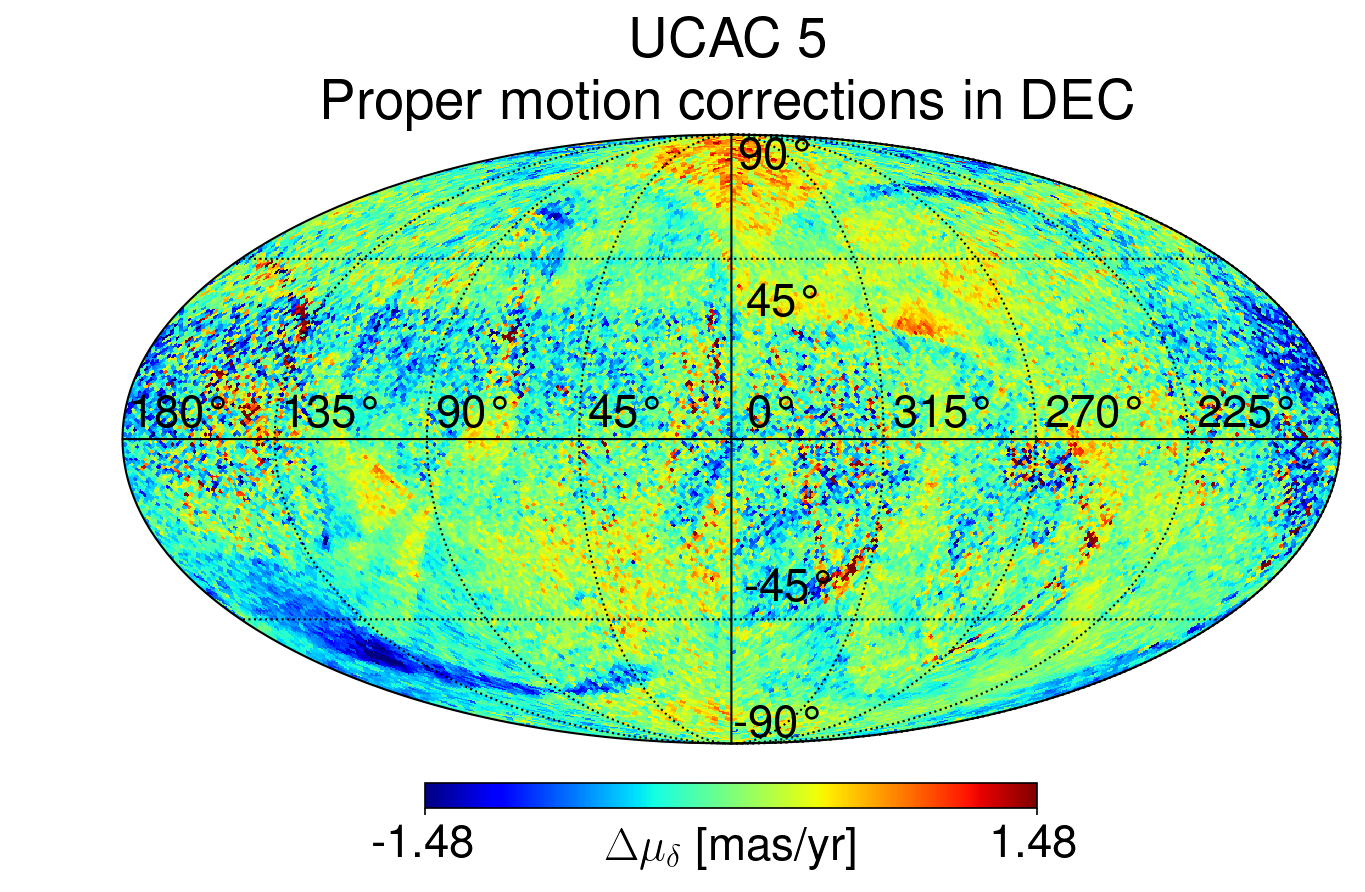}
 \end{tabular}
 \caption{Corrections in stellar positions and proper motion for the UCAC 5 catalog. \label{fig:ucac5}}
 \end{figure}

\begin{figure}
 \begin{tabular}{ll}
 \includegraphics[width=0.5\linewidth]{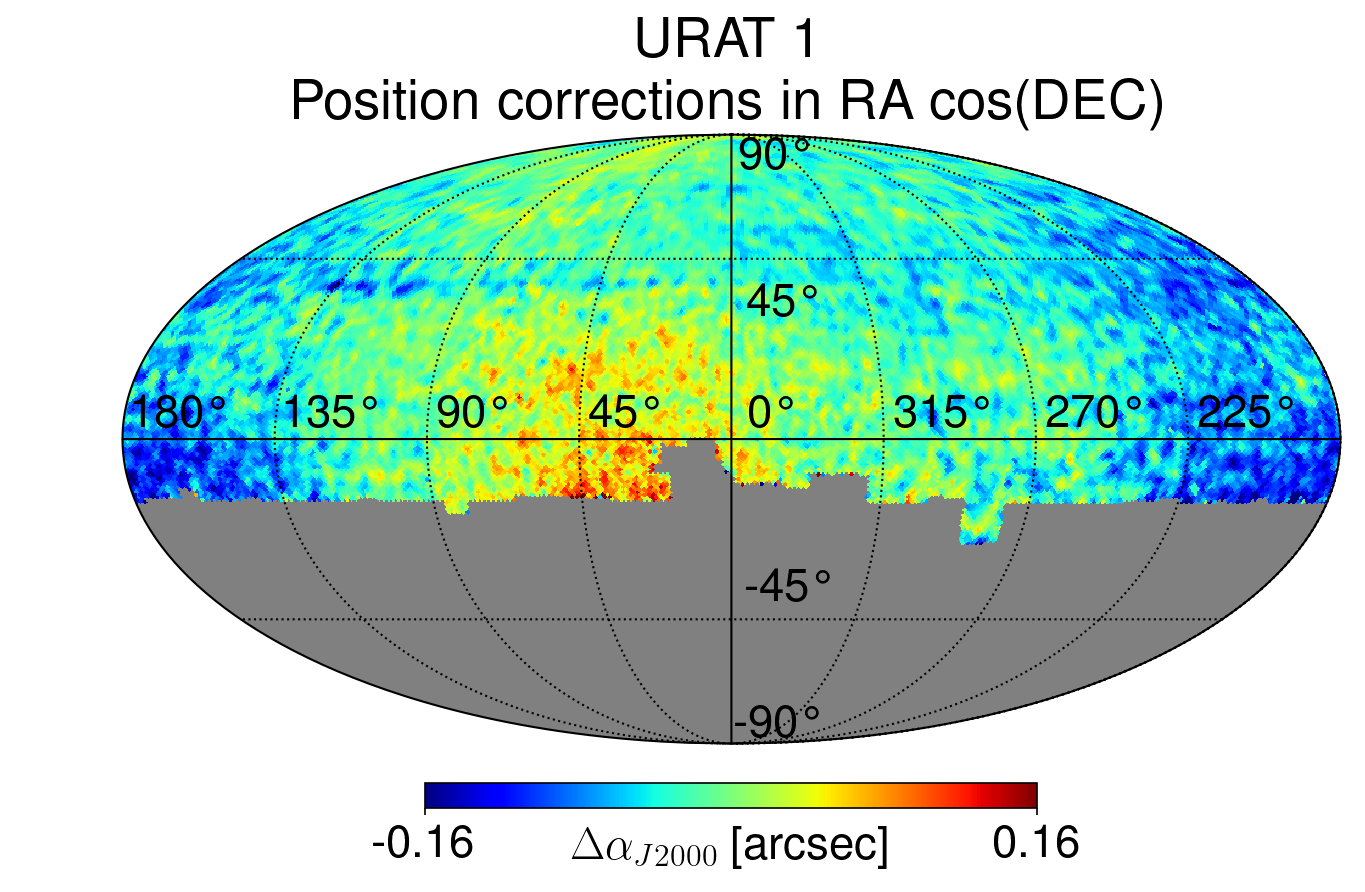} &
 \includegraphics[width=0.5\linewidth]{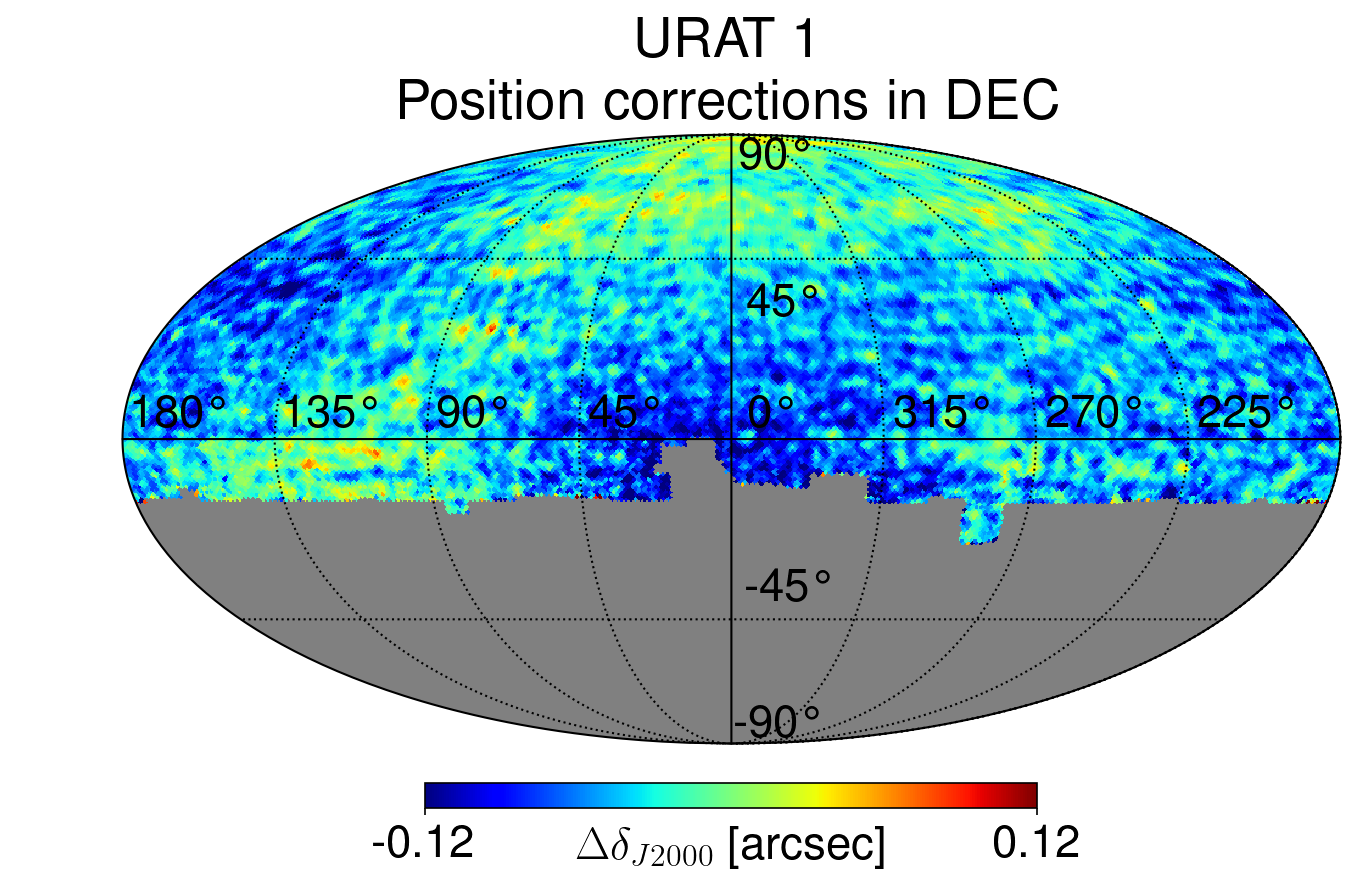} \\
 \includegraphics[width=0.5\linewidth]{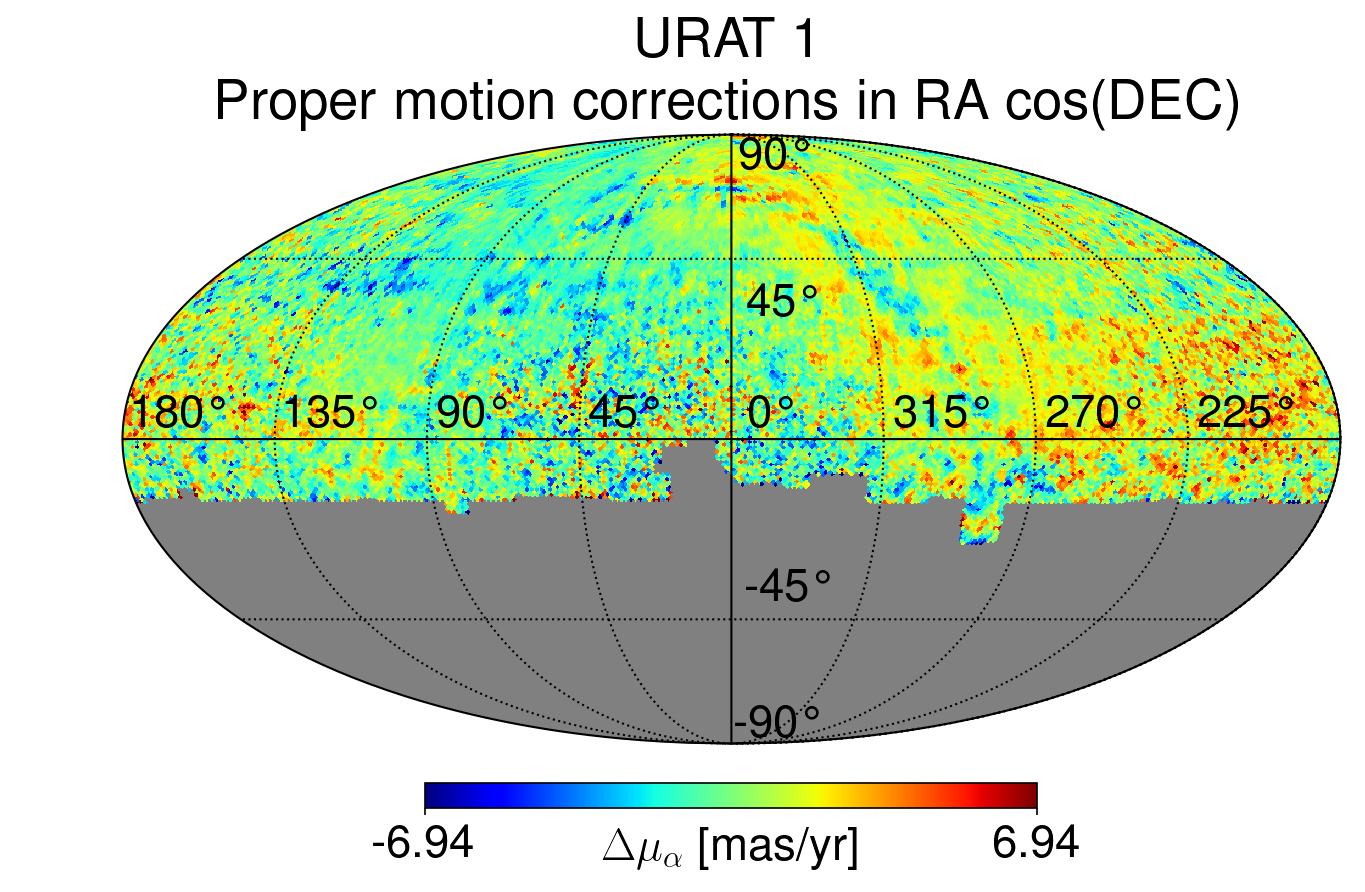} &
 \includegraphics[width=0.5\linewidth]{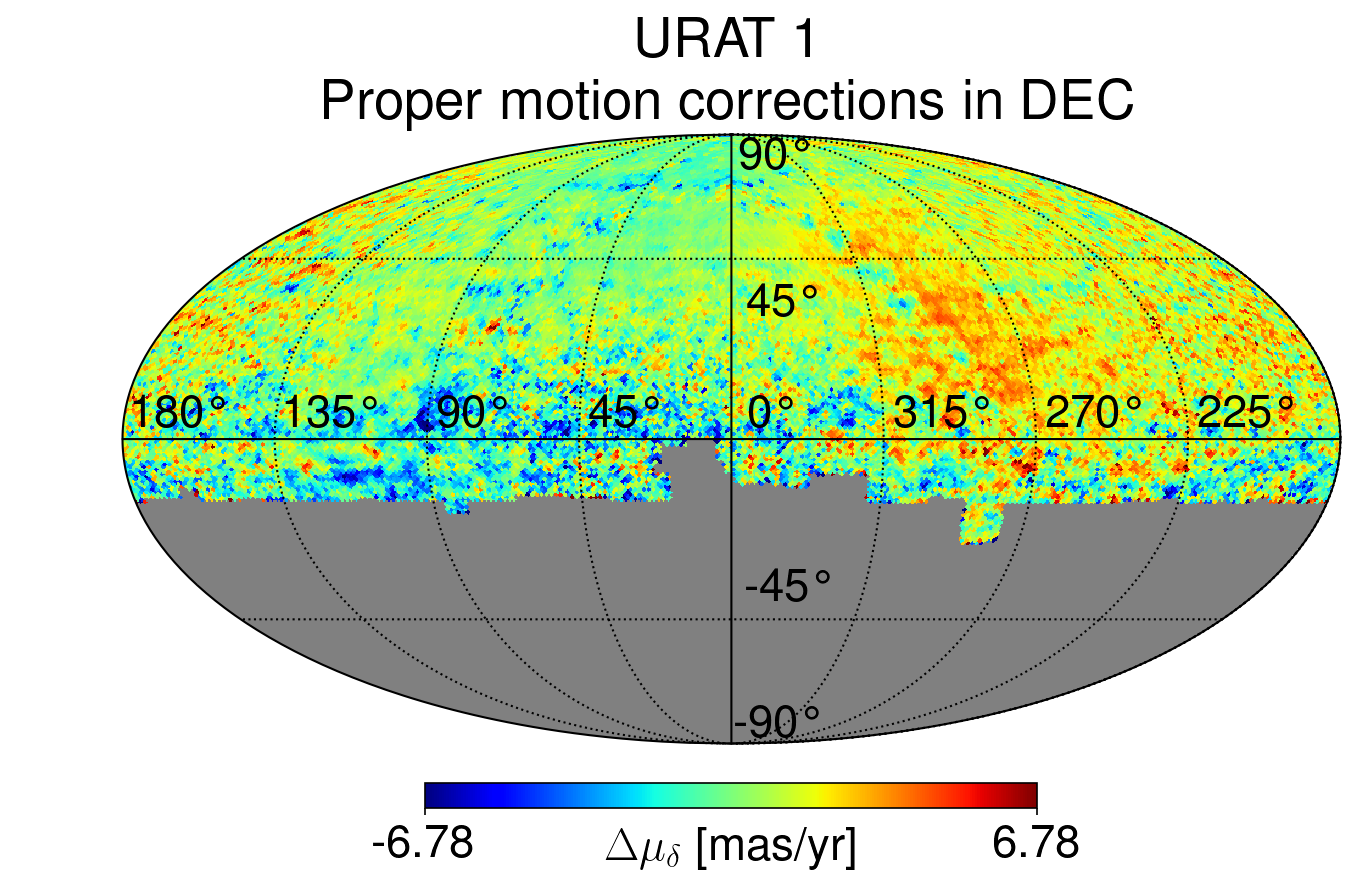}
 \end{tabular}
 \caption{Corrections in stellar positions and proper motion for the URAT 1 catalog. \label{fig:urat1}}
 \end{figure}

 The appendix shows local position and proper motion systematics for stellar catalogs not presented in the main body of the article (Figures \ref{fig:2mass}-\ref{fig:urat1}). For catalogs without proper motion only the bias values in right ascension and declination are provided. Proper motion corrections for those catalogs resemble those of USNO A2.0 displayed in Figure \ref{fig:usnoa2}.

% \begin{figure}
% \begin{tabular}{ll}
% \includegraphics[width=0.5\linewidth]{figs/hipparcos_0.png} &
% \includegraphics[width=0.5\linewidth]{figs/hipparcos_1.png} \\
% \includegraphics[width=0.5\linewidth]{figs/hipparcos_2.png} &
% \includegraphics[width=0.5\linewidth]{figs/hipparcos_3.png}
% \end{tabular}
% \caption{Corrections in stellar positions and proper motion for the Hipparcos catalog. %\label{fig:hip}}
% \end{figure}

\end{document}